\documentclass{aastex}
\usepackage{emulateapj5,apjfonts,psfig}
\newcommand{\beq}{\begin{equation}}
\newcommand{\eeq}{\end{equation}}
\newcommand{\beqn}{\begin{eqnarray*}}
\newcommand{\eeqn}{\end{eqnarray*}}

\newcommand{\ovl}{\overline}
 \def\Msun{\ifmmode{M_\odot}\else$M_\odot$\fi}

 \def\angstr{\ifmmode{\rm \AA}\else\AA\fi}
 \def\mbh{\ifmmode{M_{bh}}\else$M_{bh}$\fi}
 \def\mbhsig{\ifmmode{M_{bh}-\sigma}\else$M_{bh}-\sigma$\fi}
 \def\mbulge{\ifmmode{M_{b}}\else$M_{b}$\fi}
 \def\lbulge{\ifmmode{L_{b}}\else$L_{b}$\fi}
 \def\mlb{{\ifmmode{M_{bh}/L_b} \else $M_{b}/L_b$\fi}}
 \def\mub{{\ifmmode{M_{bh}/(\Upsilon L_{b})}
        \else$M_{b}/(\Upsilon L_b)$\fi}}
\newcommand{\parcs}{.\hspace{-0.09cm}$''$}   % \farcs in AASTEX
\newcommand{\kms}{\ifmmode \hbox{ \rm km s}^{-1} \else{ km s$^{-1} $}\fi}
\newcommand{\kmsp}{~km~s$^{-1}$.\ }
\newcommand{\kmsc}{~km~s$^{-1}$,\ }

\newcommand{\etal}{et~al.\ }

\newcommand\simlt{\lesssim}     % In amstex, works in AASTEX. Must be in "$".
\newcommand\simgt{\gtrsim}      % In amstex, works in AASTEX. Must be in "$"
\newcommand{\afour}{$a_4/a$~}

\received{2002 June 20}
\begin{document}

\title{
Kinematics of Ten Early-Type Galaxies From {\it HUBBLE SPACE TELESCOPE} and
Ground-Based Spectroscopy$^{1}$
}
     % Telescope, obtained at the Space
     % Telescope Science Institute, which is operated by the Association
     % of Universities for Research in Astronomy, Inc., under NASA
     % contract NAS 5-26555. These observations are associated with
     % proposal \# 7388. }

\author{ JASON PINKNEY\altaffilmark{2}$^,$\altaffilmark{3},
 KARL GEBHARDT\altaffilmark{4},
 RALF BENDER \altaffilmark{5},
 GARY BOWER \altaffilmark{6},
 ALAN DRESSLER \altaffilmark{7},
 S. M. FABER \altaffilmark{8},
 ALEXEI V. FILIPPENKO \altaffilmark{9},
 RICHARD GREEN \altaffilmark{10},
 LUIS C. HO \altaffilmark{7},
 JOHN KORMENDY \altaffilmark{4},
 TOD R. LAUER \altaffilmark{10},
 JOHN MAGORRIAN \altaffilmark{11},
 DOUGLAS RICHSTONE\altaffilmark{2},
 SCOTT TREMAINE \altaffilmark{12}}

\altaffiltext{1}{Based in part on observations made with the NASA/ESA 
{\it Hubble Space Telescope}, obtained at the Space
Telescope Science Institute, which is operated by the Association
of Universities for Research in Astronomy, Inc., under NASA
contract NAS 5-26555. These observations are associated with
proposal \# GO-7388. }
\altaffiltext{2}{ Department of Astronomy, University of Michigan,
 500 Church St., Ann Arbor, MI 48109; 
 jpinkney@astro.lsa.umich.edu, dor@astro.lsaumich.edu }
\altaffiltext{3}{ Department of Physics and Astronomy, Ohio Northern University,
Ada, Ohio 45810;
 j-pinkney@onu.edu}
\altaffiltext{4}{ The University of Texas at Austin, Department of
Astronomy, Austin, Texas 78712;
 gebhardt@astro.as.utexas.edu, kormendy@astro.as.utexas.edu }
\altaffiltext{5}{ Universit\"{a}ts-Sternwarte, Scheinerstra{\ss}el,
 M\"{u}nchen 81679, Germany;
 % Ludwig-Maximilians-Universitaet Scheinerstr. 1
 %               D-81679 Muenchen,
 %               Germany \\
 bender@usm.uni-muenchen.de }
\altaffiltext{6}{Computer Sciences Corporation, Space Telescope Science
Institute,
3700 San Martin Drive, Baltimore, MD 21218; bower@stsci.edu }
\altaffiltext{7}{The Observatories of the Carnegie Institution of
Washington,
 813 Santa Barbara St., Pasadena, CA 91101;
 dressler@ociw.edu, lho@ociw.edu }
\altaffiltext{8}{ UCO/Lick Observatory,
 University of California, Santa Cruz, CA 95064;
 faber@ucolick.org}
\altaffiltext{9}{ Department of Astronomy, University of California,
 Berkeley, CA 94720-3411;
 alex@astro.berkeley.edu }
\altaffiltext{10}{ Kitt Peak National Observatory, National Optical
Astronomy Observatories, P.O. Box 26732, Tucson, AZ  85726;
 green@noao.edu, lauer@noao.edu }
\altaffiltext{11}{ University of Durham, Science Laboratories, South Road,
Durham DH1 3LE, England; John.Magorrian@durham.ac.uk }
\altaffiltext{12}{ Princeton University Observatory, Peyton Hall, Princeton,
NJ
 08544;
 tremaine@astro.princeton.edu}

%\clearpage

\begin{abstract}

We present stellar kinematics for a sample of 10 early-type galaxies
observed using the Space Telescope Imaging Spectrograph (STIS) aboard the
{\it Hubble Space Telescope},
 and the Modular Spectrograph on the MDM Observatory 2.4-m telescope.
These observations are a part of an ongoing program to understand the
co-evolution of supermassive black holes and their host galaxies.
Our spectral ranges include either the calcium triplet absorption lines
at 8498, 8542, and 8662 \AA , or the Mg {\it b} absorption at 5175 \AA .
The lines are used to derive
line-of-sight velocity distributions (LOSVDs) of the stars
using a Maximum Penalized Likelihood method.
We use Gauss-Hermite polynomials to parameterize the LOSVDs
and find predominantly negative $h4$ values (boxy distributions)
in the central regions of our galaxies.
One galaxy, NGC 4697, has significantly positive central $h4$ (high tail
weight).
The majority of galaxies have a central velocity dispersion excess in the
STIS
kinematics over ground-based velocity dispersions.
The galaxies with the strongest rotational support, as
quantified with $v_{MAX}$/$\sigma_{STIS}$, have the smallest dispersion
excess at STIS resolution.

The best-fitting, general, axisymmetric dynamical models
(described in a companion paper) require black holes in all cases,
with  masses ranging from 10$^{6.5}$ to $10^{9.3}$ \Msun.
We replot these updated masses on the \mbhsig\ relation,
and show that the fit to only these 10 galaxies has a slope
consistent with the fits to larger samples. 
The greatest outlier is NGC 2778, a dwarf elliptical with
relatively poorly constrained black hole mass.
The two best candidates for pseudobulges, NGC 3384 and 7457, do
not deviate significantly from the established relation between \mbh\ and
$\sigma $.
Neither do the three galaxies which show the most evidence of a recent
merger, NGC 3608, 4473, and 4697.

\end{abstract}


%\clearpage
%\newpage

\section{Introduction}

The wealth of new data from the Space Telescope Imaging Spectrograph (STIS)
aboard the Hubble Space Telescope (HST)
is dramatically improving our understanding of the central regions of
early-type galaxies.
Its long-slit design allows for more efficient measurement of galaxy
kinematics than the previous Faint Object Spectrograph.
The $\sim$0\farcs1 spatial resolution  of STIS
resolves the influence of a 10$^{8}$ \Msun\ black hole (BH) in
an $L^*$ galaxy (velocity dispersion $\sigma = 200$\kms ) out to $\sim$ 25
Mpc.

Our team has observed a sample of 10 nearby elliptical or lenticular
galaxies using STIS in order to
address many questions concerning the demographics
of supermassive black holes.
Are BHs present in the centers of all early-type galaxies?
With which properties of the host galaxy does the BH mass correlate?
What can this tell us about the evolution of BHs and galaxies, and
the relationship of inactive BHs in nearby galaxies to more distant
active galactic nuclei?
Our data can also address several issues in the evolution of galaxies.
For example, why is there a dichotomy between power-law and core
surface brightness profiles (Faber \etal 1997)?  Were cores created by the
scouring
action of merging black holes?  Why do a few ellipticals not fit well into
either category (Rest \etal 2001)?
Why are the properties of BH in bulges and {\it pseudobulges} so similar,
when these two types of stellar systems probably formed in quite
different ways (Kormendy \etal 2002).
Our sample contains galaxies with both core and power-law profiles.
It also contains two possible pseudobulges:
the low-luminosity S0 NGC 7457 and the S0 NGC 3384.

%%%%%%%%%%%%%%%%%%%%%%%%%%%%%%%%%%%%%%%%%%%%%%%%%%%%%%%%%%%%%%%%%%%%%%%%%%
% Fig 1

\begin{figure*}[t]
\centerline{\psfig{file=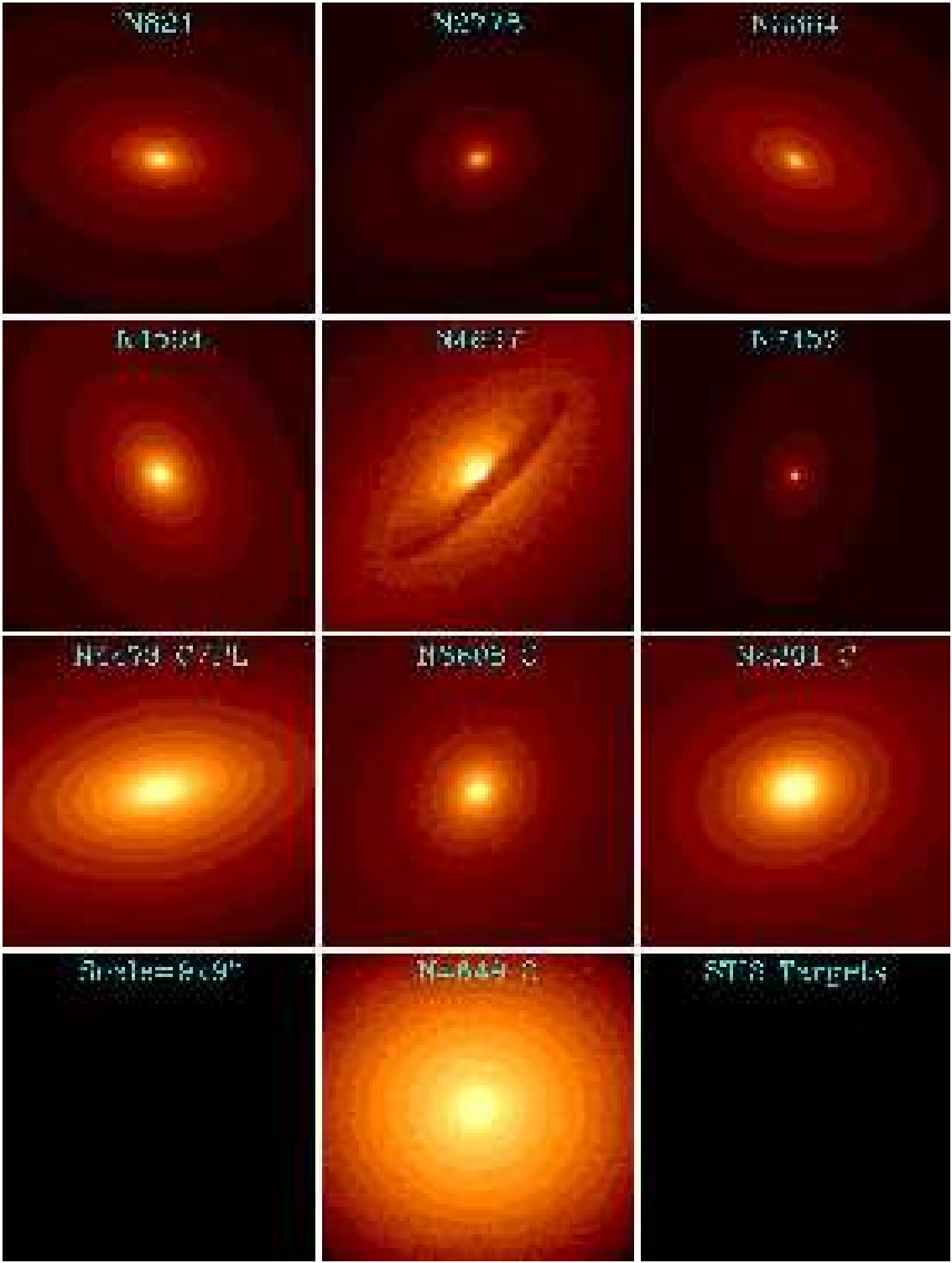,width=15cm,angle=0}}
\figcaption{{\it HST} images of our 10 early-type galaxies.  These are
$V$-band (F555W) WFPC2 images for all except NGC 4564 (F702W), and NGC
4697 (WFPC, deconvolved).  Each is a 200$\times$200 pixel subimage,
corresponding to 9\arcsec$\times$9\arcsec on the WFPC2 and WFPC
images.  The label ``C" indicates a core-type surface brightness
profile (Lauer \etal 1995), while power-law galaxies are labeled only
by their NGC number; NGC 4473 is intermediate between core and
power-law.
\label{collage}}
\end{figure*}

%%%%%%%%%%%%%%%%%%%%%%%%%%%%%%%%%%%%%%%%%%%%%%%%%%%%%%%%%%%%%%%%%%%%%%%%%%

The dataset presented in this paper has already been
instrumental in establishing a new, fundamental correlation
between the BH mass and stellar velocity
dispersion ($\mbhsig$,  Gebhardt \etal 2000 (G00);
Ferrarese \& Merritt 2000).
By combining our data with other published results, G00
demonstrate that the new correlation has less scatter than
the one between BH mass and bulge luminosity
(see Kormendy \& Richstone 1995;
Magorrian \etal 1998; Richstone \etal 1998; and van der Marel \& van den
Bosch 1998).
Here we give the details of data reduction leading to this result.
In a companion paper (Gebhardt \etal 2003), we discuss the modeling
technique
used to determine the BH masses, while Tremaine \etal (2002)
discusses the slope of the \mbhsig\ relation.

The present paper complements the modeling paper
by parameterizing the stellar line-of-sight velocity distributions
(hereafter LOSVDs)
with a Gauss-Hermite (GH) expansion.
The parameterization of LOSVDs with GH polynomials
has become common practice over the last decade as
increasing importance has been placed on the
precise shape of the stellar LOSVD
(van der Marel \& Franx 1993; Bender \etal 1994).
High-quality galaxy spectra have demonstrated that stellar LOSVDs
deviate significantly from a Gaussian distribution
(Bender 1990; Gerhard 1993; van der Marel \& Franx 1993).
For edge-on, rapidly rotating galaxies, the LOSVD
is often asymmetric. In other cases it has longer or
shorter tails compared to a Gaussian.
In the GH expansion, the first term measures the mean velocity and velocity
dispersion.
The second term measures
asymmetric deviations from a Gaussian, while the third term
measures symmetric deviations from a Gaussian.
Measuring the entire LOSVD, rather than just its first two moments
(mean velocity and velocity dispersion), constrains the phase-space
distribution of the stars and thus helps to reduce the ambiguities
in the mass distribution derived from stellar kinematics.
Although the fully general LOSVD is used in our BH modeling,
the GH parameters are a convenient means of describing an LOSVD.
They provide a check on the orbital anisotropies determined by
the modeling, and a means of comparison to previously published results.

The GH parameters can also be combined with photometric and kinematic
parameters to provide insight into elliptical galaxies.
For example, Bender, Saglia \& Gerhard  (1994, hereafter BSG) define mean
parameters
$\langle h_{3}\rangle$ and $\langle h_{4}\rangle$ and
find correlations between these and $a_{4}/a$, $v/\sigma$, $M_{B}$,
and $v/\sigma^{\ast}$ for 44 ellipticals.
The greater generality of the GH parameterization has been exploited in
deriving mass profiles from ground-based spectroscopy for
NGC 4342 (Cretton \& van den Bosch 1999),
NGC 3115 (Emsellem \etal 1999), NGC 1399 (Saglia \etal 2000),
NGC 2974 (Cinzano \& van der Marel 1994), and NGC 2434 (Rix \etal 1997).
In the case of M87, ground-based GH parameters have even been used to
constrain the mass of the central BH (van der Marel 1994a).
Many papers present GH parameters derived from ground-based
spectroscopy for large samples of galaxies (e.g., Kaprolin \& Zeilinger
2000;
Kronawitter \etal 2000; Fisher 1997).
Checks and comparisons to our ground-based GH parameters come largely from
the samples of BSG and Halliday \etal (2001), which have 5 and 3
galaxies in common, respectively.
STIS now enables us to examine the GH parameters at sub-arcsecond
resolution.
For NGC 1023 (Bower \etal 2001) and M32 (Joseph \etal 2001), STIS has
clearly revealed the influence of a BH on the stars.

This paper proceeds as follows.  We describe our galaxy sample in \S 2,
and the observations in \S 3. Section 4 details the
reduction of STIS and complementary ground-based data.  In \S 5, we
first describe how LOSVDs and GH parameters are derived from
our spectra, and then we present our kinematics for individual galaxies.
In \S 6 we will discuss our results.

\section{Sample Description}

A collage of all 10 target galaxies is shown in Figure \ref{collage}.
The galaxies were chosen from those with reliable
surface-brightness distributions at {\it HST} resolution.
We attempted to include a large range of luminosity and to
exclude problematic objects such as those containing obscuring
dust.  (The galaxy NGC 4697 shows a dust disk which extends
out to 3\farcs4, but the inner 1\farcs0 does not appear
to be seriously obscured.)

The stepped grayscales are used in Figure \ref{collage}
to show the variety of isophote shapes.  Some galaxies, like
NGC 821, have ``disky" isophotes that will produce positive
parameters $a_{4}/a$, where $a_{4}$ is the fourth cosine
coefficient in the Fourier expansion of the radial deviations, and
$a$ is the semimajor axis of the isophote (Lauer 1985; Bender 1988).
Others, like NGC 4291, have ``boxy" isophotes that will produce
negative $a_{4}/a$.  Boxy isophotes are typically found in galaxies
with ``core" surface brightness
profiles (Faber \etal 1997).  A ``core" profile has a break between inner
and outer slopes and the inner logarithmic slope,
$\gamma \equiv -d \log I/d \log r$,
must be less than 0.3.  Similarly, ``disky" isophotes are found
in ``power-law" galaxies, which have no significant break and have 
steeper slopes.  Our sample has six power-law
and four core galaxies.  We count NGC 4473 as a core because of its
surface brightness profile,
but it shares many properties with power-law galaxies (see \S 5).
The core galaxies are grouped on the bottom of Figure \ref{collage}.

The most luminous core galaxies tend to
have a large velocity dispersion and low surface brightness, making
measurement of the LOSVDs difficult. Thus, we selected three core
galaxies with relatively high
surface brightness (NGC 3608, 4291, and 4473).
However, we selected one example of a relatively low surface brightness
core galaxy, NGC 4649, which has
$\mu_{V} \approx$ 15.9 mag arcsec$^{-2}$ at 0\farcs1.  It is comparable to
M87
in luminosity, and it serves as a test case for measuring
absorption-line kinematics
in low surface brightness giants.

%\placetable{c7targets}

%\placetable{specs}

\section{Observations}

\subsection{STIS Observations}

Table \ref{specs} gives specifications for the spectrograph/grating
combinations used.   Our STIS observations used only the G750M
grating and the STIS CCD detector.  The STIS CCD is a 1024$\times$1024
pixel CCD with readout noise around 3.8 $e^-$ at a gain of 1.0.  We binned
the
CCD by a factor of 2 along the dispersion axis in all of our
observations to improve the signal-to-noise ratio (S/N).  This raised our
reciprocal dispersion to 1.1 \AA\ pix$^{-1}$.
STIS has a spatial scale of 0\farcs0507 pix$^{-1}$ in all of the
configurations used here.
 % and slightly larger (0\farcs056) perpendicular to the slit.

Table \ref{stisobs} gives the details of the {\it HST} observations.
Typically, 6 orbits were devoted to each galaxy
($\approx$ 2700 seconds per orbit).
However, the galaxies NGC 821, 3384, and 4697 had 11 orbits, and the galaxy
NGC 4649 had 18 orbits because of its low surface brightness.
Each orbit was divided (CR-split) into two, 1350-s
exposures except the first orbit of a visit, which had
shorter exposures.  For the first 2 visits of the program
(NGC 4473 and 821), the galaxy center was only dithered slightly
(at the sub-pixel level) on the chip between orbits.
Partway through the program we learned that the CCD had
a rapidly varying population of hot and warm pixels.
We then requested wider separations between dithers (20 pixels)
so that a dark frame could be constructed from the data.
Five dither positions were used.
The second visit to NGC 821 used wide dithers, so only NGC 4473
did not receive any wide-dither observations.

For one galaxy, NGC 4697, two 1050 s exposures were taken with
the G750M in setup 3 (Table \ref{specs}). Here, the spectral
range is centered near $\lambda_{cen}$ = 6581 \AA . We
see H$\alpha$+[N II] emission lines which allow a useful
comparison of gas and stellar kinematics.
The reduction and analysis of these data is described elsewhere
(Pinkney \etal 2003).

%\placetable{stisobs}

Table \ref{stisobs} includes a K3 III and a G8 III star observed by
us.  We use these as LOSVD-fitting templates,
for assessing aperture illumination corrections, and for measuring the STIS
point spread function (PSF).
We also used HR7615 from Bower \etal (2001).
The spectrum of the G8 III star (HR6770) is shown in Figure \ref{specalln1}.
The template stars were observed through the same slits that we
used for our galaxies.  We used two slit apertures in our STIS setups:
52\arcsec $\times$0.1\arcsec\ and 52\arcsec $\times$0.2\arcsec .
The aperture point-source throughputs are 64\% and 77.8\%, respectively,
at 8500 \AA .  The 0\farcs1 slit required a PEAKUP while the 0\farcs2 slit
did not.  We used the 0\farcs2 slit for the four core galaxies (NGC 3608,
4291,
4473, and 4649) and the 0\farcs1 slit for the six power-law galaxies.

\subsection{Ground-Based Observations}

 % Include stuff about the setup here.

We used the Modular Spectrograph (hereafter, Modspec) on the
2.4-m Hiltner Telescope at MDM Observatory for longslit spectroscopy
(Table \ref{gndobs}).
The camera was equipped with either a thick, frontside illuminated
LORAL 2048$^2$ CCD (``Wilbur"), or a thinned, backside illuminated
SITE 1024$^2$ CCD (``Charlotte").  The LORAL chip had 4.7 $e^-$ readout
noise,
while the SITE had 5.45 $e^-$.
The Modspec Ca II setup produces comparable spectral resolution to our
STIS + G750M setups (Table \ref{specs}).
Its spatial resolution is worse, of course, varying with seeing in the range
0\farcs6--2\farcs0.
Our CCDs had pixel scales of 0\farcs371 pix$^{-1}$ (Wilbur) and
0\farcs59 pix$^{-1}$ (Charlotte).

%\placetable{gndobs}

Multiple exposures were taken of each galaxy, with
lower surface brightness galaxies receiving more exposures.
The slit was placed along
the major axis and at least one other position angle (Table \ref{gndobs}).
Small dithers of the galaxy along the slit were used to remove CCD
defects.  The galaxy observations were bracketed by star
observations to help monitor the seeing, and to build a library of
template stars for defining the broadening functions in the galaxy spectra.
Template exposures were generally $<$ 1 minute.
Calibration frames included Ar comparison lamp spectra, continuum
lamp spectra, and twilight sky spectra.
Guiding used either starlight that was reflected off of the slit jaws or
direct guiding from stars outside of the slit-viewing area.
The root-mean-square (rms) deviations on the sky during the guiding were
typically
around 0.3\arcsec. We often have a star in the slit along
with the galaxy.
We can compare the FWHM of the star during a long exposure
with the FWHM of a star during a short exposure
to check the guiding stability. For a 20--minute exposure, the
guiding added about a 10--20\% increase in the FWHM.

\section{Data Reduction}

\subsection{G750M Ca II Spectra }

The STIS data reduction was done with our own FORTRAN programs and FITSIO
subroutines (Pence 1998).  
Our primary reasons for not using CALSTIS (Hodge et al. 1998) within IRAF 
are 1) a better dark frame can be created out of the data than the weekly 
darks used by CALSTIS, 2) many features of CALSTIS were not important to
us, such as heliocentric correction, 3) we did not want 2D rectification 
to be unnecessarily complex.

We began by extracting the raw spectra from the multi-dimensional
FITS file.  To remove the bias level, we subtracted a constant fit to
the overscan region.  The overscan was about 1110 data numbers (DN) for
galaxy spectra, and 1505 DN for flats and comparison lamps, with little
variation from one visit to the next.

Subtraction of an accurate dark current was important because the
STIS CCD has warm and hot pixels which evolve on timescales of $\simlt$ 1
day.
The mean dark current was about 0.007 s$^{-1}$ pix$^{-1}$ and this produced
about 10 DN pix$^{-1}$ in our exposures.  We used an archival dark only for
the 2 visits that were not dithered (NGC 4473 and 821);
for the rest, we created a dark out of the data.
This {\it self-dark} required many iterations.
The first dark was a biweight combination (see Beers \etal 1990)
of the individual exposures after
masking the pixels within $\sim$25 rows of the galaxy peak.
The first dark, although imperfect, was
subtracted from each spectrum.  We proceeded with the
flat-fielding, bad-pixel flagging, and un-dithering,
until a 2D spectrum can be made by combining the exposures.
This 2D galaxy spectrum was then de-flattened, dithered, and subtracted from
the raw spectra. The results contained the dark current, cosmic rays,
and some residuals from the first, rough, dark subtraction.
These were again biweight combined to form an improved dark.
We then repeated reduction with this dark frame to form a second
2D spectrum which was again subtracted from the raws for
a third {\it self-dark} iteration.  At least 5 such iterations
were performed to remove residual galaxy signal from the dark.
A slice along the spatial axis showed the galaxy profile
to dwindle to zero if low-level dark current was correctly
subtracted from the final galaxy 2D spectrum.  This dark also
corrected for any bias pattern on the chip.
The final galaxy spectrum was typically
more free of deviant pixels than if an archival dark was used.

\vskip 10pt
%%%%%%%%%%%%%%%%%%%%%%%%%%%%%%%%%%%%%%%%%%%%%%%%%%%%%%%%%%%%%%%%%%%%%%%%%%
% Fig 2
\hskip -30pt{\psfig{file=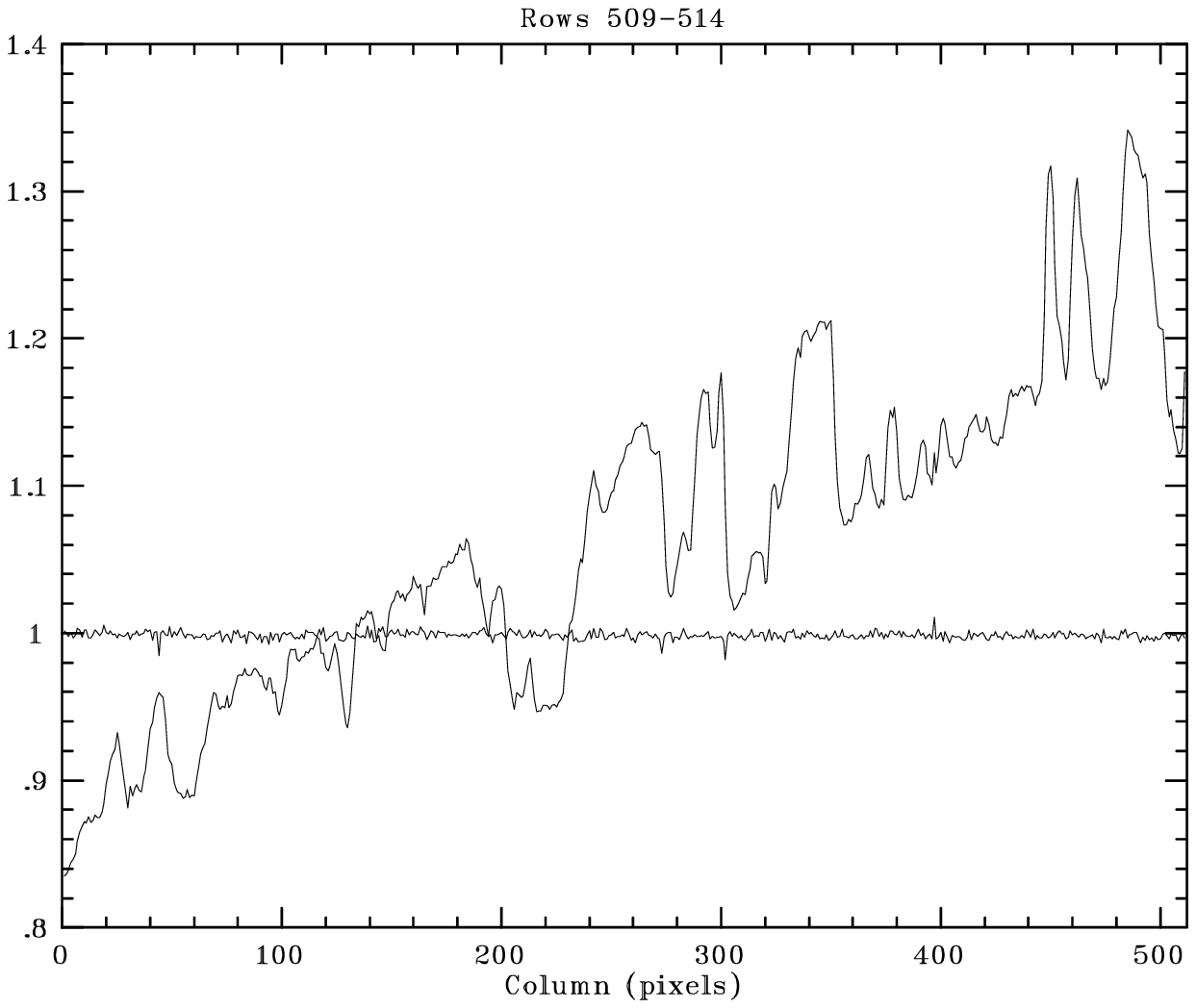,width=10.5cm,angle=0}} \figcaption{The
ratio of two CCD flats compared with the original.  The wavy line is
an average of 6 rows across the middle of the combined flat for NGC
3384. It was taken with the 52\arcsec x0.1\arcsec\ aperture and the
G750M 8561 \AA\ setup.  The relatively flat line is the result of
dividing this flat by another taken 9 months later with the same
setup.
\label{ccdflats}}
%%%%%%%%%%%%%%%%%%%%%%%%%%%%%%%%%%%%%%%%%%%%%%%%%%%%%%%%%%%%%%%%%%%%%%%%%%
\vskip 10pt

\begin{figure*}[t]
\centerline{\psfig{file=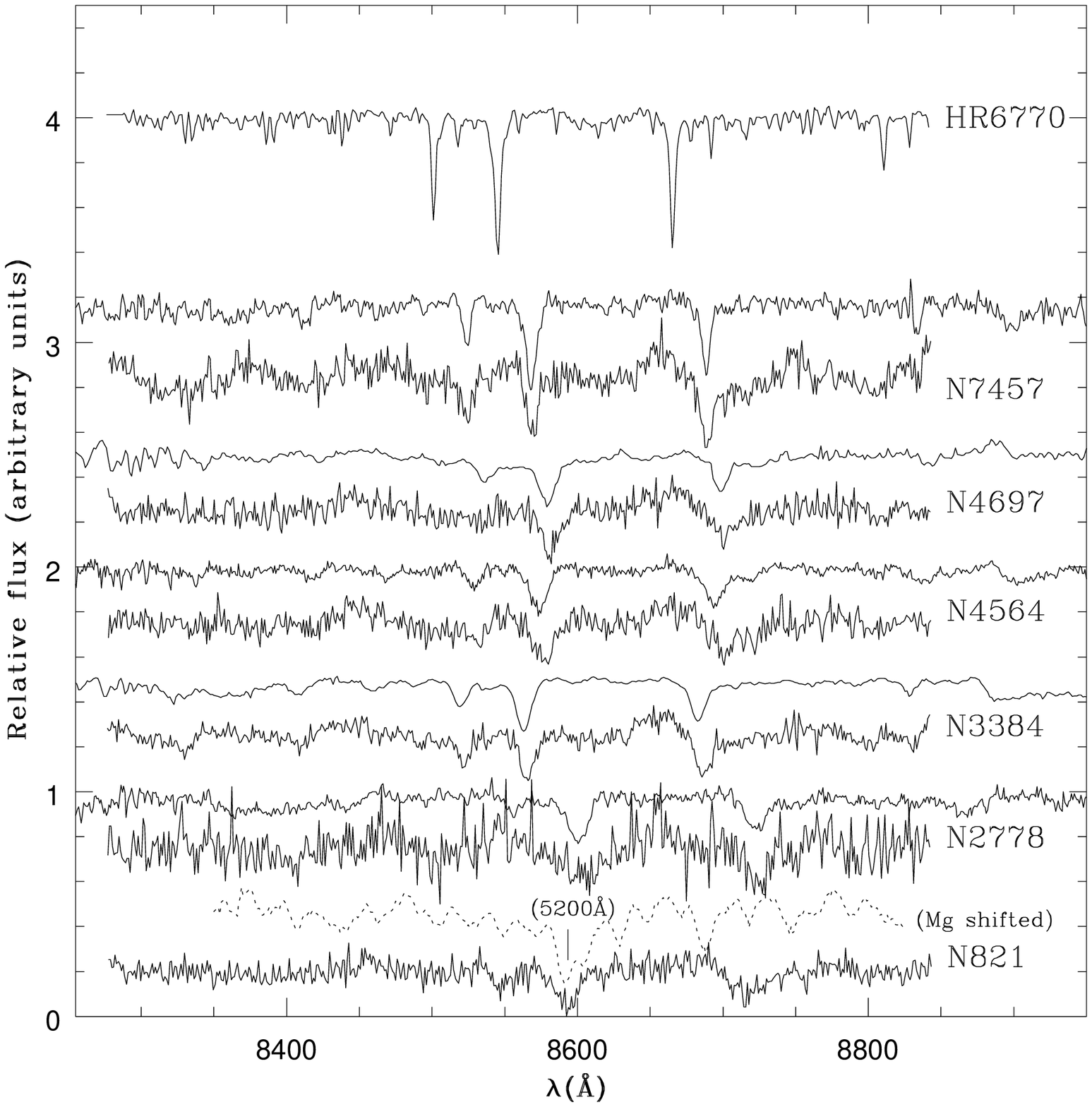,width=16cm,angle=0}}
\figcaption{Collage of spectra extracted from the central spatial bin
for each {\em power-law} galaxy.  For each galaxy, there is a pair of
spectra: the top one is from Modspec and the bottom is from STIS.  The
spectra are the average (biweight) of all exposures.  The STIS spectra
have been normalized by a linear or parabolic fit to the continuum,
while the Modspec spectra were divided by a fit to the local continuum
in 100-pixel wide regions, with linear interpolation.  The spectra are
shifted vertically to avoid overlap.  The template star HR6770 is
shown on top for comparison.  For NGC 821, there is no ground-based
calcium triplet spectrum so we have shifted the Mg (5175 \AA )
spectrum into the plotted wavelength range.
\label{specalln1}}
\end{figure*}

\begin{figure*}[t]
\centerline{\psfig{file=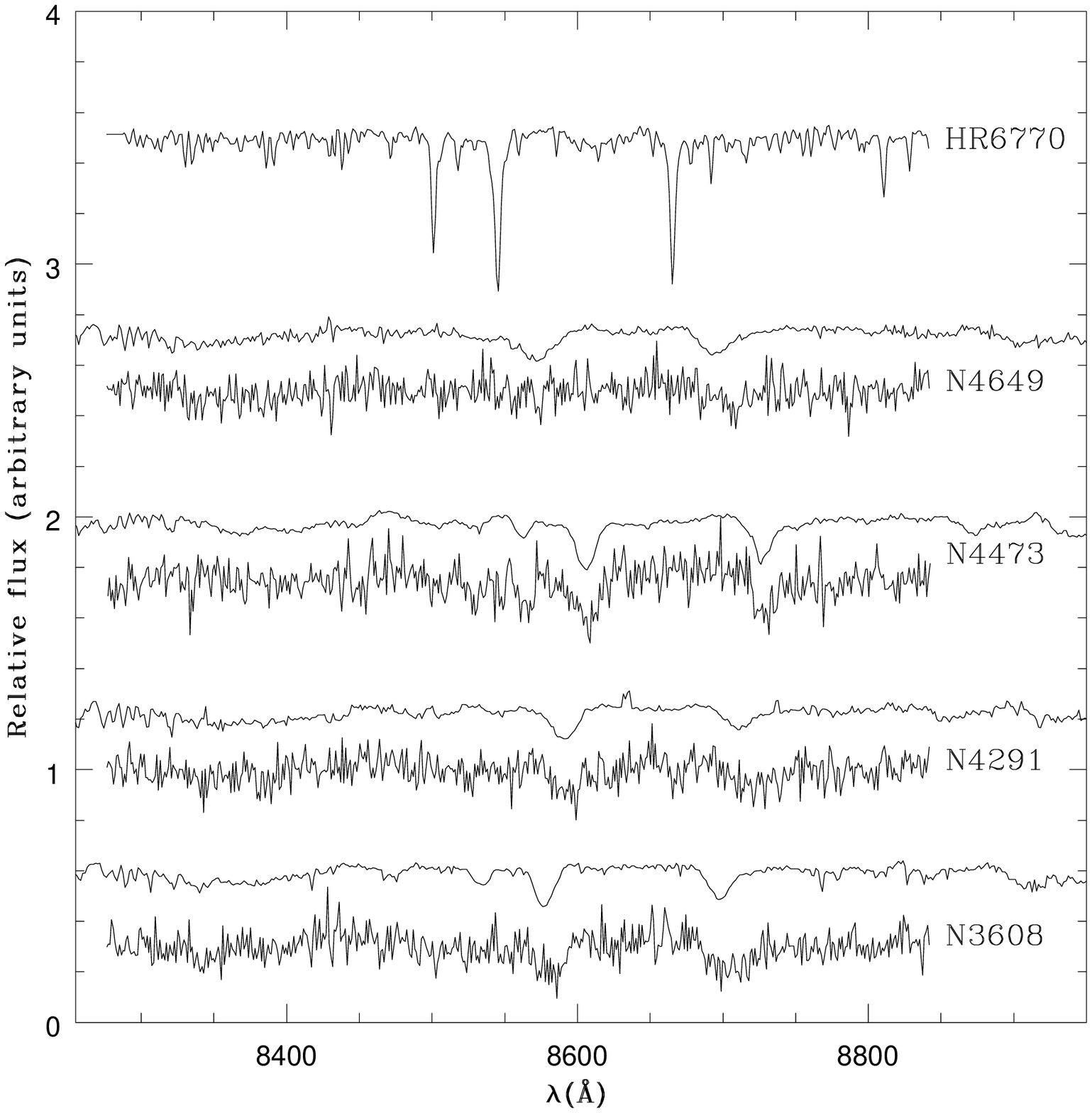,width=16cm,angle=0}} \figcaption{Same
as Figure \ref{specalln1}, except here we show spectra from each {\em
core} galaxy.
\label{specalln2}}
\end{figure*}

Within our self-dark iteration loop, the dark subtraction was
followed by flat-fielding.
The flat-fielding was important because interference fringes
become strong for $\lambda > $7500 \AA .  Our raw data had
about 10\% peak-to-peak variations due to fringing (Fig. \ref{ccdflats}).
We used a flat which was a combination of the contemporaneous tungsten
flats.  These are intended to reduce fringe effects to
the 0.9\% level  (Leitherer \etal 2001).  % p. 98 of STIS I.H.
It was difficult to measure fringe residuals in our galaxy spectra,
but ratios of flats taken within a visit showed no detectable fringe
pattern.  Also, a ratio of flats taken 9 months apart through the 52\arcsec
$\times$0.1\arcsec\
aperture contained a residual fringe pattern at the $\lesssim$1.0\%
level (Fig. \ref{ccdflats}).
The stability of these tungsten flats is one clue to their effectiveness
in our reduction.  Another is the shape of the final spectra (Figures
\ref{specalln1} and \ref{specalln2}), especially
for standard stars, which tend to show a linear continuum with only
familiar features superimposed.  The overall slope of the galaxy continuum
does not have to be perfectly accurate for our purposes, so any color
mismatch between the tungsten flat and our galaxies is not problematic.

%%%%%%%%%%%%%%%%%%%%%%%%%%%%%%%%%%%%%%%%%%%%%%%%%%%%%%%%%%%%%%%%%%%%%%%%%%
% Fig 3
%%%%%%%%%%%%%%%%%%%%%%%%%%%%%%%%%%%%%%%%%%%%%%%%%%%%%%%%%%%%%%%%%%%%%%%%%%

%%%%%%%%%%%%%%%%%%%%%%%%%%%%%%%%%%%%%%%%%%%%%%%%%%%%%%%%%%%%%%%%%%%%%%%%%%
% Fig 4
%%%%%%%%%%%%%%%%%%%%%%%%%%%%%%%%%%%%%%%%%%%%%%%%%%%%%%%%%%%%%%%%%%%%%%%%%%

After flat-fielding, the spectra had to be vertically shifted to a
common dither, combined, and rotated.  The shifting
was done by measuring the peak of the galaxy profile
in a 100-column wide region at the center of the CCD.
The peak was measured with 1/9 pixel
precision for each frame using a cubic spline interpolation.
The spectra were shifted
vertically to the same row of pixels, y=511.0.  Fractional
pixel shifts were allowed in order to provide an
accurate superposition, but this created {\it smearing}
(redistribution of counts to neighboring pixels) along
columns.  The aligned spectra were then combined using
the biweight to filter cosmic-ray hits.
Since there were five dither positions, every final pixel value
was a combination of at least five distinct pixels on the CCD chip.
The resulting spectrum has a dispersion with
a 0.6\arcdeg\ clockwise tilt relative to the CCD rows.
This was removed in the last step using a rotation
about the chip center.  The rotation adds additional
smearing which is spatially periodic, i.e., regions
of minimal smearing occur every $\sim$95 pixels.
The center of the spectrum, $\lambda$=8559 \AA , was a
smearing minimum which happens to fall near the
Ca II 8542 \AA\ line for our galaxies.

\begin{figure*}[t]
\centerline{\psfig{file=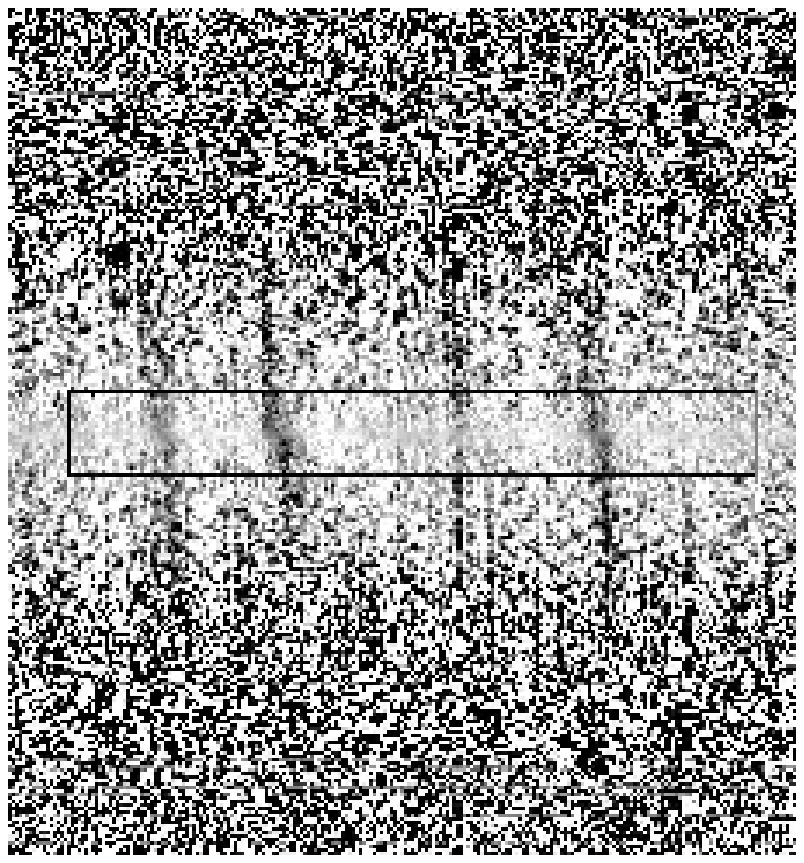,width=11.5cm,angle=0}} 
\figcaption{
Unsharp-masked 2D spectrum for NGC 3384.  Only a
subsection of the CCD is shown. The box is 2\arcsec\
tall and shows the vertical extent of the extracted 1D spectra.
The Ca II triplet absorption is visible in three vertical bands.
\label{unsh3384} }
\end{figure*}

The final 2D spectra contain imperfections.
Figure \ref{unsh3384} is an example
of a final unsharp-masked, dithered dataset for NGC 3384.
The box overlay on NGC 3384 shows the $\pm1\arcsec$ extent of our 1D
extractions
from the galaxy center.
There is a low-level ($\sigma \approx 1.0$ DN) pattern noise apparent
on
Figure \ref{unsh3384}.
It is possibly caused by residuals from dark-subtraction and/or
flat-fielding
which are then spatially repeated when the 5 dithers are combined.
The dark contains charge trails along columns which become more numerous
toward the bottom of the chip.  These features do not all subtract well
leaving linear features on the spectrogram.
Other imperfections are not easily visible.  There are typically
a dozen bright features between the fiducial bars for an average
of ten, 1350-s exposures.  These are 1-2 pixel events smeared
by shifting to cover an area of $\sim 4$ pixels.  They probably originate
in pixels that contain cosmic rays in many of the exposures.
(Cosmic ray hits occur on about 5\% of pixels on the raw frames.)

%%%%%%%%%%%%%%%%%%%%%%%%%%%%%%%%%%%%%%%%%%%%%%%%%%%%%%%%%%%%%%%%%%%%%%%%%%
% Fig 5
%%%%%%%%%%%%%%%%%%%%%%%%%%%%%%%%%%%%%%%%%%%%%%%%%%%%%%%%%%%%%%%%%%%%%%%%%%

We extracted 1--dimensional (1D) spectra from the final spectrogram
using a biweight combination of rows.  Our standardized
binning scheme (Table \ref{bins}) used 1-pixel wide bins
near the galaxy center to optimize spatial resolution.
All of the bins were contained within 1\farcs0 (20 pixels) of the galaxy
center.
Our dither pattern positioned the galaxy center on the rows y=471, 491, 511,
531
and 551, so that the residuals of the fiducial bars fell as near
as 9\farcs2 to the galaxy center.
These bars did not interfere with
our spectroscopy because the S/N generally becomes too
poor to measure kinematics by $r$=1\farcs5
for our galaxies.
Since our spectral extractions went out
20 pixels from the galaxy center and our maximum dithers were 40 pixels,
all of our spectra originate between y=450 and 570 on
the chip.  This simplified several other aspects of the
reduction, including wavelength calibration.

A wavelength solution was obtained for the final galaxy spectra
by summing the rows between y=450--550 on each of the Pt-Cr/Ne
comparison lamp frames.  There were typically six unique comparison
spectra per visit.
The final solution was an average of these six solutions.
For each solution, we fitted a line to wavelength vs.~pixel for 12 
lamp lines and got residuals with rms$\approx $0.12 \AA .
The lines were the 12 brightest after omitting a close
double.  A typical solution was y(\AA )=1.10896$\pm0.00004$x +
8276.26$\pm0.04$, the errors are the standard deviation from
the six solutions.
Over the region occupied by the galaxy spectra (y=450--570),
the lamp lines exhibit a slight bowing on the CCD which is asymmetric
with respect to the central row.
We considered errors caused by this bowing.  The top dither at y=551
has a zeropoint that differs by about +0.17 \AA\ from that of the
central dither (y=511), and the bottom dither at y=471 differs by
about $-0.11$ \AA .  Thus, by combining vertically shifted dithers
we artificially broaden the lines.  The width of the lines
increases by less than 0.5\% as a result of this co-addition.  Since
the Ca absorption lines are broader than the lamp lines, their
fractional change in width should be even less.
The errors in the lamp-line centroids caused by the co-addition are
small because the shift
caused by the top dither will be nearly compensated by the
shift from the bottom dither.  Any remaining zeropoint error
from the asymmetry is $ \simlt $ 0.05 \AA .  Finally,
the galaxy spectrum was rotated by $\sim$ 0.6\arcdeg\ while the
comparison spectrum was not. This introduced only negligible
errors ($ \simlt $0.01 \AA ) in line positions.

%%%%%%%%%%%%%%%%%%%%%%%%%%%%%%%%%%%%%%%%%%%%%%%%%%%%%%%%%%%%%%%%%%%%%%%%%%
% Fig 6
\begin{figure*}[t]
\centerline{\psfig{file=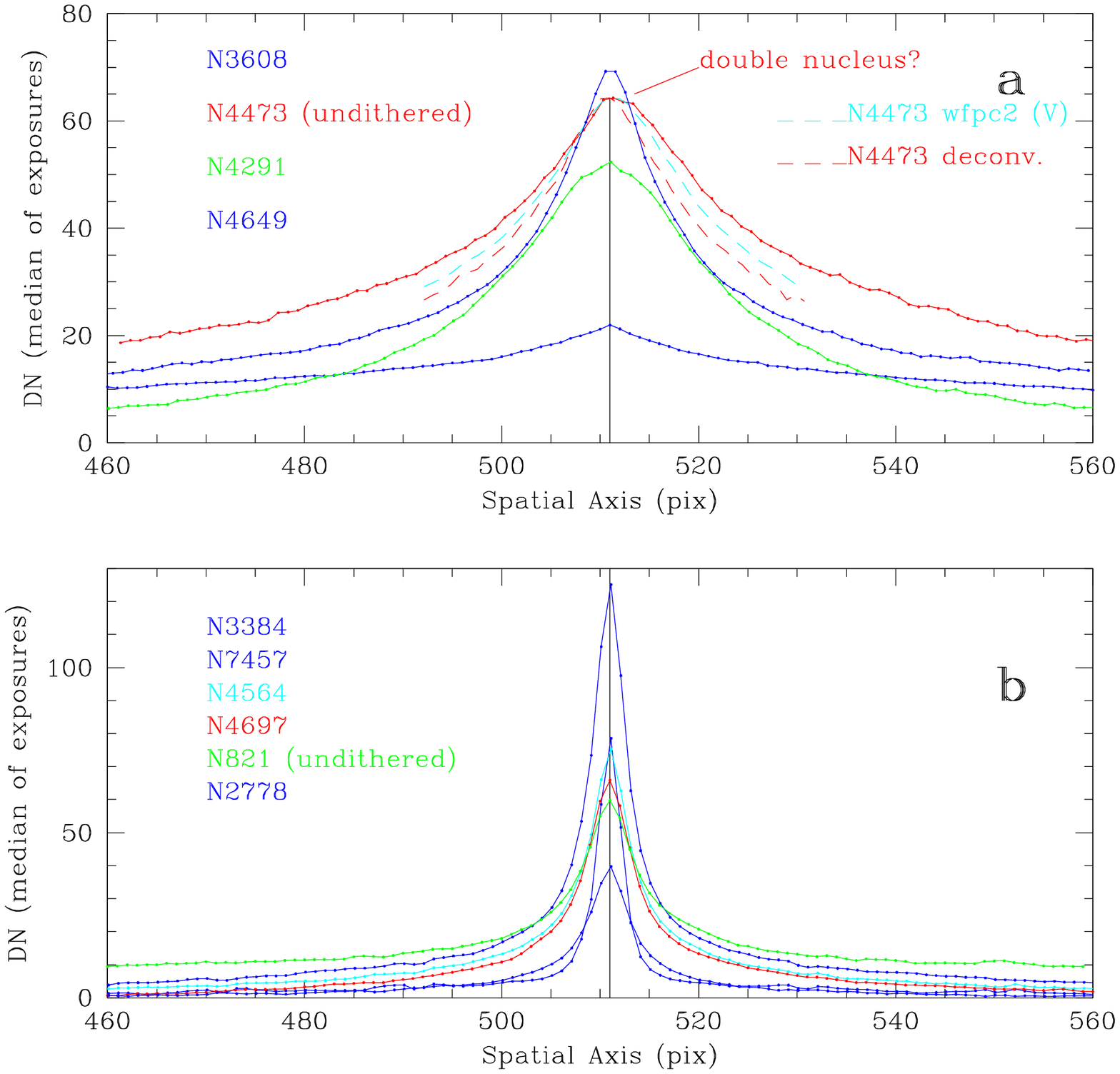,width=16cm,angle=0}} \figcaption{Galaxy
light profile along the STIS slit (roughly major axis).  Columns
200-300 on the STIS chip were averaged, corresponding roughly to the
bandwidth 8495 - 8605 \AA\ .  Panel (a) shows the core galaxies.
These were observed with the 52\arcsec x0.2\arcsec\ STIS slit.  The
``shoulder" in NGC 4473 is labelled (see \S 5.6).  Also, dashed lines
are overlayed showing the light in a simulated STIS slit over the
WFPC2 data for NGC 4473.  Panel (b) shows the power-law galaxies.
These were observed with the 52\arcsec x0.1\arcsec\ STIS slit.
\label{plotprofall}}
\end{figure*}
%%%%%%%%%%%%%%%%%%%%%%%%%%%%%%%%%%%%%%%%%%%%%%%%%%%%%%%%%%%%%%%%%%%%%%%%%%

\subsection{Template Spectra}

The template-star observations required a different reduction
than the galaxies.  Five exposures were taken through each of
the 52\arcsec x0.2\arcsec\ and 52\arcsec x0.1\arcsec\ slits.
The exposure times were 2.2 s,
so dark subtraction was unnecessary.  Between exposures, the star was
stepped
across the slit by 0\farcs04 and 0\farcs025 for the 0\farcs2
and 0\farcs1 slits, respectively.
Combining each set of five spectra allowed us to simulate the observation of
a diffuse source through both slits.
The combined spectrum was then rotated by 0.6\arcdeg.
Two lamp flats were obtained
with exposure times of 142 s (0\farcs1 slit) and 84 s (0\farcs2 slit).
The flats were also combined and rotated.
Next, 1D spectra were extracted from both the star and flatfield
spectrograms by averaging the 6 central rows.  The extracted flat
was normalized and then divided into the extracted stellar spectrum.
The resulting spectra (e.g., HR6770 in Fig. \ref{specalln1}) appear to be
corrected of fringing.

 The stepping of the stars across the slits allowed
us to measure the instrumental broadening of the Ca II triplet
absorption lines.
The first and last exposures in the 0\farcs2 slit had the star
positioned at opposite extremes in the slit, a pointing
difference of 0\farcs16 implying a 1.6 \AA\ centroid shift.
However, the measured difference
in centroids was only 1.07 \AA\ (38\kms ) in line centroids.  The
difference
presumably originates in asymmetric vignetting of the stellar PSF by the slit.
The Ca lines had widths of about 3.4, 5.5 and 3.9 \AA, for 8498 \AA, 8542 \AA\
and 8662 \AA, respectively, in the individual 2.2-s exposures.
The combined spectra showed only $\sim$0.1 \AA\ broadening when
summed in the 0\farcs2 aperture.  Thus, there is a $\simlt 5\% $
broadening of the narrowest stellar Ca line due to the wider of the
two slit widths.  The broadening of Ca lines in our target galaxies
will be even less.
Finally, the STIS PSF was also derived from the template star
exposures.  The method and results are identical to those described
by Bower \etal (2001).

\subsection{Ground-Based Spectra}

The Modspec data were also reduced with FITSIO routines.
The CCD frames were overscan corrected, bias subtracted,
and trimmed.  The flat frames were constructed out of the twilight
sky frames and the continuum-lamp frames as follows.  We derived a
small-scale structure flat from the continuum lamp.
The large-scale structure along the dispersion axis
was also taken from the continuum lamp.  The structure
along the spatial axis, however, was taken from the twilight frames.
The final flat was the product of these three individual
flats.

After flat-fielding, the spectra were rectified.  We used wavelength
solutions from neon and argon comparison lamps and the traces of
several standard-star spectra to define our geometric transformations.
Pixel smearing was not as much of a concern with the Modspec data as
with STIS because we do not rely on Modspec for high spatial
resolution.  Therefore, nonlinear transformations were used and
spectral images were shifted, aligned, and dispersion-corrected as
needed before combining into final 2D spectra.  Sky spectra were
defined on the edges of the chip, where galaxy light was minimal, and
subtracted from the 2D spectra.

Table \ref{gndobs} shows our exposure times for each position
angle (PA) for each galaxy.
We took between 2 and 23 exposures for a given PA to reach the
desired S/N.
As an example, a 1200-s exposure of an elliptical galaxy
(NGC 4564 is used here) with an {\it HST}-measured
surface brightness at 1\farcs0 of
$\mu_{V}$=16.0 mag arcsec$^{-2}$
will have a Ca spectrum with
about S/N$\approx$20.0 per \AA\ for a spectrum extracted from a 0\farcs37-wide
bin (one pixel on the {\it Wilbur} CCD).
For such a galaxy, 7 exposures will allow a S/N $\approx$ 50 per \AA\
to be reached in the central extraction.
This is easily sufficient for measuring the
$h3$ and $h4$ parameters of the LOSVD.
The outer parts of the galaxy can be measured with adequate S/N by using
broader extraction bins (see Table \ref{bins}).
In the case of NGC 4564, to reach S/N $\approx$ 25 (giving uncertainties
in $H3$ and $H4$ of $\pm$0.05)
at a radius where $\mu_V$=19 mag arcsec$^{-2}$, a 4-pixel wide bin and
7 exposures are sufficient.

\subsection{Imaging}

A luminosity density distribution is needed for the
modeling of each galaxy's mass distribution (see Gebhardt \etal 2003).
A non-parametric fit is made to the surface brightness
profile, and this is deprojected into luminosity density
following Gebhardt \etal (1996).
We used primarily WFPC2 images in filters F555W ($V$) and
F814W ($I$) from {\it HST} proposals PID 5512, 6099, 6587, and 6357.
For NGC 4697, we used pre-COSTAR WFPC1 data in F555W.
The images were typically sums of 4 or more exposures.
Starting with PID 6587, we used sub-pixel dithers for improved
spatial sampling.
All WFPC2 images were deconvolved with 40 iterations of Lucy-Richardson
deconvolution, while 80 iterations were applied to the WFPC images.
Our images are shown out to $r=4\farcs5$ in Figure \ref{collage}
and surface brightness profiles are shown in Gebhardt \etal (2003).
The profiles are actually a composite of {\it HST} and
ground-based data, where the ground-based profile is shifted by
a constant to match {\it HST} at a radius where seeing effects are
minimal.   The ground-based CCD photometry
was taken primarily from Peletier (1989).

For reference, the parameters of the best fit of the
photometry to the five-parameter Nuker law (Lauer \etal 1995)
are given in Table \ref{profiles}.
We show published values from Rest \etal (2001),
Ravindranath \etal (2001), and Faber \etal (1997), or, if
these are not available, our own fits to the $I$ and $V$
band WFPC2 images.
The Nuker law is a double power law given by
\beq
I(r)=2^{(\beta - \gamma)/\alpha}I_{b}\left ( \frac{r_b}{r} \right )^{\gamma}
\left[ 1+\left(\frac{r}{r_b}\right)^{\alpha} \right]^{(\gamma - \beta)/\alpha},
\eeq
where  $\gamma$ is the slope as $r \rightarrow 0$, $\beta$ is the slope
at large $r$, and $\alpha$ determines the sharpness of the break
between the two power laws at $r_b$.
These parameters are correlated (see Byun \etal 1996), and the
slippery nature of the 5-parameter fit allows for occasional
sporadic values.  These are fits to the major axis surface
brightness distribution.  If two sets of parameters are given for a
galaxy, we adopt the average for each parameter for the purpose
of plotting (\S 6).

%\placetable{profiles}

\subsection{STIS Light Profiles}

Our STIS spectra also provide information on
the near-infrared surface brightnesses of our galaxies.
Figure
\ref{plotprofall} shows intensity as a function of slit position.
The slits were approximately aligned with the major axes
(Table \ref{stisobs}).
We produced intensity profiles in order to check the shifts
of our galaxy peaks to a common position, row 511.0.  The profiles
are slices along the spatial axis of the 2D spectrum, which are
each a biweight of at least 10 exposures.  The slices are
101 pixels wide (columns 200-300).  The peak was
defined as the maximum of a smooth, cubic spline fitted to the
profile, thus allowing sub-pixel interpolation.
The final 2D spectra were always peaked within 0.4 pixels of row 511.0.
For the creation of Figure \ref{plotprofall}, the
fitted peaks were shifted to fall exactly on 511.0
to best demonstrate any asymmetry in the galaxies.

It is worth noting that these STIS datasets are valuable
as probes of surface brightness features because total exposure
times are over 10 times greater then typical WFPC2 surveys,
the near-IR wavelengths are less dust sensitive, and
saturation is avoided by dispersing the light.
The ordinate of Figure \ref{plotprofall} gives
the counts obtained in 1350 seconds (``DN" are equivalent to the
number of electrons because gain = 1).
However, at least 1010 pixels are averaged to determine
each point so that the Poisson errors are actually very small:
at 60 DN, the 68\% error is only 0.2 DN, or
about one-half the size of the plotted points.
The kinematics are derived only out to $\pm$ 20 pixels (1\farcs0).
% Figure \ref{plotsym} folds the profiles about the center and
% reveals an especially significant asymmetry in NGC 4473.
A point is labelled ``double nucleus?" in reference to the
note given in Table 1 of  Byun \etal (1996).  Here
we find a significant asymmetry in the two sides of NGC 4473:
one side is $\sim 5$ DN higher than the other at $|r|=$2 pixels.
Also, the labelled point deviates from a line connecting
the adjacent points by 1 DN, which is also marginally significant.
As can be seen in the steep profiles of the
power-law galaxies (Figures \ref{plotprofall}b),
the asymmetric PSF of STIS does not create asymmetries in
these light profiles as strong as the one in NGC 4473.

%%%%%%%%%%%%%%%%%%%%%%%%%%%%%%%%%%%%%%%%%%%%%%%%%%%%%%%%%%%%%%%%%%%%%%%%%%
% Fig 7
\begin{figure*}[t]
\centerline{\psfig{file=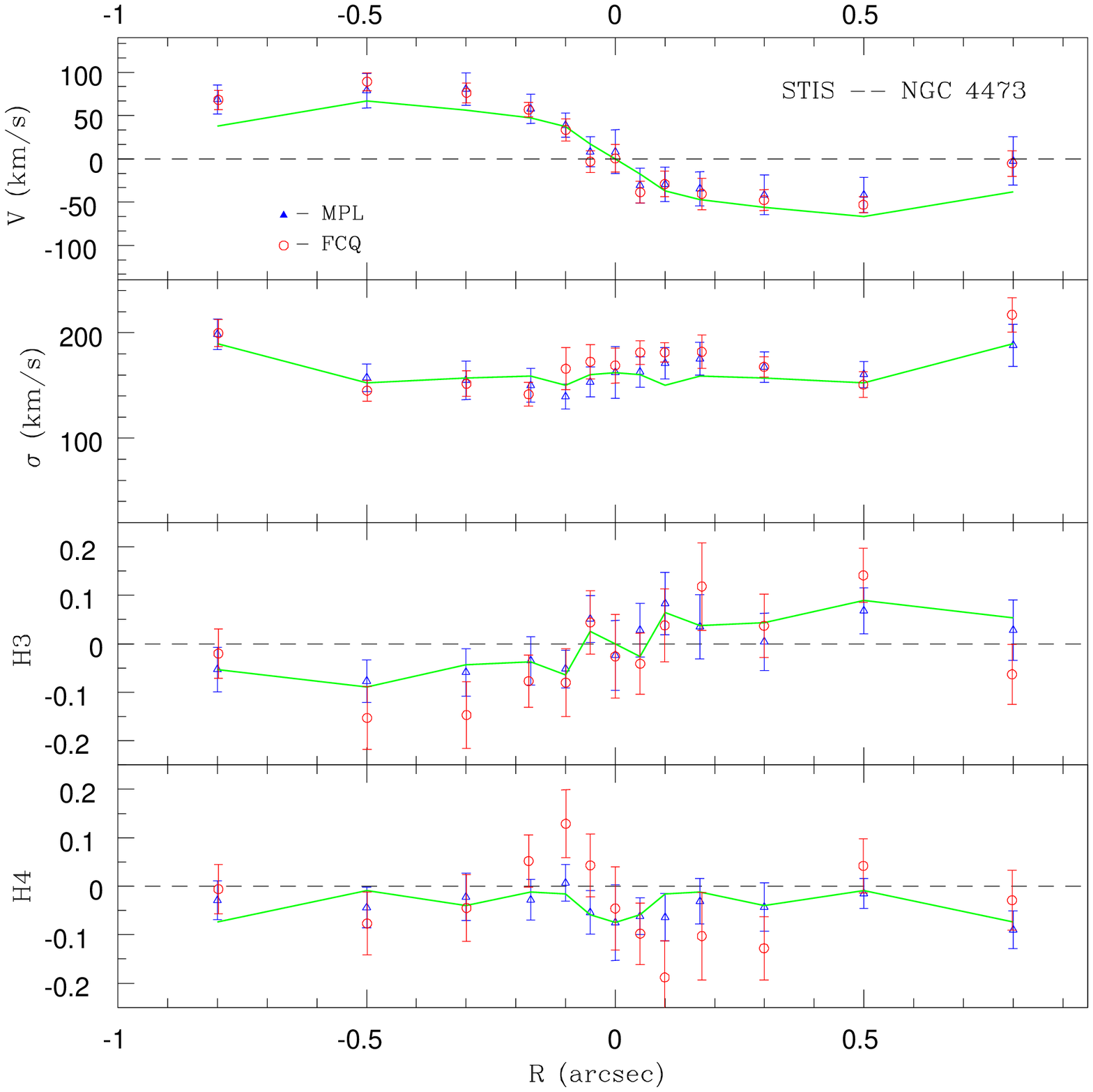,width=16cm,angle=0}} \figcaption{A
comparison of FCQ (circles) and MPL (triangles): kinematic profiles
for NGC 4473.  These parameters were derived from unsymmetrized
spectra from STIS.  From top to bottom, we show mean velocity,
velocity dispersion, $h3$ (3rd coefficient from the Gauss-Hermite
expansion), and $h4$ (4th coefficient).  The solid lines are derived
from the symmetrized LOSVD from MPL.
\label{kpcomp}}
\end{figure*}
%%%%%%%%%%%%%%%%%%%%%%%%%%%%%%%%%%%%%%%%%%%%%%%%%%%%%%%%%%%%%%%%%%%%%%%%%%

%%%%%%%%%%%%%%%%%%%%%%%%%%%%%%%%%%%%%%%%%%%%%%%%%%%%%%%%%%%%%%%%%%%%%%%%%%
% Fig 8
\begin{figure*}[t]
\centerline{\psfig{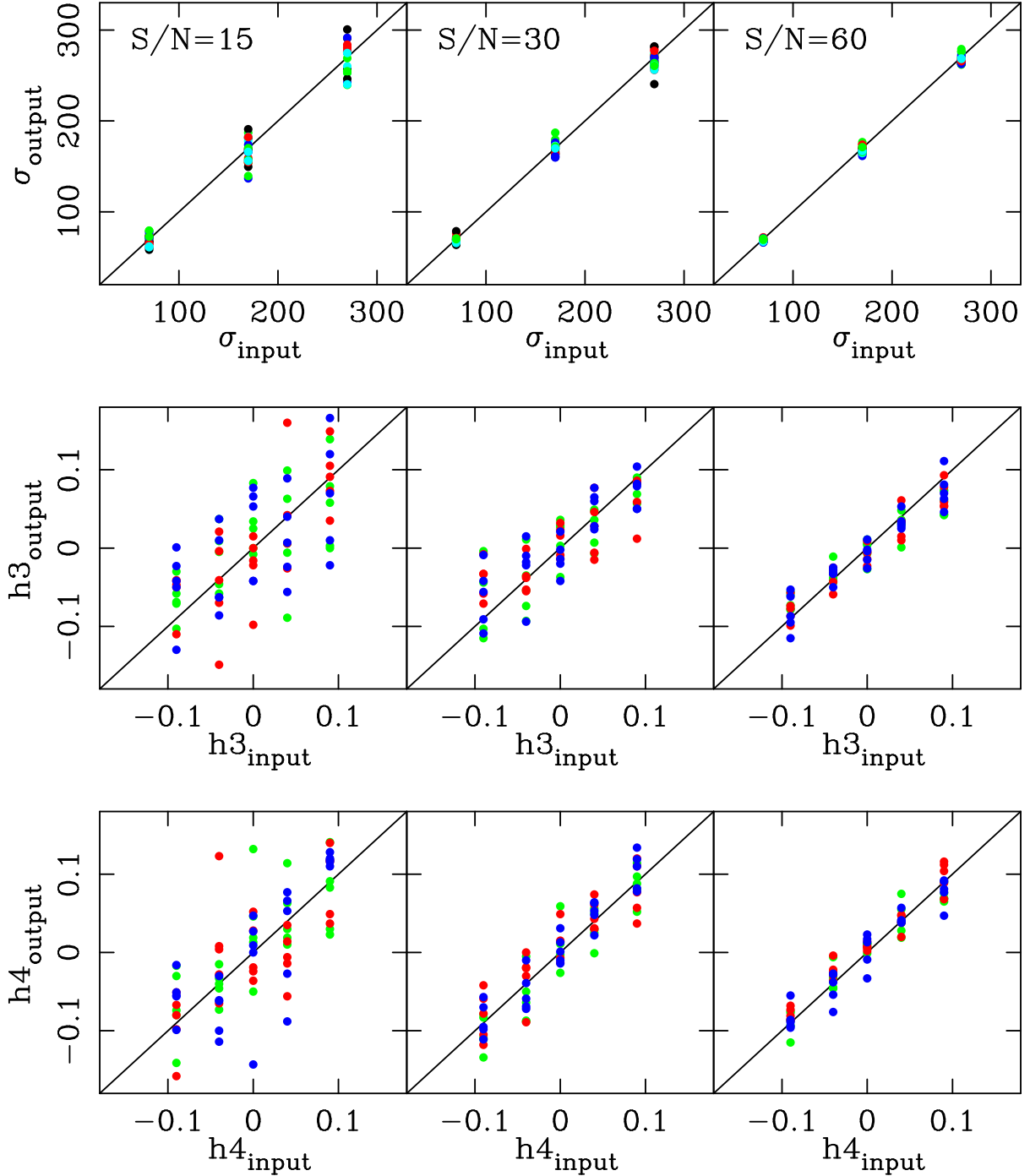}}
\figcaption{Investigation of bias in MPL estimates of Gauss-Hermite
parameters.  From top to bottom, we show velocity dispersion, $h_3$
(3rd coefficient from the Gauss-Hermite expansion), and $h_4$ (4th
coefficient).  Each point is a different realization of a broadened
template spectrum with noise added.  The signal-to-noise of the input
spectra increases to the right in each row of panels.  In the top
three panels, the colors light blue, blue, black, red and green
represent $h_4$s of -0.09, -0.04, 0.0, 0.04 and 0.09, respectively.
In the bottom six panels, green stands for $\sigma_{input} = 70$ \kmsc
red for 170 \kmsc and blue for 270 \kmsp
\label{mplbias}}
\end{figure*}
%%%%%%%%%%%%%%%%%%%%%%%%%%%%%%%%%%%%%%%%%%%%%%%%%%%%%%%%%%%%%%%%%%%%%%%%%%

\section{Kinematics From Stars: LOSVD Fitting}

The modeling of stellar kinematics requires the derivation of LOSVDs
at numerous positions on the galaxy. The STIS spectra are used to
sample only $\pm$1\farcs0 along the major axis, while our Modspec
observations sample to 1--2 half-light radii (ranging from
10-70\arcsec) along the major axis, and other position angles.  After
some experimentation, the binning shown in Table
\ref{bins} was chosen.  It provides the best spatial resolution while
maintaining adequate S/N. The first three bins are each 1 STIS pixel wide.

%\placetable{bins}

Obtaining internal kinematic information requires a deconvolution of
the observed galaxy spectrum using a template spectrum composed of
several representative stellar spectra.  Both the deconvolution
process and the template library are significant issues for obtaining
the LOSVD.  We deconvolve each spectrum using two different
techniques: a maximum-penalized likelihood (MPL) estimate that obtains
a non-parametric LOSVD, and a Fourier correlation quotient technique
(FCQ; Bender 1990).  MPL proceeds as follows. We choose an initial
velocity profile in bins (the choice of initial profile has no effect on
the result). The initial velocity profile convolved with a
weighted-average template (discussed below) provides a galaxy spectrum
that is used to calculate the residuals to the observed spectrum.  The
program fits the shape of the losvd by subdividing it into 13 velocity
bins, and varying the counts in these bins while adjusting the
template weights to provide the best match to each galaxy
spectrum. The exact number of bins has little influence on the final
results. Thus, the template mixture can vary from one radius to the
next, just as the velocity, velocity dispersion, etc. can.  The MPL
technique is similar to that used by Saha~\& Williams (1994) and
Merritt (1997). However, our method differs from theirs in that we fit
simultaneously for the velocity profile and template weights.

We continuum-divide all template and galaxy spectra.  We use a local
continuum for the estimate as opposed to a global $n$-parameter fit to
the full spectrum.  The continuum is estimated by dividing the
spectrum up into about 10 wavelength windows, estimating a robust mean
(the biweight; see Beers, Flynn, \& Gebhardt 1990) in each window, and
interpolating between these values.  This mean is measured from the
highest 1/3 of the points to compensate for absorption features.  We
have tried a variety of continuum estimates (varying the number of
local windows and number of points used in the local averaging) and
find insignificant differences in the kinematic results.

We use Monte Carlo simulations to measure the uncertainties on the
velocity profile bins. For each realization, we generate a simulated
galaxy spectrum based on the best-fit velocity profile and an estimate
of the rms residual of the initial fit. The initial galaxy spectrum
comes from the template star convolved with the measured LOSVD. This
provides a galaxy spectrum with essentially zero noise (the noise in
the template is insignificant for our purposes). From that initial
galaxy spectrum, we then generate 100 realizations and determine the
velocity profile, and hence the velocity dispersion, each time. Each
realization contains flux values at each wavelength that are chosen
from a Gaussian distribution, with the mean given by the initial
galaxy spectrum and the standard deviation given by the rms of the
initial fit. The 100 realizations of the velocity profiles provide a
distribution of values from which we estimate the 68\% confidence
bands. These velocity profiles and their 68\% confidence bands are
used directly in the dynamical modeling (Gebhardt et al. 2003).

We wish to compare the MPL-derived LOSVD with those derived using FCQ
(see above).  To do this, we must convert the non-parametric LOSVDs
into Gauss-Hermite polynomials.  We use the least-squares estimator
MRQMIN from Press \etal (1992) to find the best fit to this
parameterization:
\beq
f(y) = I_{0}\exp(-y^2/2)(1+h_{3}H_{3}(y)+h_{4}H_{4}(y) ),
\eeq
where the mean velocity $v_{fit}$ and velocity dispersion
$\sigma_{fit}$ are contained within $y = (v-v_{fit})/\sigma_{fit}$.
$I_0$ is the amplitude of the LOSVD at $y=0$.  The coefficient $h_3$
multiplies the asymmetric deviations from a Gaussian (i.e., skewness),
while $H4(y)$ parameterizes symmetric deviations from a Gaussian
(kurtosis).  The definitions of $H_3$ and $H_4$ are given in BSG.  The
first and second moments of the LOSVD will differ from $v_{fit}$ and
$\sigma_{fit}$ when $h_3$ and $h_4$ are non-zero.  For example, the
$\sigma_{fit}$ is $\sim$ 10\% smaller than the second moment about the
mean when $h_4 = 0.1$, and $\sim$ 5\% smaller for $h_3 = 0.1$.
Similarly, $v_{fit}$ will be $\sim$ 15\% larger than the first moment
for $h_3=-0.1$ (BSG).

%%%%%%%%%%%%%%%%%%%%%%%%%%%%%%%%%%%%%%%%%%%%%%%%%%%%%%%%%%%%%%%%%%%%%%%%%%
% Fig 9
\begin{figure*}[t]
\centerline{\psfig{file=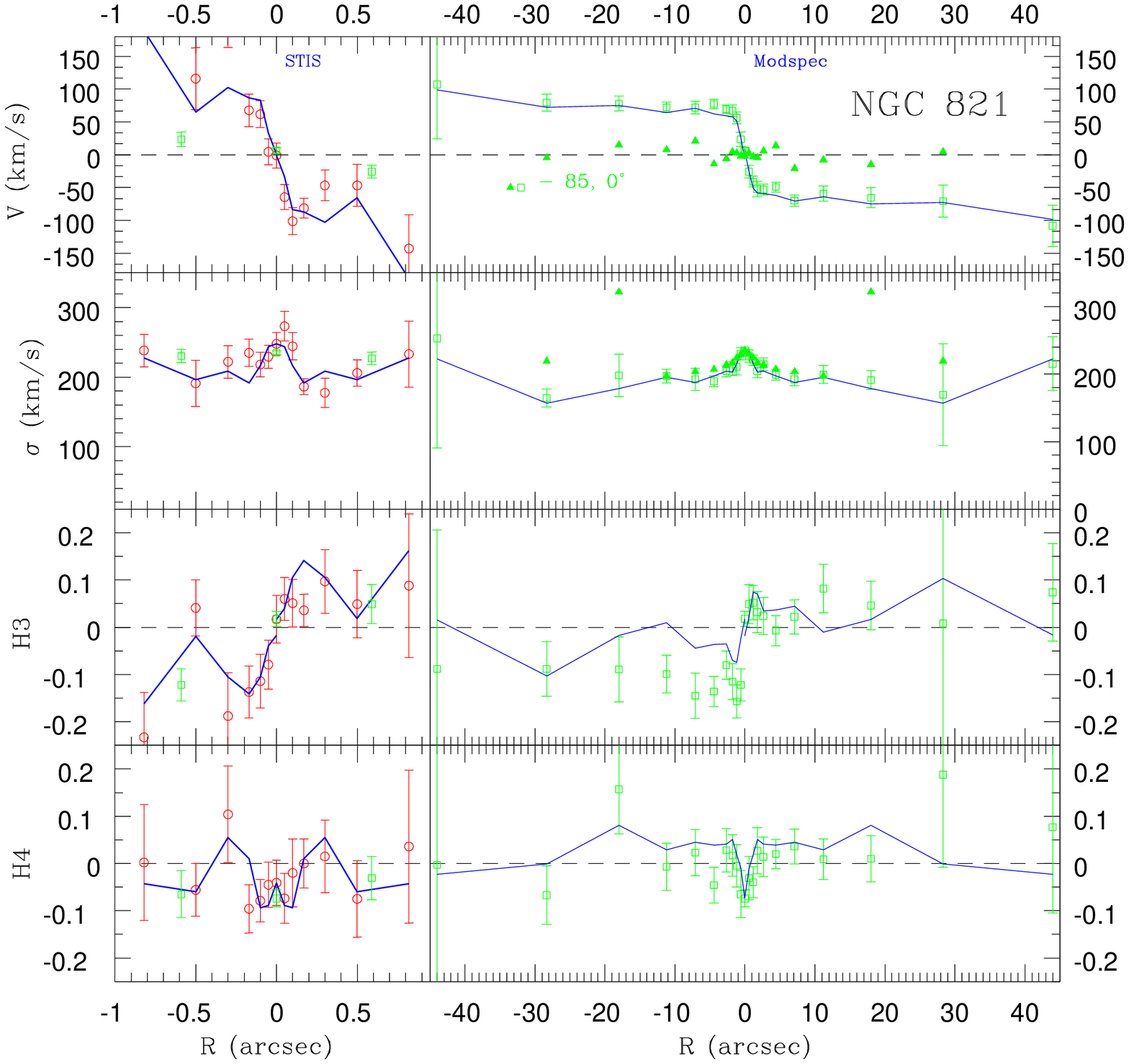,width=16cm,angle=0}}
\figcaption{Kinematic profiles for NGC 821. From top to bottom, we
show mean velocity, velocity dispersion, $h3$ (3rd coefficient from
the Gauss-Hermite expansion), and $h4$ (4th coefficient).  {\it Left:}
major axis data from $r<$ 1\farcs0.  The solid lines are derived from
the symmetrized LOSVD from STIS Ca II data.  The circles are
unsymmetrized data from STIS Ca II.  The squares are ground-based,
unsymmetrized data.  {\it Right:} the entire radial extent of only the
ground-based ``Modspec" data.  Again, a solid line connects the
symmetrized, major-axis data, and the squares represent unsymmetrized,
major axis (PA = 0\arcdeg) data.  Other symbols are labelled by the
rotation of the slit with respect to the galaxy's major axis, in
degrees.  For plotting purposes, the prograde side is always on the
right ($r>0$).  For NGC 821, both ground-based PAs are from spectra
centered on Mg b.
\label{kp821}}
\end{figure*}
%%%%%%%%%%%%%%%%%%%%%%%%%%%%%%%%%%%%%%%%%%%%%%%%%%%%%%%%%%%%%%%%%%%%%%%%%%

%%%%%%%%%%%%%%%%%%%%%%%%%%%%%%%%%%%%%%%%%%%%%%%%%%%%%%%%%%%%%%%%%%%%%%%%%%
% Fig 10
\begin{figure*}[t]
\centerline{\psfig{file=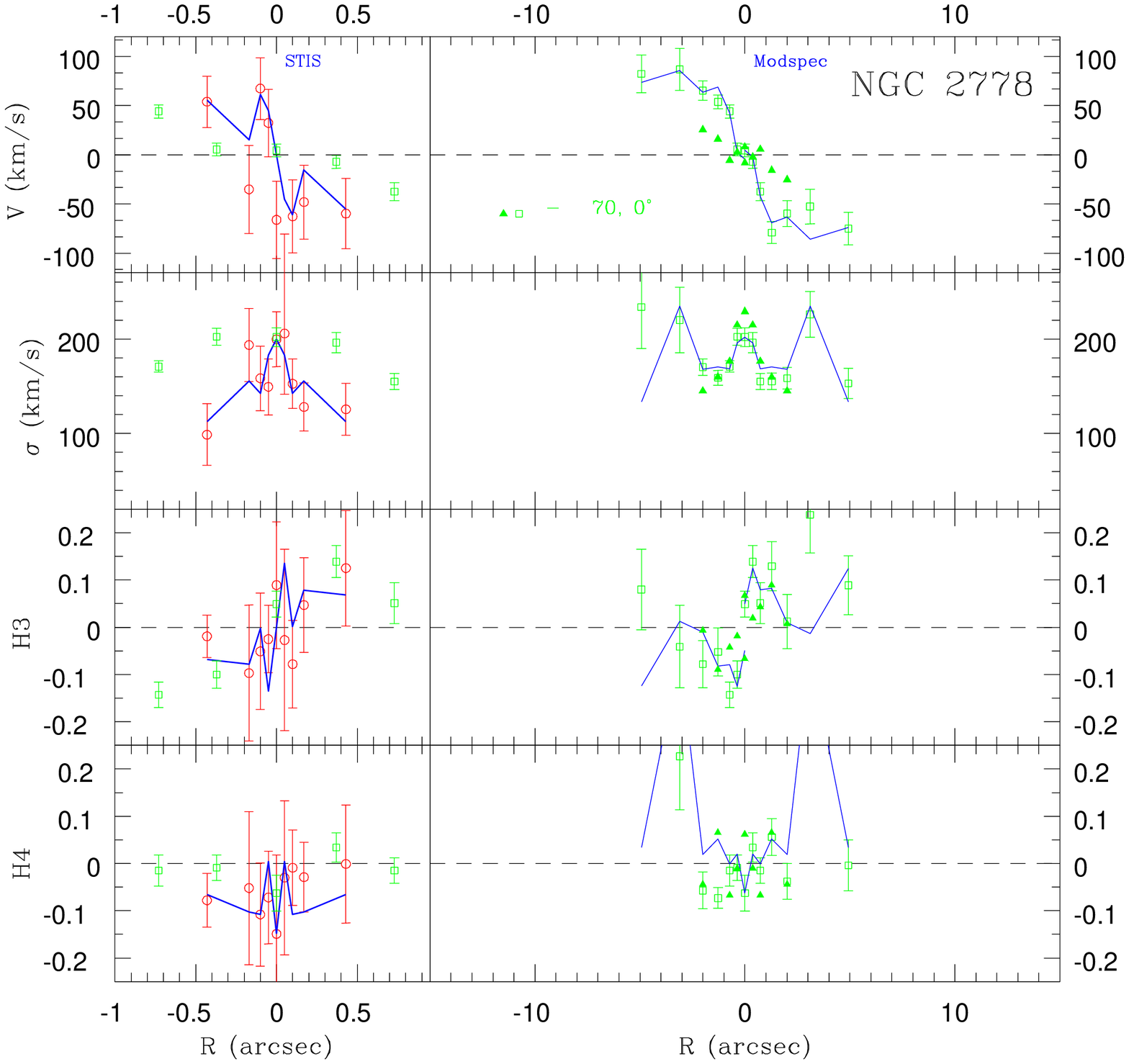,width=16cm,angle=0}}
\figcaption{Kinematic profiles for NGC 2778 (see Fig.~\ref{kp821} for
the meaning of the symbols).  For NGC 2778, all ground-based PAs are
from Ca spectra.
\label{kp2778}}
\end{figure*}
%%%%%%%%%%%%%%%%%%%%%%%%%%%%%%%%%%%%%%%%%%%%%%%%%%%%%%%%%%%%%%%%%%%%%%%%%%

%%%%%%%%%%%%%%%%%%%%%%%%%%%%%%%%%%%%%%%%%%%%%%%%%%%%%%%%%%%%%%%%%%%%%%%%%%
% Fig 11
\begin{figure*}[t]
\centerline{\psfig{file=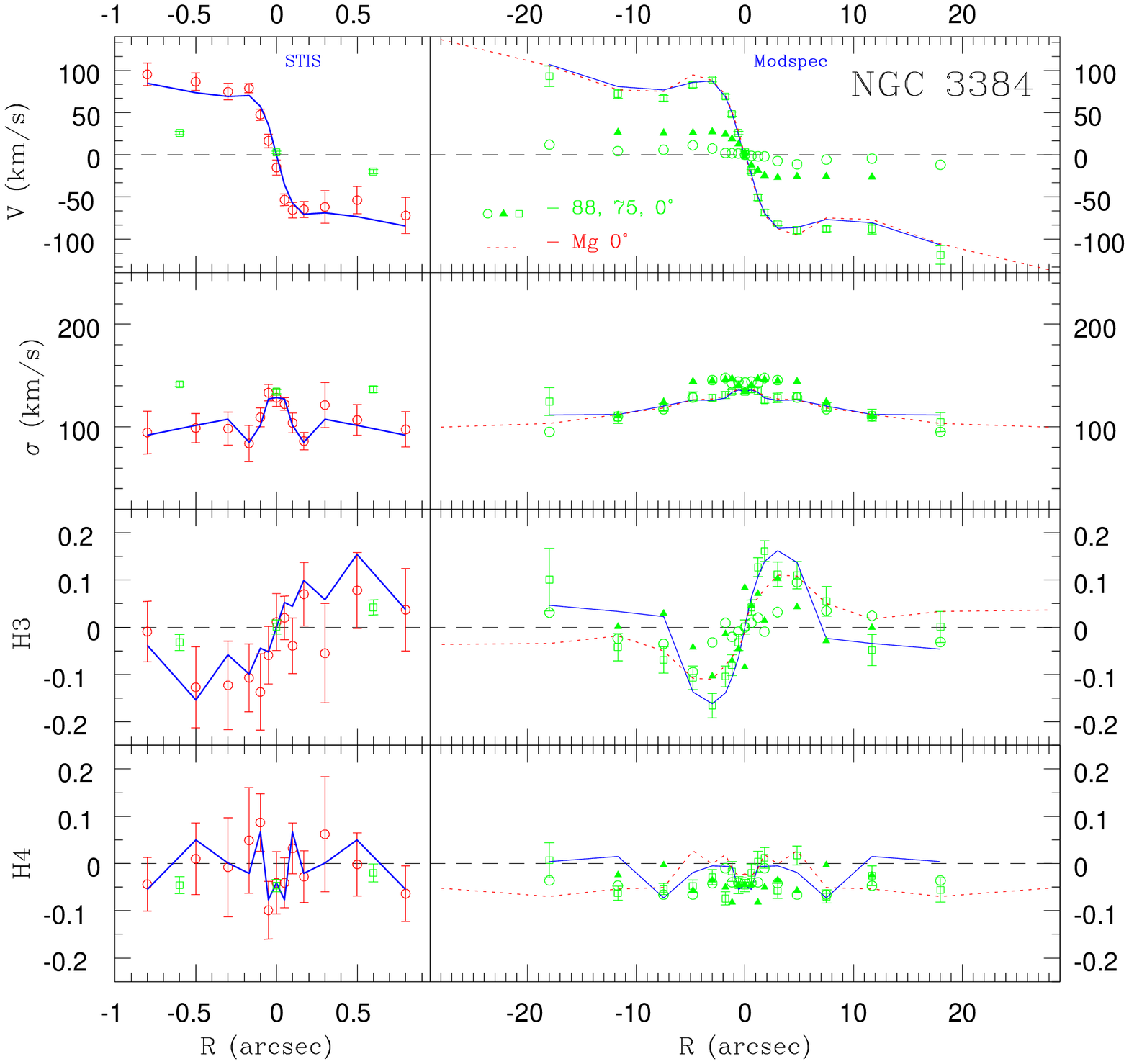,width=16cm,angle=0}}
\figcaption{Kinematic profiles for NGC 3384 (see Fig.~\ref{kp821} for
the meaning of the symbols).  For N3384, the dashed line (PA =
0\arcdeg ) and PA = 88\arcdeg\ circles are Mg spectra.  All others are
Ca spectra.
\label{kp3384}}
\end{figure*}
%%%%%%%%%%%%%%%%%%%%%%%%%%%%%%%%%%%%%%%%%%%%%%%%%%%%%%%%%%%%%%%%%%%%%%%%%%

%%%%%%%%%%%%%%%%%%%%%%%%%%%%%%%%%%%%%%%%%%%%%%%%%%%%%%%%%%%%%%%%%%%%%%%%%%
% Fig 12
\begin{figure*}[t]
\centerline{\psfig{file=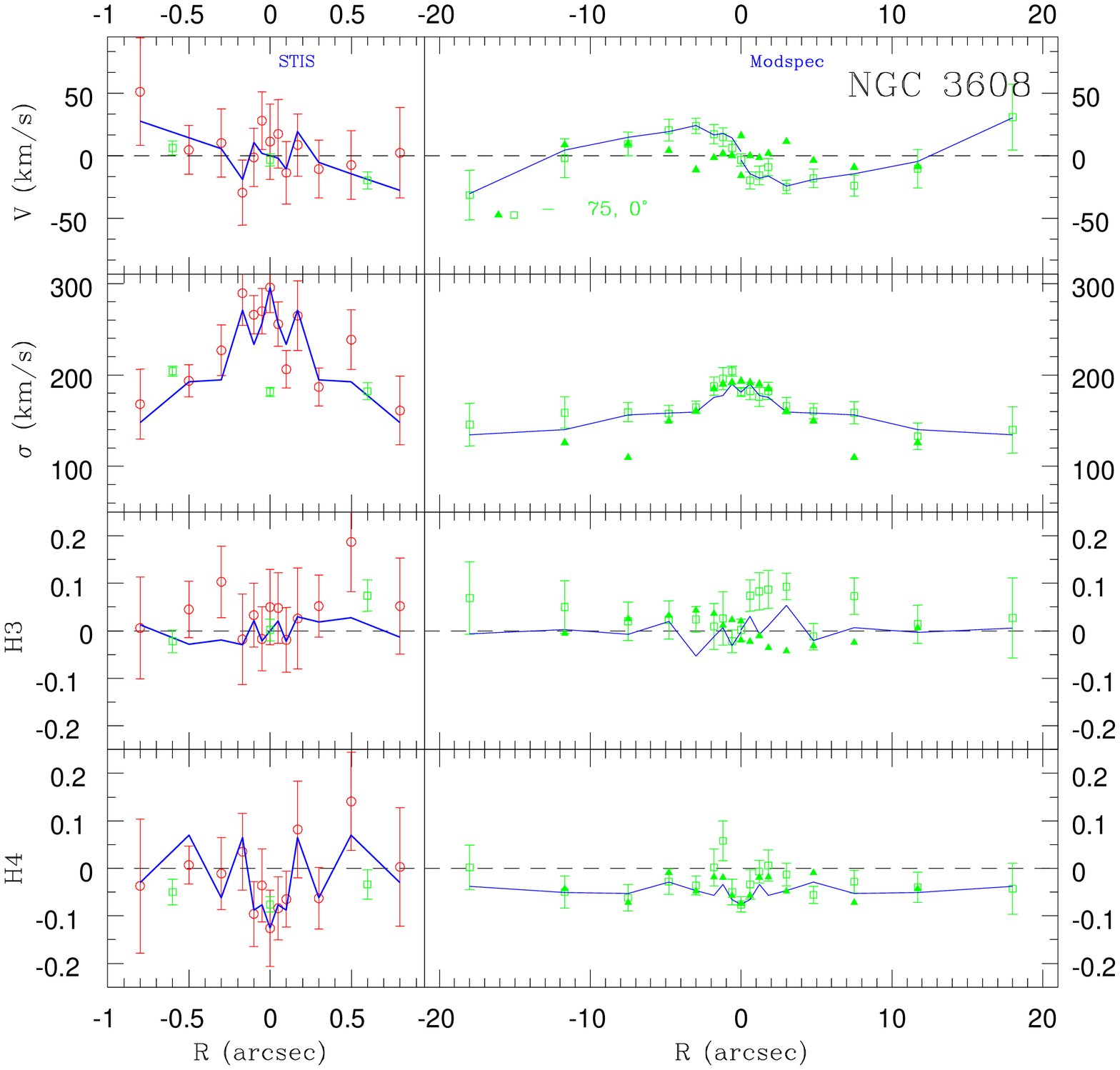,width=16cm,angle=0}}
\figcaption{Kinematic profiles for NGC 3608 (see Fig.~\ref{kp821} for
the meaning of the symbols).  For NGC 3608, all ground-based PAs are
from Ca spectra.
\label{kp3608}}
\end{figure*}
%%%%%%%%%%%%%%%%%%%%%%%%%%%%%%%%%%%%%%%%%%%%%%%%%%%%%%%%%%%%%%%%%%%%%%%%%%

%%%%%%%%%%%%%%%%%%%%%%%%%%%%%%%%%%%%%%%%%%%%%%%%%%%%%%%%%%%%%%%%%%%%%%%%%%
% Fig 13
\begin{figure*}[t]
\centerline{\psfig{file=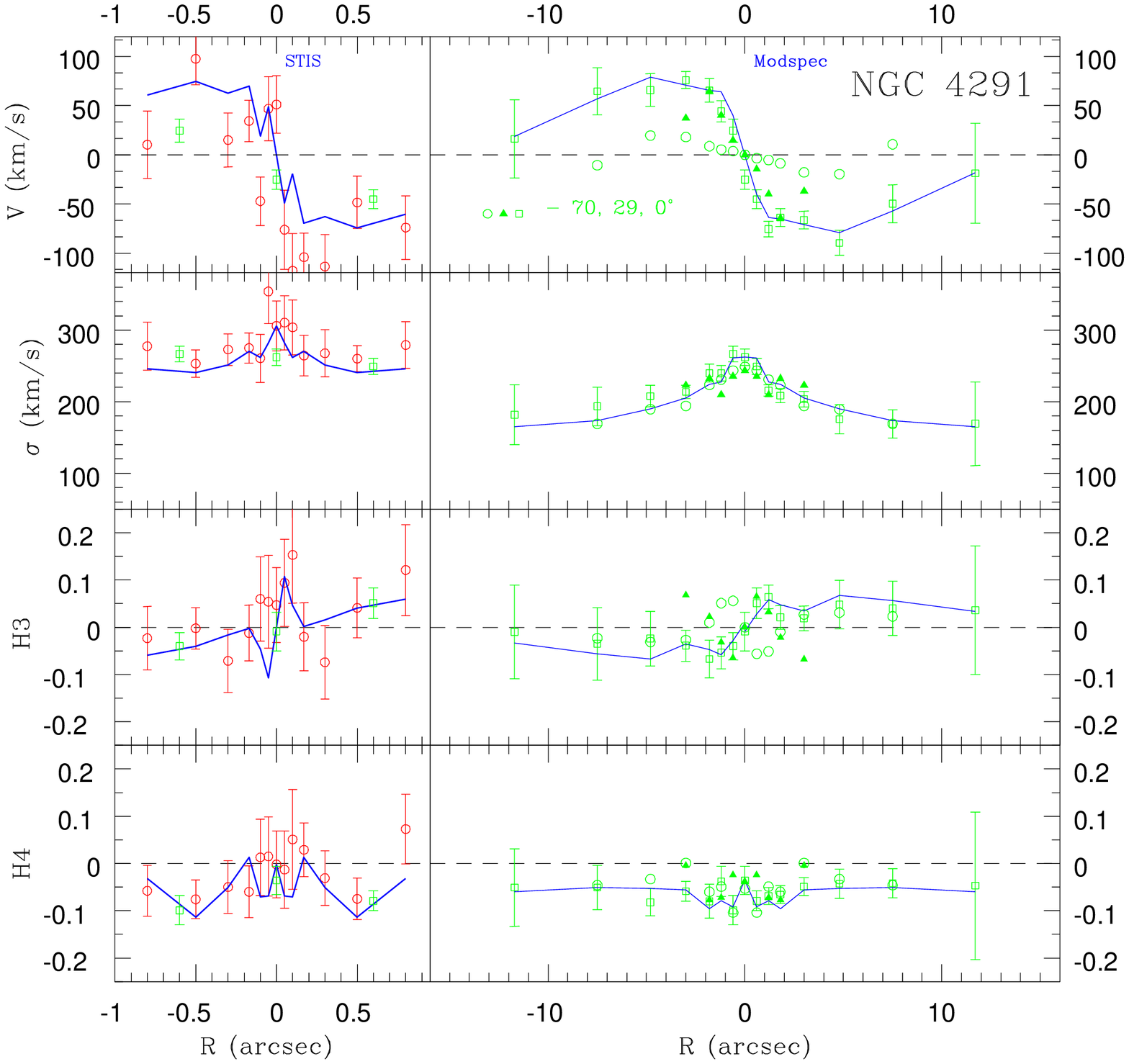,width=16cm,angle=0}}
\figcaption{Kinematic profiles for NGC 4291 (see Fig.~\ref{kp821} for
the meaning of the symbols).  For NGC 4291, all ground-based PAs are
from Ca spectra.
\label{kp4291}}
\end{figure*}
%%%%%%%%%%%%%%%%%%%%%%%%%%%%%%%%%%%%%%%%%%%%%%%%%%%%%%%%%%%%%%%%%%%%%%%%%%

%%%%%%%%%%%%%%%%%%%%%%%%%%%%%%%%%%%%%%%%%%%%%%%%%%%%%%%%%%%%%%%%%%%%%%%%%%
% Fig 14
\begin{figure*}[t]
\centerline{\psfig{file=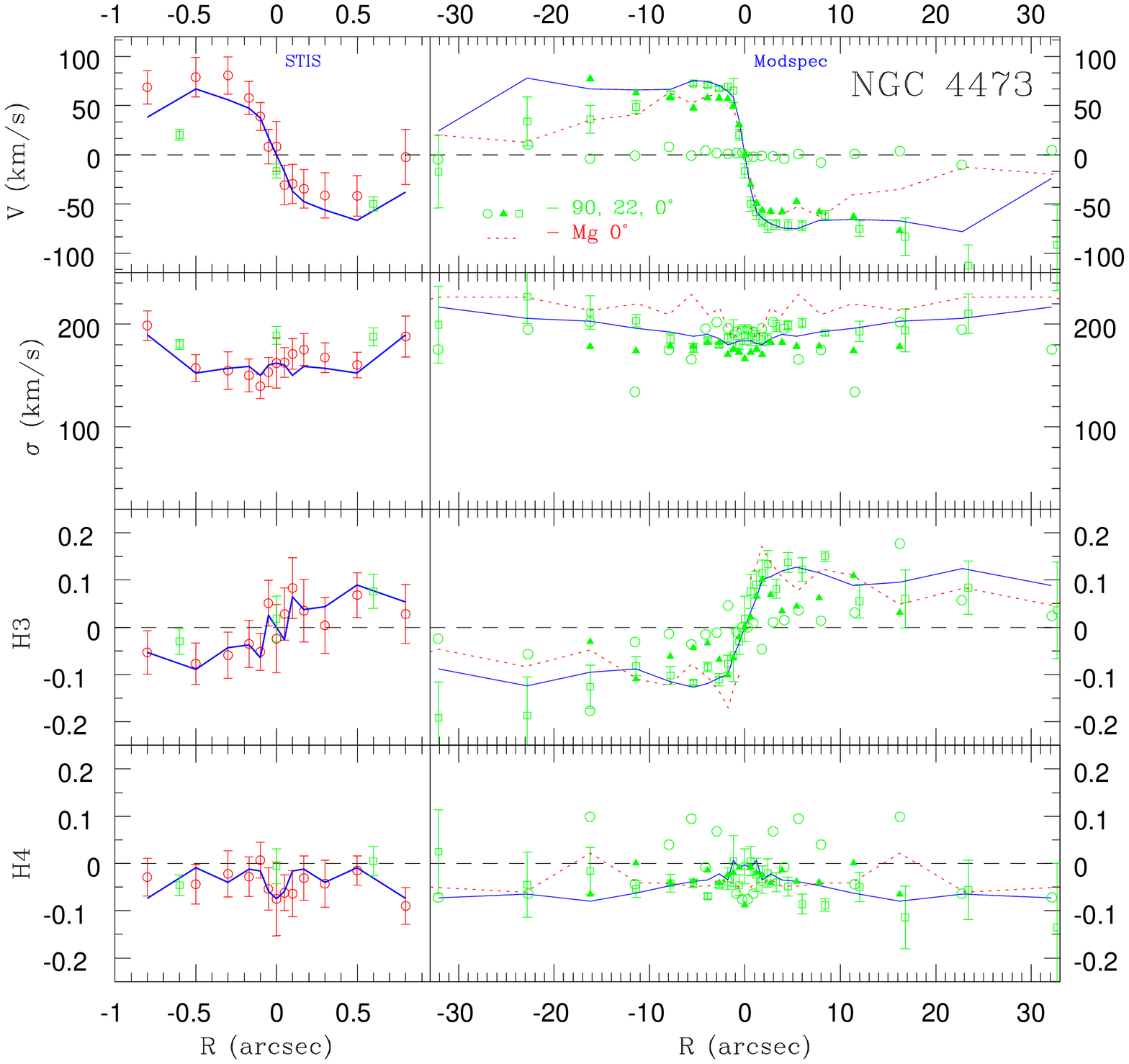,width=16cm,angle=0}}
\figcaption{Kinematic profiles for NGC 4473 (see Fig.~\ref{kp821} for
the meaning of the symbols).  For NGC 4473, the dashed line (PA =
0\arcdeg ) and the circles (PA = 90\arcdeg ) are from Mg spectra.  All
others are Ca spectra.
\label{kp4473}}
\end{figure*}
%%%%%%%%%%%%%%%%%%%%%%%%%%%%%%%%%%%%%%%%%%%%%%%%%%%%%%%%%%%%%%%%%%%%%%%%%%

\begin{figure*}[t]
\centerline{\psfig{file=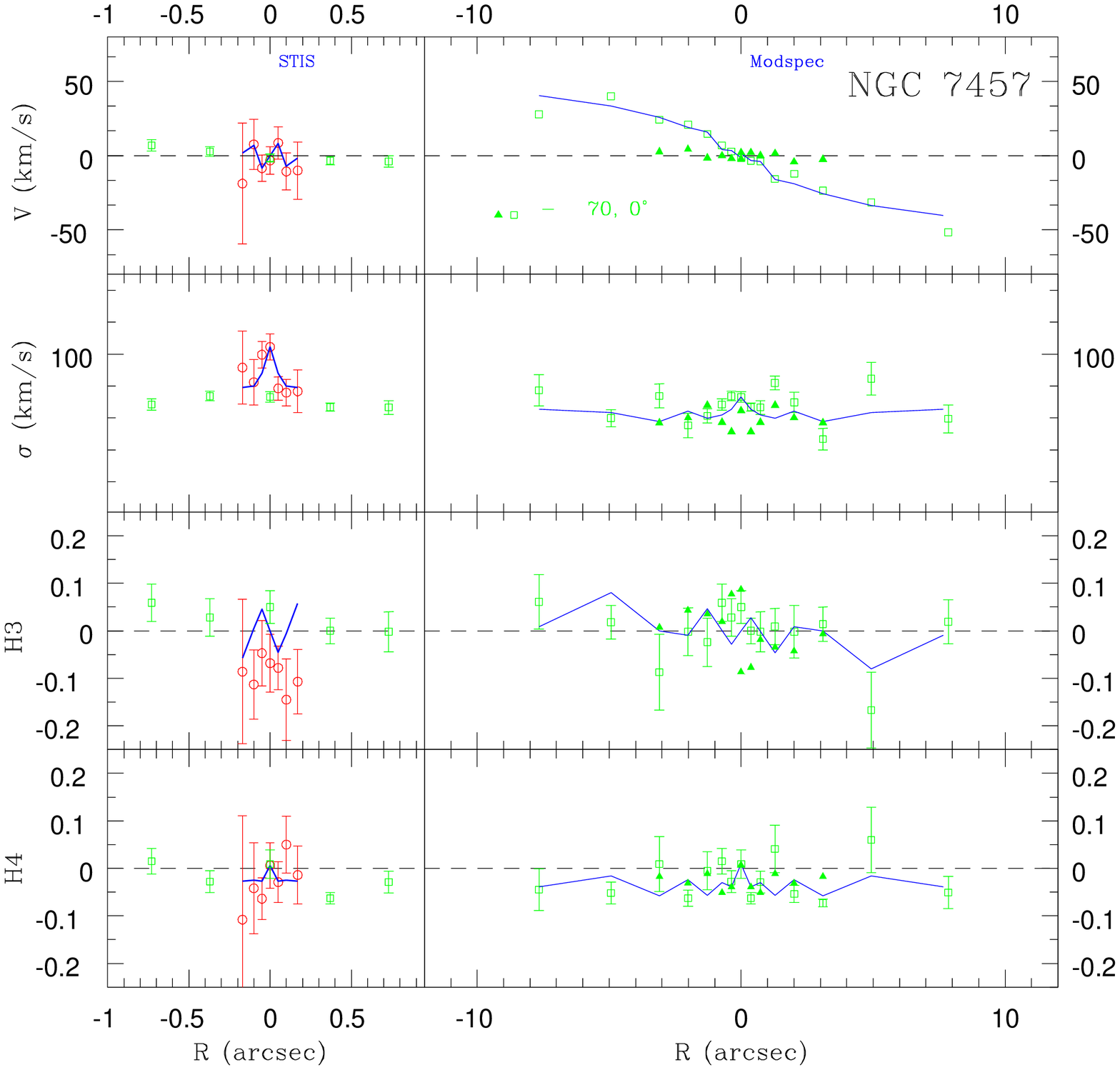,width=16cm,angle=0}}
\figcaption{Kinematic profiles for NGC 7457 (see Fig.~\ref{kp821} for
the meaning of the symbols).  For NGC 7457, all ground-based PAs are
from Ca spectra.
\label{kp7457}}
\end{figure*}

We tabulate the first four terms of the Gauss-Hermite expansion of the
LOSVD for each galaxy in Tables \ref{kintabGND} and \ref{kintabSTIS}.
We use $v_{fit}$ and $\sigma_{fit}$ rather than corrected values
(e.g., $v_{0,c}$ and $\sigma_{0,c}$ used in Joseph \etal 2001) for
reasons listed in BSG, and for the most direct comparison to published
results.  Hereafter, we refer to $v_{fit}$ and $\sigma_{fit}$ as $v$
and $\sigma$.  We found that FCQ and MPL gave consistent profiles and
errors. Figure \ref{kpcomp} plots the first four Gauss-Hermite moments for 
both FCQ and MPL for NGC~4473.  FCQ showed more scatter than MPL in galaxies
with poorly defined lines (e.g., NGC~4649).  We plot only the
MPL-derived $v$, $\sigma$, $h3$, and $h4$ against $r$ in Figures
\ref{kp821}-\ref{kp7457}.

The Monte Carlo simulations used to measure the uncertainties also
provides an estimate of the estimator biases. This bias may be
important for the dynamical models since the shape of the velocity
profile at some level determines the internal orbital
structure. Furthermore, Joseph et al. (2001) argue that $H_4$
estimated from FCQ is biased to lower values when the galaxy
dispersion is low (less than 100 km/s). The Monte Carlo simulations
provide the most accurate estimate of any biases since they simulate
the exact instrumental setup and spectral sampling. We find little to
no bias in our MPL estimates of the first four Gauss-Hermite
moments. Another way to check biases and confidence bands is to fit to
simulated datasets. We have done this for both FCQ and MPL.  The FCQ
results are similar to what is discussed in Joseph et al. (2001); that
is, for galaxies with dispersion below 100 km/s, $H_4$ is biased to
lower values by about 0.04 for S/N=30 per angstrom, which is typical
for these data.  Figure \ref{mplbias} shows the results for MPL run on
simulations using the same setup we used on the STIS galaxies, and for
S/N ratios of 15, 30 and 60 per angstrom.  The MPL results show no
bias in either the $H_3$ or $H_4$. Also, the uncertainties determined
from the simulated datasets agree with those determined from the
galaxy spectra at the same S/N.  I.e., at S/N=40 per angstrom, we
obtain MPL uncertainties on $H_3$ and $H_4$ to around 0.02. After
correcting for the bias in FCQ, the uncertainties on $H_3$ and $H_4$
are similar.

While we report the first four Gauss-Hermite moments in Table
\ref{kintabSTIS}, we use the non-parametric velocity profile for the
dynamics. These moments are reported only to provide a simple
parameterization of the velocity profile for comparison and
correlation studies. An important difference is that the
non-parametric velocity profiles are not allowed to be negative at any
point. Using only the first four moments of a Gauss-Hermite expansion
will allow negative portions of the velocity profile. This is
unphysical, but the effect should be small (at least when $h_3$ and
$h_4$ are small).  Thus, the optimal comparison for the dynamical
models is to use non-negative velocity profiles.

We now discuss each individual galaxy.  We will include special points
of interest, and a comparison to published kinematics, where
available.

\subsection{NGC 821}

NGC 821 is classified as {\verb+.E.6.$.+} (de Vaucouleurs \etal 1991,
hereafter, RC3).  The uncertainty is likely to originate from its
resemblance to an S0.  NGC 821 is a clear example of an elliptical
with disky isophotes (Lauer 1985; Bender \etal 1988 (BDM); Nieto \etal
1991).  As is typical of such galaxies, it has a power-law surface
brightness profile at {\it HST} resolution \cite{Rav}.

NGC 821 is not detected in X-rays (Pellegrini 1999), or in H$\alpha$
(Macchetto \etal 1996).  It also does not appear to contain dust
(Forbes 1991) although it is detected by IRAS in 100 $\mu$m (Roberts
\etal 1991).  Trager \etal (2000) estimate the average age of its
stellar population to be 8 Gyr.

Kinematics show NGC 821 to be a fairly rapid rotator; Nieto \etal
(1988, 1994) measure ($v_{max}$/$\sigma$)$^* \simeq $ 0.71 and we find
$\simeq$ 0.54 for our ground-based data (Table \ref{kinresGND}) and
0.75 using our higher STIS $v_{max}$ (Table \ref{kinresSTIS}).  We get
$v_{max}$/$\sigma$ $\simeq$ 0.4 (Tables \ref{kinresSTIS} and
\ref{kinresGND}) which is lower than previous values (0.49 in Bender
1988; 0.65 in Busarello \etal 1992).

Our kinematic profiles from Modspec are similar to those of BSG.  Both
show that the rotation rises quickly to $\sim$60\kms\ at 2\arcsec.
The STIS kinematics show the largest velocity gradient at 0\farcs17 in
the sample (Table \ref{kinresSTIS}).  $h3$ nicely mirrors the velocity
profile, as is typical in a rotating, edge-on elliptical.  Our
ground-based $h4$ shows a different behavior than in BSG: we find
positive values at $r >$ 2\arcsec and negative values in the central
bin, whereas BSG finds only $h4\simlt0$.  Our values are probably more
reliable in this case because we have a longer effective exposure
time.  The main difference between our data and others is the central
rise in velocity dispersion.  We measure a higher ground-based value,
235$\pm $2\kmsp McElroy's (1995) weighted average of published
velocity dispersions was 207\kmsc and Hypercat \cite{hypercat} gives
the average of 12 velocity dispersions to be 209\kms .  Our Modspec
value does not exceed our central STIS measurement (248$\pm 16$\kms ),
and may simply be a result of good seeing.  When we measure the rms
dispersion within a slit aperture of length 2$r_e$ (i.e., $\sigma_e$),
we find 209\kmsc in agreement with the published averages.

\subsection{NGC 2778}

NGC 2778 is the second least luminous elliptical in the sample
with $M_{B} = -18.6$ mag. It is located close to NGC 2779 in a group.
Trager \etal (2000) estimate the ages of its
stars to be only 6.0 Gyr.  Its isophotes indicate an E2 morphology
with a power-law profile (Peletier 1990; Lauer 2002, private communication).

Our STIS observations for this galaxy were the only ones
taken in Backup Guiding Mode during the {\it HST} visit.  
During galaxy acquisition, only one of the two
guidestars was found so that drifting about the roll
axis was only constrained by gyros.  Our parallel observations
with the WFPC2 show virtually no change in the centroid of a galaxy
image between the first and last (6th) orbit.  Since the
PC chip is about as far from the guide star as the STIS slit,
there was probably no significant error in the STIS guiding.

Davies \etal (1983), Gonz\'{a}lez (1993) and Fisher \etal (1995)
all provide kinematic profiles for NGC 2778.
These rotation curves extend beyond ours to $\sim$25\farcs0.
The maximum rotation velocity, $\sim$ 100\kmsc is reached by
about 5\arcsec .  No rotation was observed along the minor
axis; the galaxy is approximately an oblate, rapid rotator
(Jedrzejewski \& Schechter 1989).
All dispersion profiles show a steep rise from
$\sigma \approx $  100 to 200\kms in the inner 5\arcsec .  This galaxy
has the greatest deviation from the \mbhsig\ relation in our sample:
it has a small BH mass for its effective dispersion.
The low surface brightness (see Fig.~\ref{plotprofall}) and poorly resolved 
sphere of influence ($G\mbh /\sigma^2 \approx$ 0.02\arcsec ) make 
the BH detection in NGC 2778 relatively uncertain (Gebhardt \etal 2003).

\subsection{NGC 3384}

NGC 3384 is classified as {\verb+ SB(s)0-:+} (RC3).  It is in the Leo-I
group,
neighbored closely by the elliptical
M105 (NGC 3379) and the spirals M95 and M96.  It appears to have
been named redundantly in the NGC as NGC 3371.
Several H I clouds orbit around NGC 3384 and 3379
in a 200 kpc diameter ring formation (Schneider 1985).
The H I has been interpreted as gas stripped during an encounter
between NGC 3384 and the spiral NGC 3368 (Rood \& Williams 1985).
The inner regions of NGC 3384 appear largely free of interstellar matter (ISM):
Tomita \etal (2000) find no dust in WFPC2 color-excess images.

In the WFPC2 {\it V} and {\it I} images (Fig. \ref{collage}), NGC 3384
appears consistent with a lenticular morphology.
The {\it Nuker}-law profile shows a rapid transition
($\alpha = 11.2$) between inner cusp region ($\gamma = 0.6$)
and the outer power-law ($\beta = 1.8$) at the break
radius $r_b$ = 2\farcs9 (Lauer 2002, private communication).
There also appears to be light in excess of the {\it Nuker} law at
$r\simlt$ 0\farcs3.
Tomita \etal (2000) use the {\it V} and {\it I} images to search for dust
and blue nuclei.
They suspect a blue nucleus in NGC 3384, but claim that it is uncertain
because the center is saturated on the PC.
High resolution, ground-based imaging also hints at a nucleus (Kormendy
1985).
Our deconvolved surface brightness profiles show a nearly flat
{\it V-I} profile.
Our near-infrared light profile along the STIS slit (which is {\em not}
saturated) shows a strong central peak, but not necessarily a
stellar nucleus (Figure \ref{plotprofall}).
Also, we do not see Paschen absorption in our Ca II spectra
which would indicate a young stellar population.

NGC 3384 has the highest S/N ($\simgt$ 40 in central
bins) among our STIS datasets owing to its high surface brightness
($\mu_{V,0.1}$=13.7 mag arcsec$^{-2}$) and 11 orbits of 
exposure (Table \ref{stisobs}).
Our ground-based kinematics are also of high quality (Fig. \ref{kp3384}).
We have both Ca II triplet (solid line) and Mg b (dashed line) data along
the major axis.  The results we obtain from these two wavelength ranges
are consistent.

Our Modspec kinematics for NGC 3384 can be compared to the integral
field spectroscopy of de Zeeuw \etal (2002), and to long-slit
spectroscopy by Fisher (1997), and Busarello \etal (1996).
All rotation velocity curves peak
at $r\sim$ 3\farcs0 followed by a decrease.
Then, beyond 10\arcsec , the rotation climbs again.
Such an ``overshoot" is seen in other S0 galaxies, most notably NGC 4111,
and in about 5-10 of the 18 S0 galaxies studied by Fisher (1997) display
some sort of overshoot or kink in their rotation at $\sim$ 5\arcsec .
De Zeeuw \etal and Fisher plot minor-axis profiles which indicate that
the rapidly rotating
disk component has a lower velocity dispersion than
the bulge.  This is also seen in our 75\arcdeg\ and 88\arcdeg\ rotation
curves which show a higher dispersion beyond 2\arcsec\
than the major axis data (Fig. \ref{kp3384}).  Thus, it appears that a
cold stellar disk contributes differentially to the total
light, increasing the fitted rotation velocity and decreasing the dispersion
at small radii.  At STIS resolution, however, the major axis dispersion
profile shows a distinct peak at $r<$0\farcs2.
The stellar disk also contributes to the strongly asymmetric LOSVDs:
our data show a peak $h3$ value of 0.15 along the major axis,
the highest $h3$ in our sample after NGC 4697.
Our ground-based $h4$ values are systematically lower than those of
de Zeeuw \etal and Fisher, but all show a dip at $r<2''$.

    NGC 3384 has some of the observed characteristics of a {\it
pseudobulge\/}, i.{\thinspace}e., a high-density central mass
component that resembles a bulge but that is thought to have been
built by the inward transport of disk gas by (in the present case) a
bar.  This secular process is qualitatively different from the
classical picture in which bulges form on approximately a collapse
timescale by violent relaxation and dissipation in a galaxy merger.
Pseudobulge characteristics of the present galaxies are discussed in
Kormendy et al.~(2002), and more general reviews of pseudobulges are
given in Kormendy (1993) and Kormendy \& Gebhardt (2001).  In essence,
pseudobulges are left with some memory of their origin -- they have
disky structural properties.  In NGC 3384, the pseudobulge dominates
the light at radius $r < 6^{\prime\prime}$, i.{\thinspace}e., interior
to the bar, which is most important at $r \sim 10^{\prime\prime}$ to
20$^{\prime\prime}$ (Busarello \etal 1996; Jungwiert \etal 1997).  In
NGC 3384, the evidence for a pseudobulge includes the highest
$v/\sigma$ value in Table 7, an unusually high $(v/\sigma)^*$ value in
Table 8, and the observation that the ``bulge'' is almost as flattened
as the disk (its ellipticity $\epsilon \simeq 0.45$ in Table 8 as
compared with $\epsilon \simeq 0.50$ for the outer disk (Davoust \etal
1981; Busarello et al.~1996).  Consistent with this, Bureau \etal
(2002) and de Zeeuw et al.~(2002) find evidence for a ``disk embedded
in the bulge'' (high rotation and a flattened light distribution) from
SAURON observations.

\subsection{NGC 3608}

NGC 3608 is a core galaxy in the Leo Group, separated by only 5.8\arcmin\ (39
kpc) from
NGC 3607.  It is undetected at 6 cm (Wrobel \& Heeschen 1991)
but contains extended X-ray emission ($\log(L/erg s^{-1})=40.4$,
Pelligrini 1999).
Surface photometry shows an E2 elliptical with slightly
boxy isophotes beyond 5$''$.  {\it Nuker}-law parameter fits are published
for the WFPC1 data in Lauer \etal (1995) and Faber \etal (1997),
and we tabulate the
parameters for the WFPC2 {\it V} image in Table \ref{profiles}.
NGC 3608 has a definite ``core" profile with $\gamma \approx $0.06.
There is evidence for patchy dust in the WFPC2 data.
Carollo \etal (1997) note an elongated dust feature in the
innermost 0\parcs6 (67 pc for $d=23$ Mpc) that looks like an off-centered
ring.
Tomita \etal (2000) estimate the dust feature to be 190 pc across.
Singh \etal (1994) suggest that the interacting neighbor, NGC 3607, has
acquired gas and dust from NGC 3608.

Our ground-based kinematics can be compared to those in
Jedrzejewski \& Schechter (1988, 1989).
Our mean velocity and velocity dispersion profiles
are consistent with those published by
these authors, although they plot a few more points beyond 20\arcsec .
The outstanding feature of NGC 3608 is the kinematically distinct
core. The slope of the rotation curve changes sign at about 5$''$
(coincident with the onset of boxy isophotes),
and the rotation itself changes sign at $r\approx 13''$.
The maximum rotation is only 24\kmsc
giving NGC 3608 the lowest $v/\sigma$ in our sample.
Our STIS velocity dispersion profile shows a central peak which is higher
than the ground-based value by about 100\kms .  Also notable
is the significantly negative central $h4$ values in the ground-based data.

\subsection{NGC 4291}

NGC 4291 has a low luminosity for a core galaxy, M$_{B}=-19.6$ mag.
Michard \& Marchal (1994) suspected a peculiar asymmetric
envelope, but it is difficult to distinguish because of
nearby stars.  It is neighbored by NGC 4319 only 7.4$'$ (56 kpc at $d=26$ Mpc) away,
but does not appear to be interacting (RC3).
Like other core galaxies, it has slightly boxy isophotes ($100a_{4}/a$
= $-0.3$,
BSG), and an excess of X-ray emission above that expected from
binary stars alone (Pellegrini 1999).

Our STIS light profile for NGC 4291 (Figure \ref{plotprofall}) shows a slight
excess at $r<$ 0\farcs1.  No strong evidence exists for a central
excess in the light profile of the WFPC2 images.

Our major axis kinematics for NGC 4291 can be compared to those of
BSG and Jedrzejewski \& Schechter (1989).
All datasets show a centrally rising velocity dispersion
which peaks at 270-300\kms .
The rotation peaks at $\sim$80\kms\ at about 5\arcsec\ and then
declines to about 40\kms\ by 12\arcsec .
This overshoot is reminiscent of disky galaxies like N3384,
but there is no photometric evidence for a disk in NGC 4291.
The other authors also show minor axis spectra, whereas we
have a spectrum offset by 29\arcdeg .
NGC 4291 shows very little minor axis rotation
(e.g., $-3.3\pm 1.2$ \kms ; Jedrzejewski \& Schechter 1989; BSG).
The velocity profile shows an asymmetric ``overshoot" at $\sim$5\arcsec\
suggestive of a co-rotating core.
We find lower $h4$ values in our ground-based data than do BSG.
The BH mass estimate from the 2-integral modeling of Magorrian \etal
(1998) is 1.9$\times 10^{9}$ \Msun, while 3-integral modeling applied
to the data herein give 3.1$\times 10^{8}$ \Msun (Gebhardt \etal 2003).
 % (1.9$\times 10^{9}$ \Msun).
This is the largest such discrepancy in our sample.  We believe it results
from the Magorrian \etal (1998) assumption of isotropy.
Wrobel \& Herrnstein (2000) observed NGC 4291 in 8.5 GHz and placed an upper
limit on its ADAF accretion rate using the Magorrian \etal (1998) BH
mass estimate.

\subsection{NGC 4473}

NGC 4473 is an E5 galaxy in the Virgo Cluster.
As seen in other E5 galaxies, the isophotes
are primarily disky (BDM; Michard \& Marchal 1994).
Van den Bosch \etal (1994) note some irregularities in the
higher-order deviations from ellipses between 2\arcsec\ and
4\arcsec .
But ground-based surface photometry generally reveals an ordinary, 
disky elliptical.

Some searches for an ISM, dust, and emission-line gas have found none
(Roberts \etal 1991; Michard \& Marchal 1994; Ho \etal 1997),
while Macchetto \etal (1996) detect a small amount (1200 \Msun ) of
excited gas in H$\alpha$, and Ferrari \etal (1999) report
a similarly sized dust disk at their limit of detection.

Our kinematic measurements are very good for NGC 4473 because of
its relatively high surface brightness (see Fig.~\ref{plotprofall}).
In particular, the signal in the
higher-order moments, $h3$ and $h4$, is relatively strong (Figure
\ref{kp4473}).
Our ground-based values of $h3$ are in good agreement with BSG, but our
$h4$ and $\sigma$ values are systematically lower.  We find
$h4$ to be negative at most radii and gradually increasing toward the
center. $h4$ then shows a central dip at STIS resolution.
Our $\sigma$ profile shows a gradual decrease toward the center
where our ground-based values are about 10\kms\ lower than those of BSG.
However, our central, Modspec dispersion (183\kms ) agrees with the average
of 11 sources in Hypercat, 179\kms .
Young \etal (1978) published long-slit spectroscopy out
to 45$''$. They measure a $v_{max}$ of 60\kms\ and $\sigma$ =
180\kmsp  Morton \& Chevalier (1973) measured a higher
rotation velocity, 100 km/s at $r = 10''$.
Our ground-based $v_{max}$ is 75\kmsp
Michard \& Marchal (1994) point out that this rotation is surprisingly
small for a disky elliptical.

%%%%%%%%%%%%%%%%%%%%%%%%%%%%%%%%%%%%%%%%%%%%%%%%%%%%%%%%%%%%%%%%%%%%%%%%%%
% Fig 15
\hskip -30pt{\psfig{file=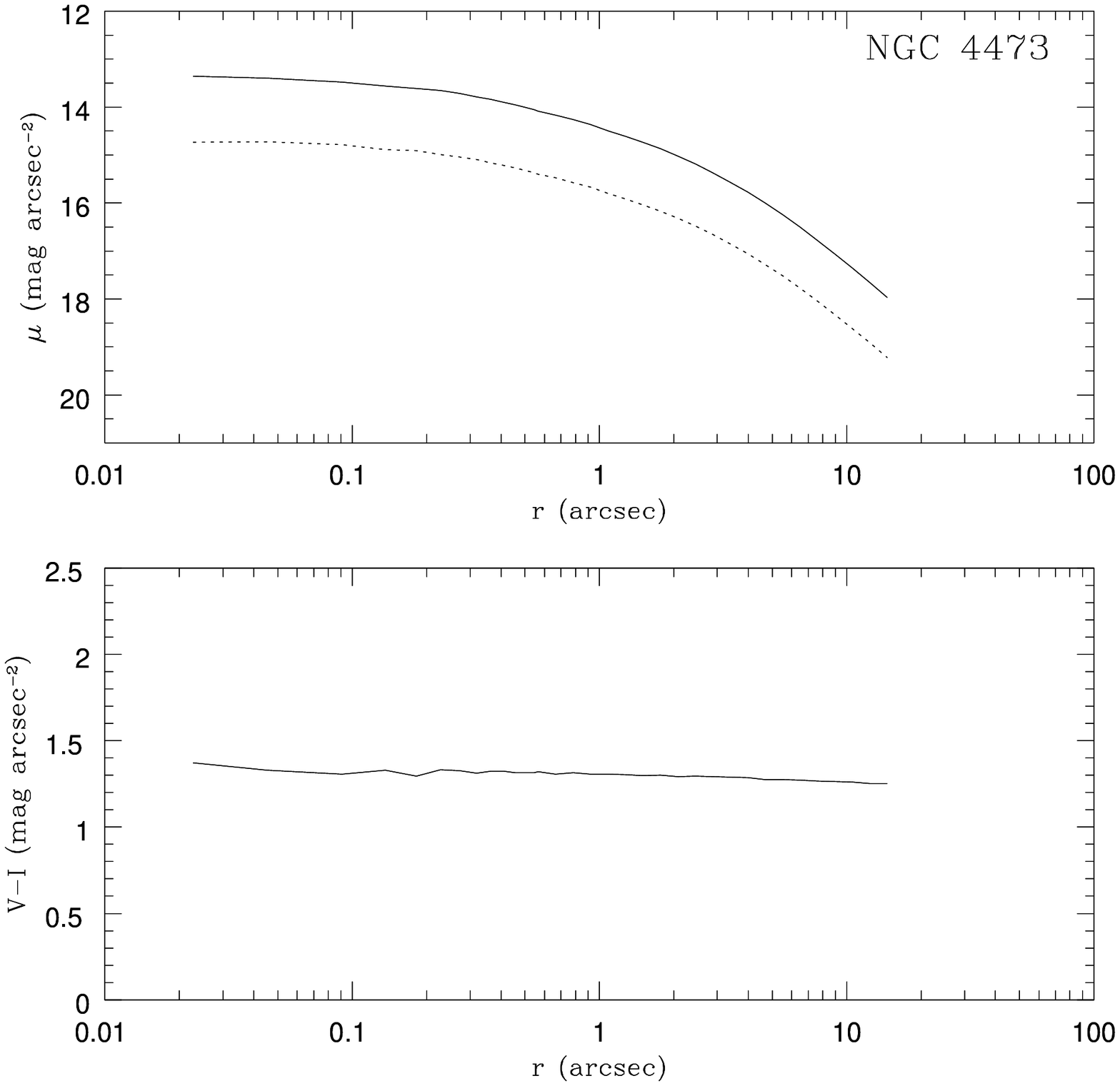,width=9cm,angle=0}} \figcaption{Top:
deconvolved, WFPC2, {\it V} and {\it I} surface brightness profiles of
NGC 4473.  Bottom: ($V-I$) color profile.
\label{lp4473}}
%%%%%%%%%%%%%%%%%%%%%%%%%%%%%%%%%%%%%%%%%%%%%%%%%%%%%%%%%%%%%%%%%%%%%%%%%%
\vskip 10pt

%%%%%%%%%%%%%%%%%%%%%%%%%%%%%%%%%%%%%%%%%%%%%%%%%%%%%%%%%%%%%%%%%%%%%%%%%%
% Fig 16
\begin{figure*}[t]
\centerline{\psfig{file=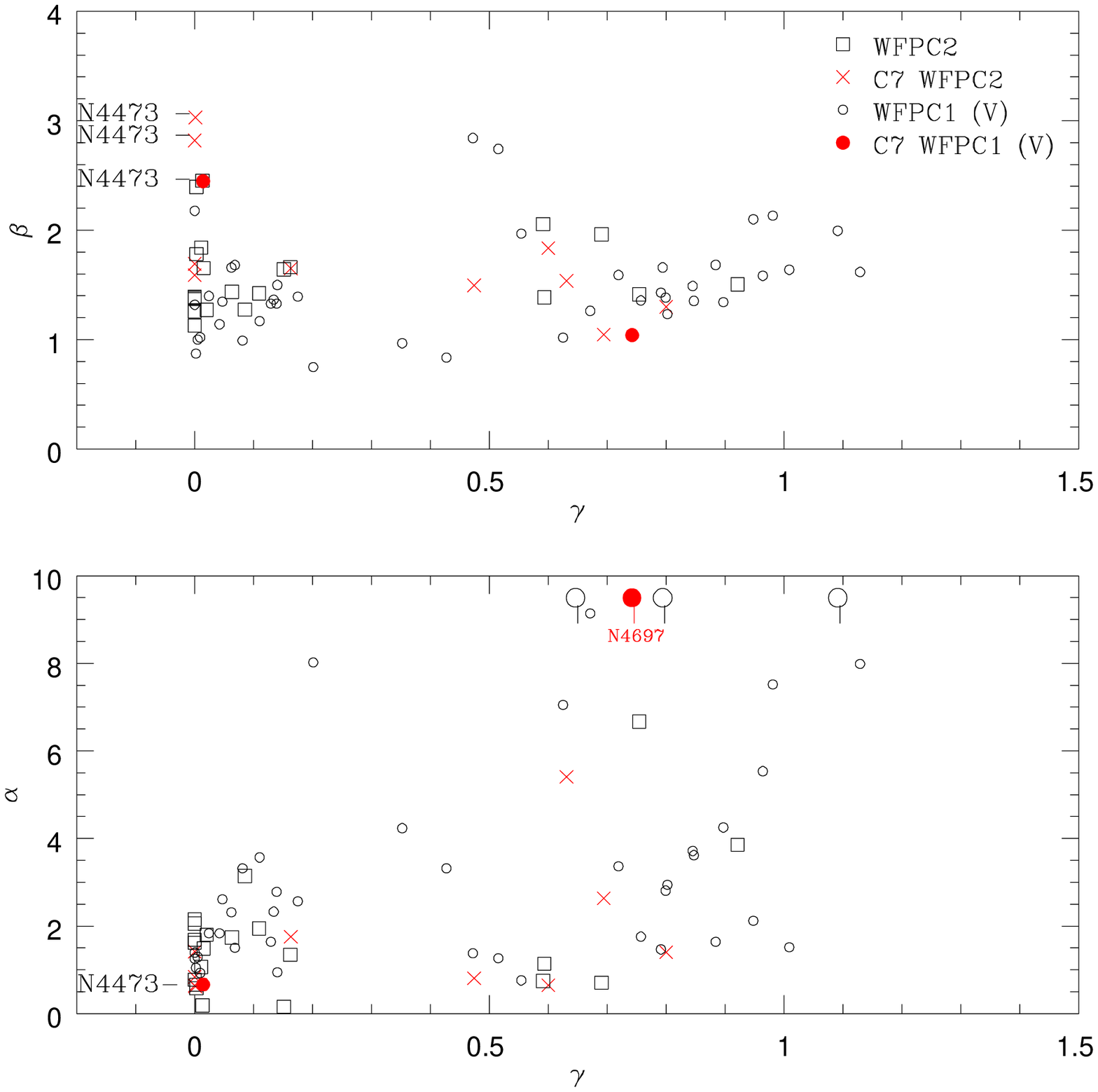,width=16cm,angle=0}} 
\figcaption{Plot of the {\it Nuker}-law parameters $\beta$
vs.~$\gamma$ (top), and $\alpha$ vs $\gamma$ (bottom) for 72 galaxies.
% two samples. We chose the {\it I} band WFPC2 data whenever possible. 
The open circles are the WFPC1 data tabulated in Byun \etal (1996).
The squares are mostly {\it I} band WFPC2 data ({\it V} or {\it R} if
{\it I} was not available) from Lauer (2002, private communication).
The $\times$'s and filled circles represent the 10 galaxies of this
paper.  For NGC 4473 we show three points: WFPC2 {\it I}, {\it V}, and
WFPC1 {\it V}.  The four galaxies with very large $\alpha$ values that
fall off of the plot are marked by circles with tails.
\label{gvb}}
\end{figure*}
%%%%%%%%%%%%%%%%%%%%%%%%%%%%%%%%%%%%%%%%%%%%%%%%%%%%%%%%%%%%%%%%%%%%%%%%%%

The high-resolution {\it HST}~imaging reveals NGC 4473 to be an unusual
galaxy (see Byun \etal 1996 for WFPC1 results, Table \ref{profiles}
for WFPC2 $V$ and $I$ results).
Its surface brightness profile (Fig.~\ref{lp4473}) has an asymptotic inner
slope, $\gamma = 0.01$ (WFPC2 $I$-band, $\gamma = 0.014$ in WFPC1).
This makes it a core galaxy, by definition ($\gamma < 0.3$; Faber \etal
1997).
However, the transition between the inner and outer slopes is very gradual
($\alpha =0.66$ for WFPC1; 0.70 for WFPC2), while all the other cores have
$\alpha$ = 1.0-8.0 (the next smallest being NGC 3608, with the
counter-rotating core).
Furthermore, the outer slope of the profile ($\beta = 2.45$ and 2.7 for
WFPC1
and WFPC2 $I$-band, respectively)
is very steep -- it has the highest $\beta$ among the cores in Byun \etal
(1996).
Figure \ref{gvb} demonstrates how NGC 4473 has unusual {\it Nuker}
parameters.
Finally, the absolute magnitude, M$_B\sim -19.9$ mag, is consistent
with power-law galaxies, falling faintward of the region
of overlap between power laws and cores (Faber \etal 1997).
Why does a relatively faint, E5, disky elliptical have the central profile
of a
core galaxy?

A recent merger can provide several explanations for these
peculiarities.  First, the merger could contribute a secondary 
nucleus which could artificially flatten the inner surface brightness profile
and account for the gradual transition between the inner and outer slopes.
Second, the merger could deposit dust which masks the
central surface brightness peak.  Third, the merger may be
accompanied by an inflow of gas which forms into a 
stellar disk or torus which, in turn, makes the original peak less distinct.

There is some evidence for each of these scenarios.  
For example, a disky stellar structure at $r < 2.0$\arcsec\ 
is suggested by the increasing ellipticity of isophotes towards the center;
WFPC1 surface photometry shows isophotes increasing in
ellipticity from 0.36 at 15\arcsec\ to 0.5 at 0\farcs5 (van den Bosch \etal 1994).
Evidence for the {\em recent} formation of a stellar disk includes the 
blue color of the stellar disk detected in ground-based images 
(Goudfrooij et al. 1994).  
The presence of dust is supported by the aforementioned observation
of central dust and gas (Ferrari \etal 1999).
Finally, in support of double nuclei or tori,
Byun \etal (1996) noted ``double nucleus?" for NGC 4473 in their Table 1.
Perhaps it is no coincidence that NGC 4473 also shows the most
conspicuous asymmetry in its STIS light profile for our entire sample -- 
a shoulder located $\approx$0.2\arcsec\ from the 
peak (see \S 4.5, Fig. \ref{plotprofall}).

%%%%%%%%%%%%%%%%%%%%%%%%%%%%%%%%%%%%%%%%%%%%%%%%%%%%%%%%%%%%%%%%%%%%%%%%%%
% Fig 17
\begin{figure*}[t]
\centerline{\psfig{file=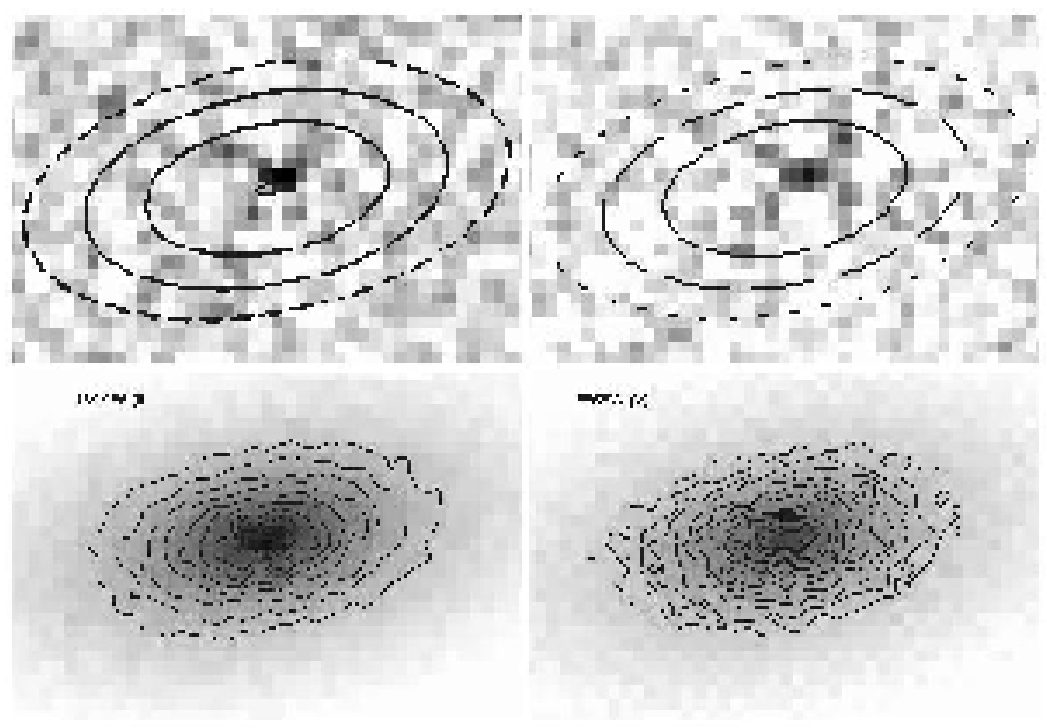,width=16cm,angle=0}}
\figcaption{WFPC2 greyscale images of the inner part ($r<$0\farcs7) of
NGC 4473.  Dark greylevels represent large pixel values in all panels.
Top: $I$-band (left) and $V$-band (right) residuals after subtraction
of a model with purely elliptical isophotes.  Some isophotes from the
ellipse fitting are shown.  The center was a free parameter.  Bottom:
$I$-band (left) and $V$-band (right) images before model subtraction,
displayed with the same orientation and scale.  Each has 40 iterations
of Lucy-Richardson deconvolution.  Minimally-smoothed contours are
overlayed with linearly-spaced levels.  The inner isophotes become so
boxy as to appear ``peanut"-shaped.
\label{N4473cntrs}}
\end{figure*}
%%%%%%%%%%%%%%%%%%%%%%%%%%%%%%%%%%%%%%%%%%%%%%%%%%%%%%%%%%%%%%%%%%%%%%%%%%

The above conclusions that dust, blue colors, and secondary
nuclei are present are all based on only marginal detections.
Thus, we have inspected the deconvolved $HST$ WFPC2 images for 
corroborating evidence.
The $I$ and $V$ band images (Figure \ref{N4473cntrs})
have 2000 s and 1800 s total exposure, respectively.
They provide higher resolution than the WFPC1 data used
by van den Bosch et al. (1994).
Our {\it HST} ($V-I$) color profile in Figure \ref{lp4473}
does not confirm the blue color of the disk,
although blue and red light may be mixed within each elliptical isophote.
A double nucleus is not present in the form of two, distinct maxima.
However, the contours appear compressed along
the minor axis so as to make isophotes ``peanut" shaped
at $r\approx $ 0.2\arcsec .
Ellipse fitting reveals that the strong diskiness
in isophotes at $r>$1.6\arcsec\ (peaking at 100\afour = $2.3 \pm 0.5$ 
at $r=$4.3\arcsec )
suddenly gives way to very strong boxiness at $r<1.6$\arcsec ;
the 100\afour drops below $-5.0$ at $r = $0.23\arcsec\
and then returns to values consistent with 0 at
the resolution limit ($r \approx 0.1\arcsec $).
The noisiness of the deconvolution process should be
taken into consideration; the peak
pixel is not in the same position in the two bottom images
and the minimally-smoothed contours are noisy.  
However, the position of the boxiest isophotes is robust, 
occurring at $r\approx $ 0.2\arcsec\ in both the $V$ and $I$ band 
deconvolved images
and reaching magnitudes as low as 100\afour = -5 in both datasets.
The top panels of Figure \ref{N4473cntrs} shows that subtracting
a purely elliptical model leaves behind an "$\times$"--shaped
residual in both cases.  The offset of the peak residual from
the best-fit ellipse centroids is also robust, suggesting an
asymmetric enhancement about 0.1\arcsec\ from those centroids.
At $r > 1$\arcsec\ (not shown), the residual pattern 
inverts to a dark ``+", indicating the diskiness previously reported
by others.

%%%%%%%%%%%%%%%%%%%%%%%%%%%%%%%%%%%%%%%%%%%%%%%%%%%%%%%%%%%%%%%%%%%%%%%%%%
% Fig 18
\hskip -30pt{\psfig{file=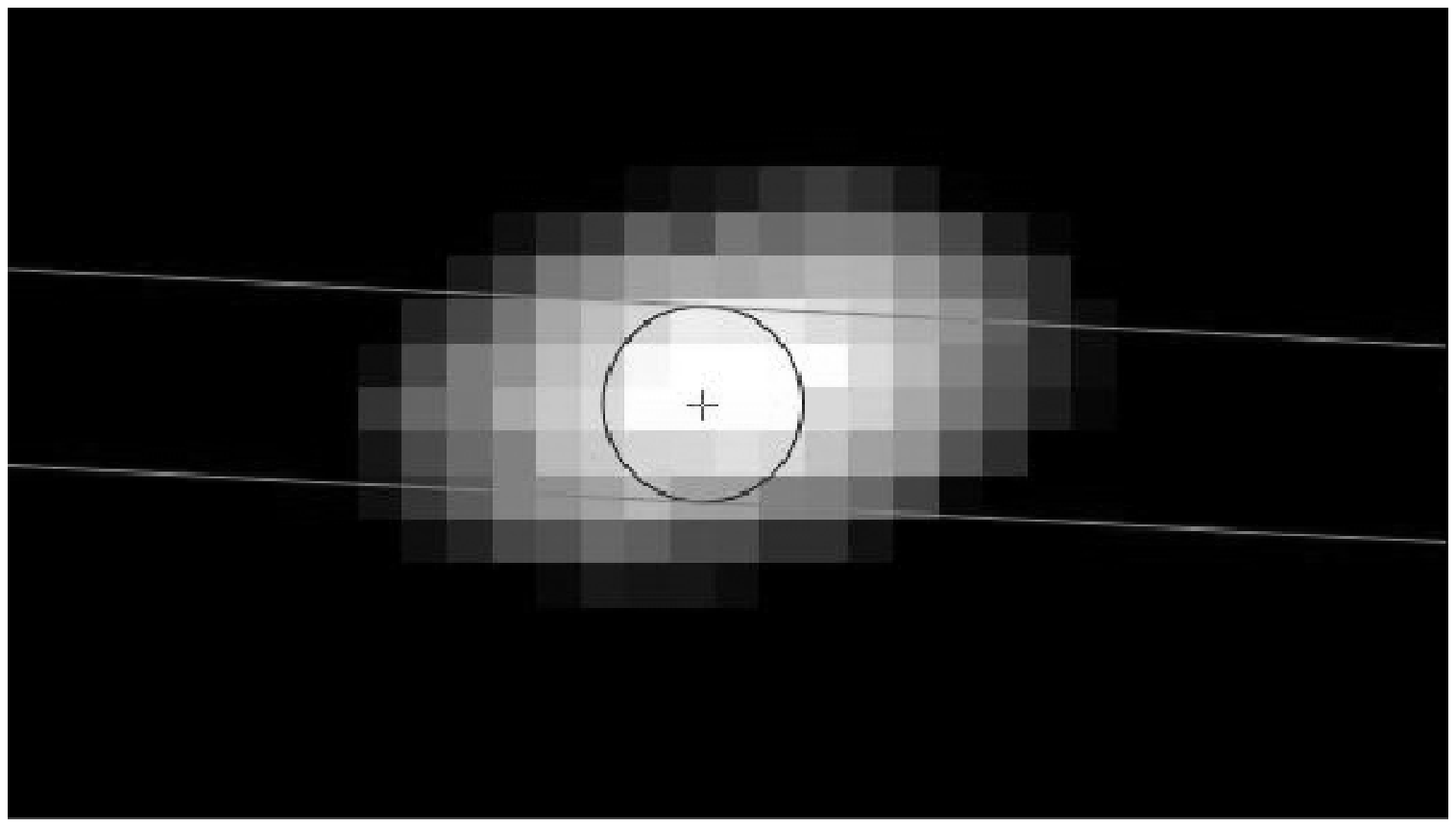,width=10.5cm,angle=0}}
\figcaption{$V$-band WFPC2 data shown with the STIS
slit overlayed to match the observation.  The circle is 0\farcs2 wide.
North is 173\arcdeg\ clockwise from up, and the position along the
slit used elsewhere in this paper increases towards the left.
\label{N4473slit}}
%%%%%%%%%%%%%%%%%%%%%%%%%%%%%%%%%%%%%%%%%%%%%%%%%%%%%%%%%%%%%%%%%%%%%%%%%%
\vskip 10pt

The asymmetry we see in the residuals after subtracting
an elliptical model are one way to explain the shoulder
in the light profile along the STIS slit.  A more direct
test is to simulate the STIS slit aperture on the WFPC2
data and create light profiles.  These profiles are shown
in Fig. \ref{plotprofall}.  The light is more concentrated
near the center because of the greater spatial resolution 
in the WFPC2 data, especially in the deconvolved data.  
Nevertheless, the profiles are higher to the right of the
center than to the left under the ``double nucleus" label.
% Can this surface photometry explain the shoulder in the
% light profile along the STIS slit?  
Figure \ref{N4473slit} is an overlay of the 0\farcs2 STIS slit
onto the $V$-band WFPC2 image.  
Notice that the STIS slit is skewed with respect to the major
axis by 20\arcdeg , more than any other galaxy in our sample
(this was necessary to acquire the guide stars).
Nevertheless, the position of the slit allows the passage
of some light from the region which created the "$\times$"--shaped
residuals.

There are no obvious signs of a recent merger in the kinematics.  No
counter-rotation is seen in the velocity profile, and the velocity
``overshoot" seen at 5\arcsec\ is typical of ellipticals with embedded
disks.
The photometry
and STIS light profile (Fig. \ref{plotprofall}) suggest a possible secondary 
nucleus at $r\approx 0.2''$,
so a kinematic deviation might be expected at STIS resolution.
The unsymmetrized STIS dispersion in Figs. \ref{kp4473} and \ref{kpcomp}
does, in fact, show a smooth gradient across the center, but it is only 
marginally significant.  NGC 4473 , like NGC 3384,
has a lower central dispersion in the STIS data than in the
ground-based data, indicating a kinematically cold stellar population
near the center.  
These two galaxies are interesting to contrast:
both show kinematic and photometric evidence for a cold stellar disk
all the way into the center, yet NGC 4473 has much lower $(v/{\sigma})$
than NGC 3384.  Moreover, NGC 4473 has a core profile with an indistinct
center,
while NGC 3384 has a power-law profile with the strongest central peak in our
sample.

We conclude that NGC 4473 is a peculiarity -- a disky
galaxy lacking a strong central cusp -- and that its
peculiarities probably originated with a merger.
The occurrence rate of such `disky cores' is low, judging by
the study of Rest \etal (2001) which classifies only
1 of 9 cores as ``disky" (within a sample of 57 galaxies).
Any {\it obvious} evidence for a recent merger is absent; there
is no counter-rotating stellar or gas system, 
no erratic dust, and no multiple nuclei.
But current work on the formation of cores 
(e.g., Faber \etal. 1997; Milosavljevi\'{c} \& Merritt 2001; Makino 1997) 
involves the merger and 
coalescence of two galactic nuclei with BH, and these scenarios
predict that observability of the merger will diminish 
continuously.
Some cores have been identified with easily resolved double
nuclei (e.g., NGC 4486B, Lauer \etal 1996), and recently, six
cores have been identified with subtle ``central inversions"
in their surface brightness profile
which are likely to be toroidal stellar systems (Lauer et al. 2002).
NGC 4473 may be another case of subtle central structures
resulting from coelescing binary nuclei.  
Like most of those in Lauer \etal (2002),
it has a poorly defined central peak and ``peanut"-shaped isophotes.
NGC 4473 differs in that it has disky isophotes at $r \simgt $1\arcsec .
Alternatively, no BHs were contributed by the last merger,
only gas which has subsequently formed a substantial disk at $r>$1.5\arcsec
and a non-cuspy structure (possibly a torus) near the limits of
resolution.
Future imaging at slightly higher resolution should clarify the central 
structure.

%%%%%%%%%%%%%%%%%%%%%%%%%%%%%%%%%%%%%%%%%%%%%%%%%%%%%%%%%%%%%%%%%%%%%%%%%%
% Fig 19
\begin{figure*}[t]
\centerline{\psfig{file=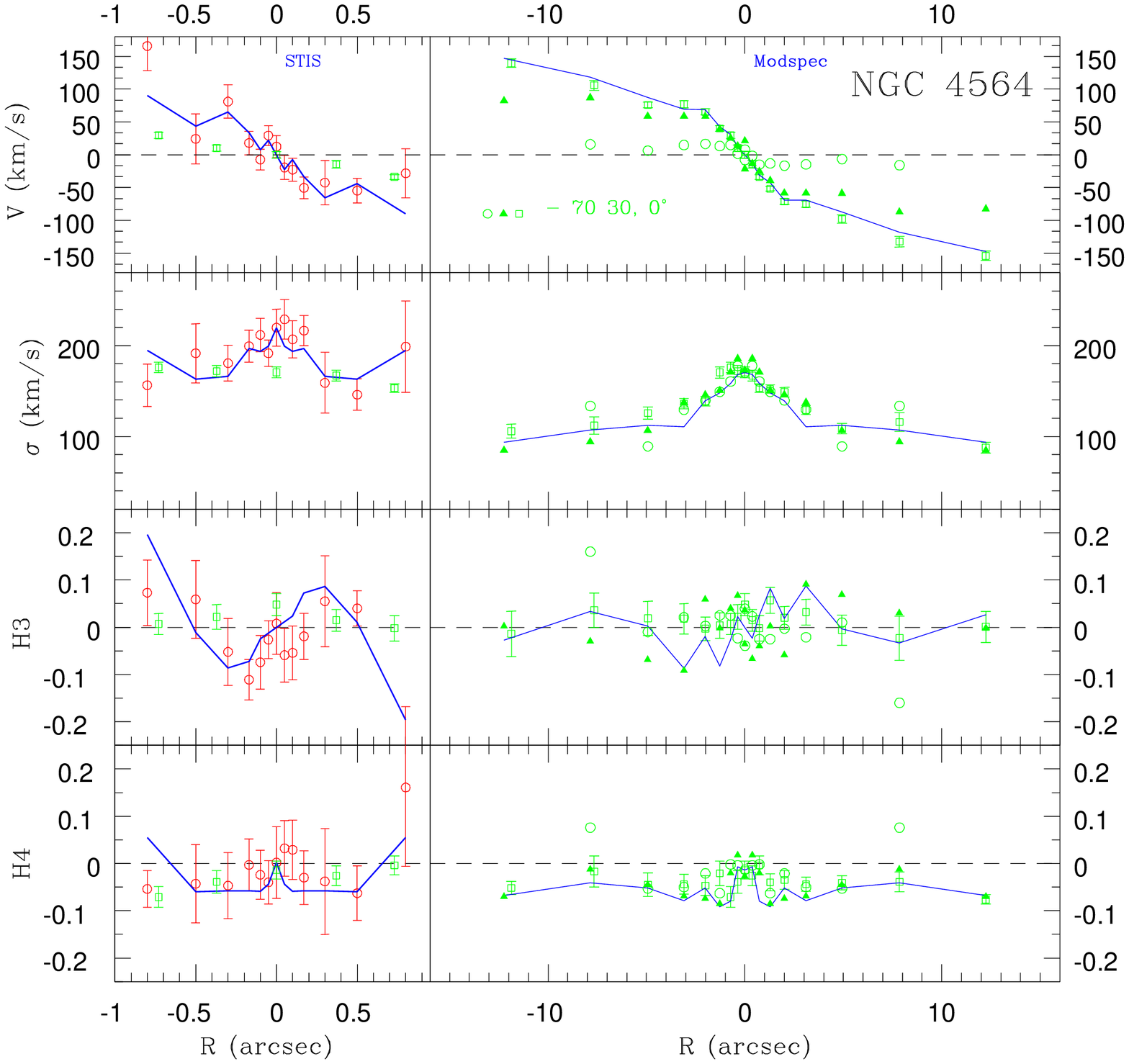,width=16cm,angle=0}}
\figcaption{Kinematic profiles for NGC 4564 (see Fig.~\ref{kp821} for
the meaning of the symbols).  For NGC 4564, all ground-based PAs are
from Ca spectra.
\label{kp4564}}
\end{figure*}
%%%%%%%%%%%%%%%%%%%%%%%%%%%%%%%%%%%%%%%%%%%%%%%%%%%%%%%%%%%%%%%%%%%%%%%%%%

%%%%%%%%%%%%%%%%%%%%%%%%%%%%%%%%%%%%%%%%%%%%%%%%%%%%%%%%%%%%%%%%%%%%%%%%%%
% Fig 20
\begin{figure*}[t]
\centerline{\psfig{file=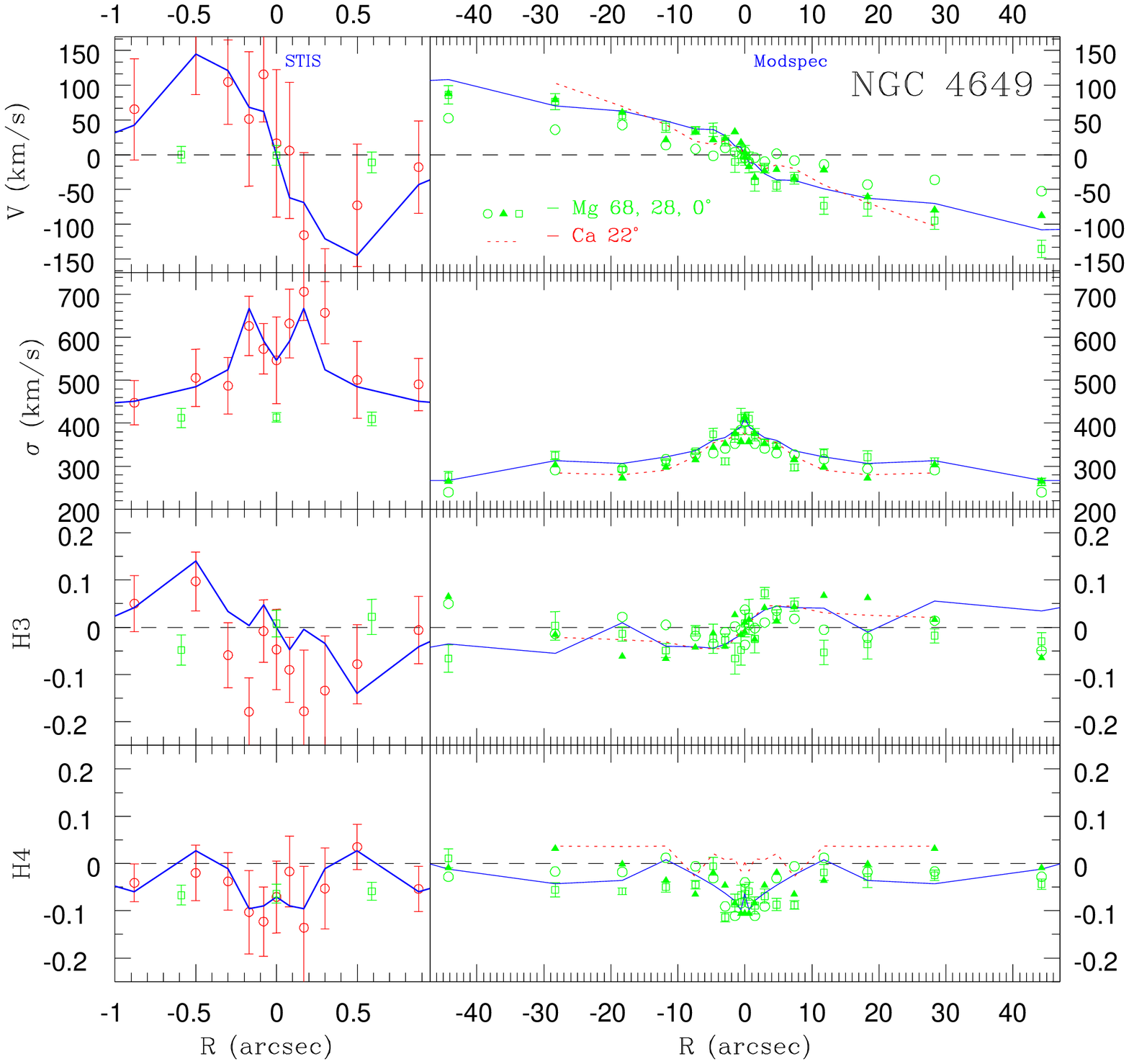,width=16cm,angle=0}}
\figcaption{Kinematic profiles for NGC 4649 (see Fig.~\ref{kp821} for
the meaning of the symbols).  For NGC 4649, all of the ground-based
data are from Mg spectra except the dashed line (PA = 22\arcdeg )
which is Ca spectra.
\label{kp4649}}
\end{figure*}
%%%%%%%%%%%%%%%%%%%%%%%%%%%%%%%%%%%%%%%%%%%%%%%%%%%%%%%%%%%%%%%%%%%%%%%%%%

%%%%%%%%%%%%%%%%%%%%%%%%%%%%%%%%%%%%%%%%%%%%%%%%%%%%%%%%%%%%%%%%%%%%%%%%%%
% Fig 21
\begin{figure*}[t]
\centerline{\psfig{file=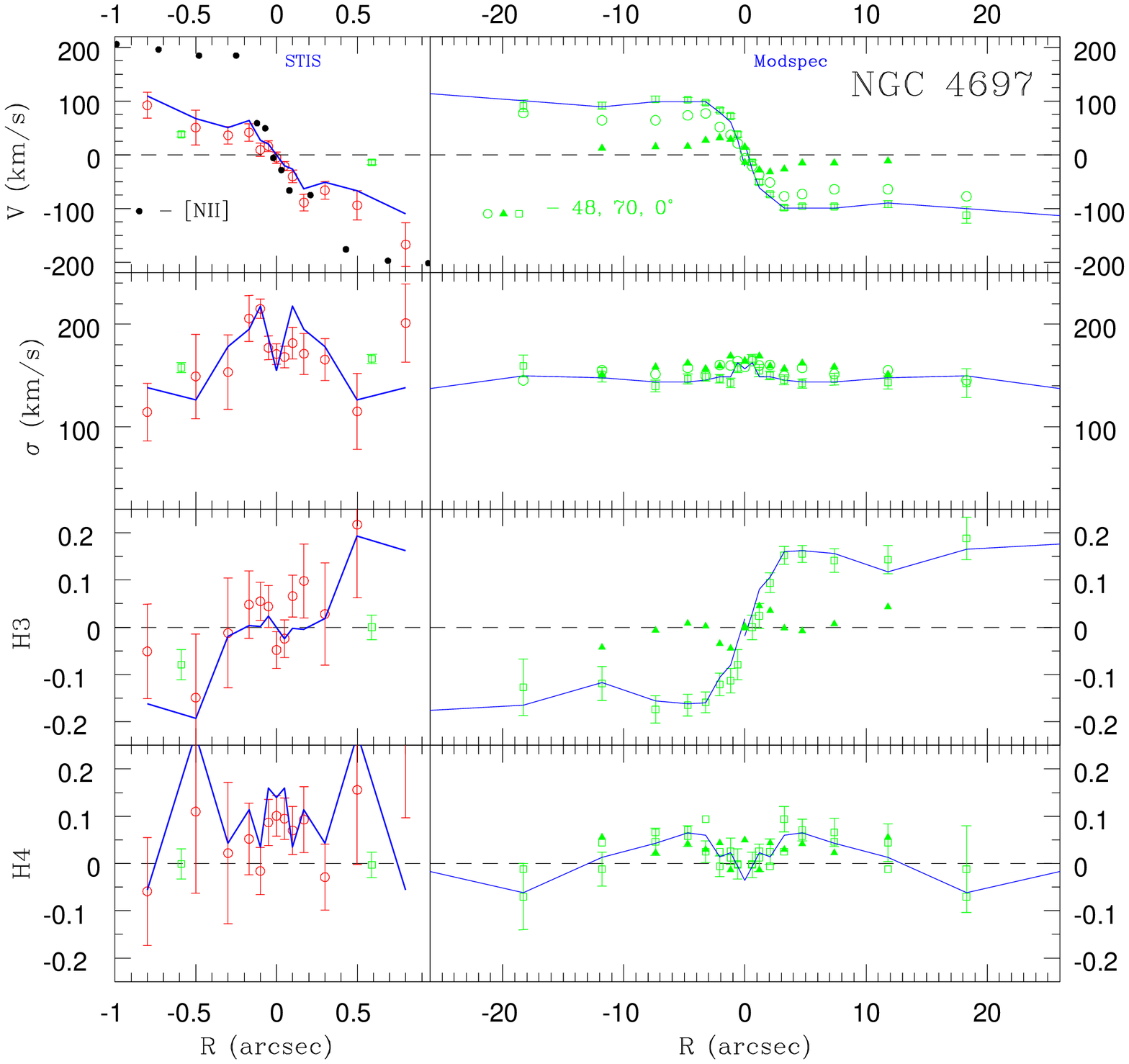,width=16cm,angle=0}}
\figcaption{Kinematic profiles for NGC 4697 (see Fig.~\ref{kp821} for
the meaning of the symbols).  For NGC 4697, all ground-based PAs are
from Ca spectra.  In the top left subpanel, we include filled circles
for [N II] emission line kinematics (see Pinkney \etal (2003)).
\label{kp4697}}
\end{figure*}
%%%%%%%%%%%%%%%%%%%%%%%%%%%%%%%%%%%%%%%%%%%%%%%%%%%%%%%%%%%%%%%%%%%%%%%%%%

%%%%%%%%%%%%%%%%%%%%%%%%%%%%%%%%%%%%%%%%%%%%%%%%%%%%%%%%%%%%%%%%%%%%%%%%%%
% Fig 22
%%%%%%%%%%%%%%%%%%%%%%%%%%%%%%%%%%%%%%%%%%%%%%%%%%%%%%%%%%%%%%%%%%%%%%%%%%

\subsection{NGC 4564}

NGC 4564 is classified by Sandage \& Tammann (1980) as an E6, but it has
characteristics of an S0.
First, its isophotes show disky deviations from an ellipse at large radii.
The parameter \afour begins to climb at 10\arcsec , reaching about 0.025
at 20\arcsec\ and then tapering down to 0.01 (BDM).  Such diskiness at
intermediate radii is also seen in NGC 821.
Second, its ellipticity increases gradually from 0.2 at $r=3$\arcsec\ 
to 0.6 at 30\arcsec\ (BDM).
Sahu \etal (1996) decompose their $R$ band image into an elliptical
bulge (obeying a de Vaucouleurs law) and a disk.  Their scale length for
the disk
is $r_{s} =28.14 "$ with axial ratio $b/a=0.3$.
Third, Sil'chenko (1997) finds a younger stellar population in the center
of the disk (8 Gyr) compared to the rest of the galaxy.

High-resolution images were obtained with WFPC1 (PID 2607,
Byun \etal 1996) and WFPC2 (Jaffe, PID 6357; Faber \etal 1997).
In the center, NGC 4564 appears more bulge-dominated, with
rounder isophotes and boxy deviations from ellipses.
For example, van den Bosch \etal (1994) find
that the ellipticity is only 0.16 at 0\farcs5
and that \afour is significantly negative around 2\farcs0.
They also find sharp variations in PA and $\epsilon$ near 2\farcs0.
This is indicative of dust; however, Roberts
\etal (1991) do not find evidence for an ISM.

Our ground-based kinematics for NGC 4564 (Figure \ref{kp4564})
can be compared to those of BSG and Halliday \etal (2001).
Our spectra have relatively low S/N (Figure \ref{specalln1}).
The published profiles are in good agreement with ours,
but are plotted to greater radii (r = 35\arcsec and 40\arcsec,
respectively).
The velocity dispersion declines
by a factor of 2 from 0\arcsec to 10\arcsec , so
one should expect large variations in published values of 
$\sigma$-dependent quantities.
For example, Bender 1988 give a $v_{max}$/$\sigma$ of 1.11
($v_{max}=150\pm$5\kmsc $\sigma$=135$\pm$5\kms ) while we have  1.3 $\pm
0.2$
($v_{max}=147\pm$4\kmsc $\ovl{\sigma}$= 112$\pm$17\kms , see
Table \ref{kinresGND} ).
The average of 6 central velocity dispersions in Hypercat is
158\kmsc while we measure $\sigma_0 \approx $170\kms\ in all of our Modspec
PAs.
Asymmetric ($h3$) deviations of the LOSVDs from a Gaussian are not
significant, but $h4$ appears to be negative at $r <$ 10\arcsec .
Halliday \etal (2001) show $h3$ becoming negative beyond 10\arcsec , 
coincident with an increase of $a_4$ above 0.
The sudden changes in the PA, ellipticity, $\cos 3\theta$, and $\cos
4\theta$ coefficients
at 2\farcs0 (van den Bosch 1994) are coincident with
subtle bends in our profiles of symmetrized $v$, $\sigma$, and $h4$.

The STIS kinematics feature a central velocity dispersion peak
of 220$\pm$20\kmsc well above the ground-based result.  The
higher resolution also reveals rotation of $\sim$50\kms\ in
the inner 1\farcs0.  $h3$ and $h4$ are fairly noisy, but
the symmetrized curves indicate a gradient in $h3$
across the inner 0\farcs5, and predominantly negative $h4$.
The kinematics allow a secure detection of a BH (Gebhardt \etal 2003),
but little evidence for nuclear activity is found
in X-rays (Pellegrini 1999) or radio continuum (Wrobel \& Herrnstein 2000).

\subsection{NGC 4649}

NGC 4649 (M60) is a giant elliptical in the Virgo cluster,
comparable in luminosity to M87.
NGC 4649 stands out in our sample as the galaxy with the lowest central
surface brightness (Fig. \ref{plotprofall}) and highest luminosity.
It is the only galaxy with a large enough dispersion
so that the crowding of the Ca II triplet absorption lines becomes
problematic.  This will be discussed further in \S 6.

We find a $10^{9.3}$ \Msun\ BH in NGC 4649 (Gebhardt \etal 2003),
only about 30\% less massive than the BH in M87
(Macchetto \etal 1997).
Fortunately, it has far weaker nuclear activity than M87,
thus simplifying a BH search.
Nuclear activity is detected in the radio (Fabbiano \etal 1987)
and
but not in the optical or X-rays (Byun \etal 1996; Di Matteo \& Fabian 1997).
Stanger \& Warwick (1986) also find only low-level extended radio emission
out to $\sim$3\farcs0.
Di Matteo \& Fabian (1997) estimate that the total measured core
flux from radio to X-ray is lower than 10$^{41}$ erg s$^{-1}$.

NGC 4649 is separated by only 2.5\arcmin\ ($\sim 12$ kpc for a distance of
16 Mpc) from the spiral NGC 4647.  Many studies report that they are
non-interacting (e.g., Sandage \& Bedke 1994),
although the spiral is clearly asymmetric (see Koopman \etal 2001).
It is given an E2 morphology in the RC2 (de Vaucouleurs \etal 1976), but
Sandage \& Bedke (1994) think it is an S0$_1$
because of its ``prominent extended envelope."   NGC 4649 has an
extended X-ray halo (Bohringer \etal 2000), and it has been identified with
its own compact subgroup within the Virgo cluster (Mamon 1989).

CCD surface photometry has been performed by
Peletier \etal (1990), Caon \etal (1990), and Michard \& Marchal (1994).
There are no significant deviations from ellipses in the isophotes
between 1\arcsec and 8\arcsec , and then they become mildly boxy
($100a_{4}/a \approx -0.5$).
The maximum ellipticity measured by Peletier \etal (1990) is $\approx 0.2$ 
which suggests a nearly face-on inclination ($i < 36$ \arcdeg).
This makes difficult the detection of disky isophotes (Michard \& Marchal 1994).
% The $\sim$ 36\arcdeg\ inclination (E2) is not conducive to disky isophotes.
Instead, support for the S0 classification is found in the surface brightness
profiles which show an outer envelope above the $r^{1/4}$-law bulge.
There is no strong evidence for a tidal influence from nearby NGC 4647.
HST WFPC2 $V$-band photometry is discussed by Byun \etal (1996) and
Faber \etal (1997).  NGC 4649 has a core profile with a break radius
of 3.58\arcsec .

Our ground-based kinematics can be compared to Fisher \etal (1995;
hereafter FIF), De Bruyne \etal (2001), and BSG.
FIF report an average rotation of 87\kmsc
and their rotation curve
climbs from 60 to 120\kms\ between $r=$20 to 60\arcsec .
Similarly, our rotation curve is climbing at 45\arcsec\ where it
is $\sim$110\kms\ (symmetrized).
In general, NGC 4649 has strong rotational support compared to
other giant ellipticals.  Its large ($v/\sigma$)* stands apart from the 
brightest cluster galaxies in the study of FIF.
Our unsymmetrized rotation curve is also fairly {\em asymmetric}
with respect to the galaxy center.  The data of
De Bruyne \etal (2001) demonstrate this even better with a rotation
curve out to 90\arcsec , showing an
$\sim$ 70\kms\ maximum difference in rotation between the two sides.
This asymmetry along the major axis provides some support for the idea
of an interaction with NGC 4647.

Our velocity dispersion profile is consistent with that of BSG and
De Bruyne \etal , but FIF has systematically lower values.  Most dispersion
profiles peak at $\sim$ 400\kms
and then rapidly decline to $\sim$ 300\kms\ by $r=20"$;
they show nearly a factor of two decrease from the center to the
effective radius (r$_{e} \approx 80"$).
De Bruyne measures positive $h4$ values within 10$"$ whereas
we measure negative $h4$ here at all of our PAs.

The STIS Ca II triplet data reveal an enormous central velocity
dispersion.  The entire inner 1\farcs0 is above 450\kmsc and
it appears to rise to over 600\kmsp  However, the uncertainties
are high because of the ill-defined absorption lines (\S 6).
This differs from the case of IC 1459 (Cappellari \etal 2002),
another giant elliptical ($M_B \approx -21.4$), 
which has $\sigma_0 \approx $ 350\kms\ in both STIS
and ground-based observations.
Another unusual (but uncertain) feature is the strong
rotation within 1\farcs0.  The maximum STIS rotation,
$\sim$150\kmsc  exceeds the maximum ground-based rotation.

\subsection{NGC 4697}

This is an E6 galaxy with a stellar disk along the
apparent major axis (Carter 1987; Goudfrooij \etal 1994a ).
NGC 4697 has more dust and gas than the other galaxies
in this sample.  Molecular
gas was detected by Sofue \& Wakamatsu (1993), and
the high IRAS 100 $\mu m$ flux densities also imply cool gas.
The radial color gradient is larger than the metallicity gradient
from the Mg$_2$ index (Peletier 1990).
{\it HST} WFPC1 imaging reveals an organized dust
disk (Lauer \etal 1995).
This dust was not successfully imaged from the
ground (V\'{e}ron-Cetty \& V\'{e}ron 1988; Kim 1989), but was
suggested by the red nucleus in $B-I$ (Goudfrooij \etal
1994a).
The H$\alpha$+[N II] emission proved difficult to detect
(Trinchieri \& di Serego Alighieri 1991; Kim 1989).
Long-slit spectroscopy by Goudfrooij \etal (1994b) gives
an H$\alpha$+[N II] flux of 2.4$\times 10^{-13}$ erg s$^{-1}$
cm$^{-2}$.  The narrowband imaging by these authors
shows H$\alpha$+[N II] emission extending out to  35\arcsec .
Ground-based imaging (e.g., Peletier 1990) indicates embedded,
disky isophotes. The parameter $100a_{4}/a$ is over 1.8 until
$r$= 20\arcsec and then it declines to 0.

A great deal of spectroscopy exists for comparison with our
Modspec results.
Bertola \& Capaccioli (1975) first pointed
out that ellipticals are not rotationally supported using
NGC 4697.  It has also been used for case studies showing
that 3-integral model are an improvement over 2-integral models 
(Dejonghe et al 1996; Binney \etal 1990).

Carter (1987) points out that there is a large
variation in $v_{max}$ measurements for this galaxy.
Bertola \& Capaccioli (1975) find 65\kms, Illingworth (1977)
gets 90\kms , and Davies (1981) derives 146\kms. We measure
a ground-based $v_{max} = 101\pm13$, and a STIS $v_{max}$ of
$110\pm21$\kms . A likely
explanation is the presence of a cold, rapidly rotating, stellar disk
(composing more than 10\% of the light along the major axis) which
has a varying effect on the measured radial velocity for different
methods.
Figure \ref{kp4697} includes the rotation curve of the excited
gas (black points) measured from [N II] emission with STIS
(Pinkney \etal 2003).
This provides an interesting contrast to the stars in that the
gas shows less pressure support beyond 0\farcs2.

\begin{figure*}[t]
\centerline{\psfig{file=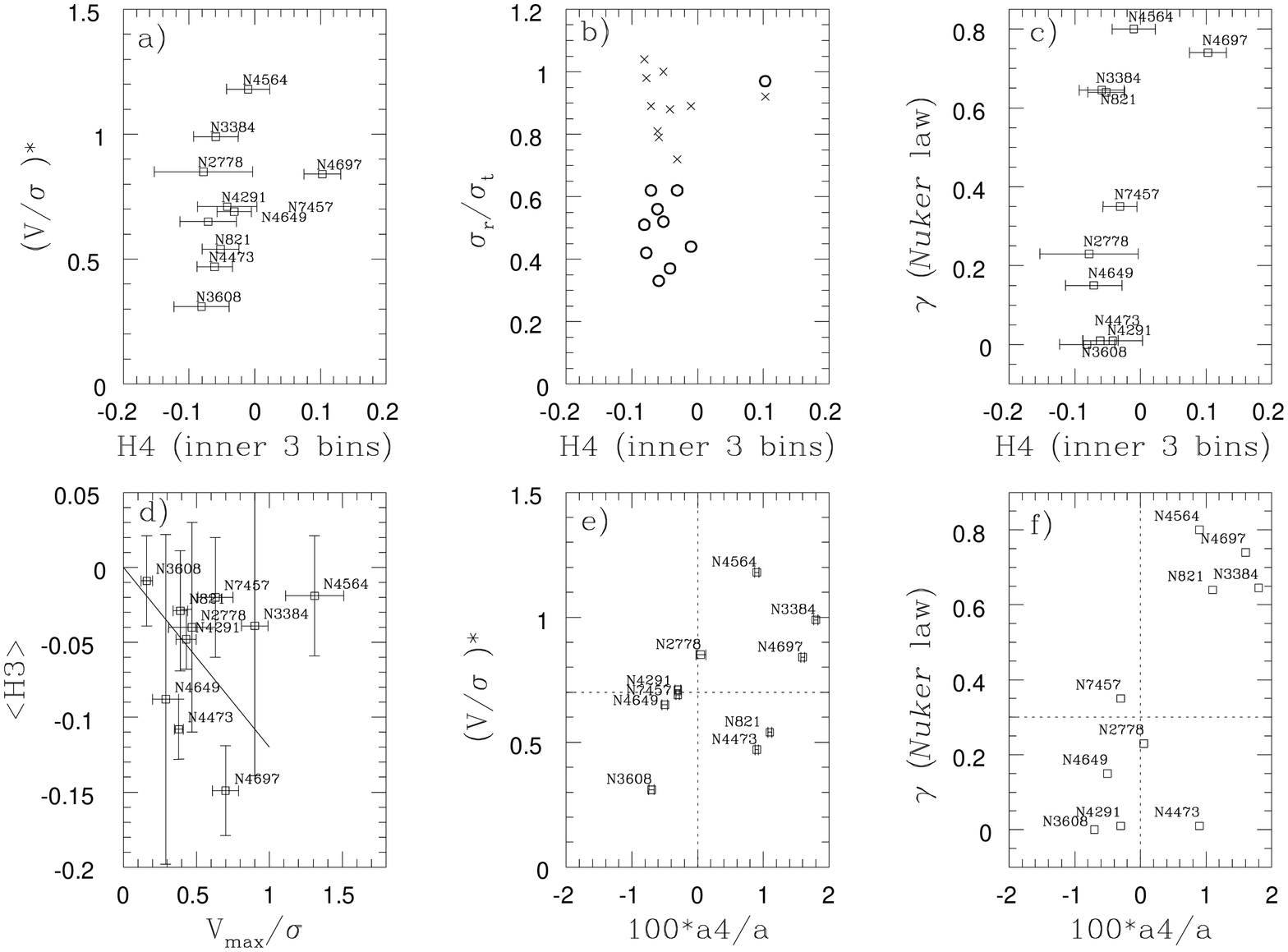,width=18cm,angle=-90}} \figcaption{Plots
of kinematic and structural parameters.  a) Anisotropy parameter
($v/\sigma$)* vs. $h4$ averaged over the innermost 3 STIS bins.
($v/\sigma$)* is calculated using
($v_{max}/\ovl{\sigma}$)/$\sqrt{\epsilon/(1-\epsilon)}$, where
$v_{max}$ and $\ovl{\sigma}$ are taken from the Table \ref{kinresGND}.
b) The ratio of the radial to tangential velocity dispersion taken
from the modeling by Gebhardt \etal (2003), $\sigma_{r}$/$\sigma_{t}$,
vs. $h4$ from the innermost 3 STIS bins.  The circles have
$\sigma_{r}$/$\sigma_{t}$ measured in the central modeling bin, while
the x-marks were measured at $R_{e}$/4 (see Gebhardt \etal 2003).  c)
Asymptotic power-law slope of the surface brightness profile,
$\gamma$, vs.  $h4$.  d) Mean $h3$ over $r>$2\farcs0 of the
symmetrized, Modspec data vs $v_{max}/\sigma$.  The solid line is the
mean relation found by BSG.  e) Ground-based anisotropy vs 100\afour ,
the deviations of isophotes from perfect ellipses (taken from the
literature).  Disky galaxies have 100\afour $>$ 0.  f) Asymptotic
power-law slope, $\gamma$, vs. 100\afour .
\label{h4plot}}
\end{figure*}

The mean central velocity dispersion given by Hypercat is 174\kmsp
This is a bit higher than our central, GB, velocity dispersions of
157, 159, and 164\kmsc for our 0, 48, and 70\arcdeg\ slit
orientations, respectively.  We measure a higher central
dispersion with STIS, 171\kmsp  This galaxy stands out
as one of the few with a local minimum at $r=0$ in the STIS
velocity dispersion profile (Fig. \ref{kp4697}).
It is also unusual in that its $h4$ parameter is significantly
positive (see Fig.~\ref{h4plot})

\subsection{NGC 7457}

NGC 7457 is the smallest, lowest luminosity galaxy in our
sample.  It is an S0 galaxy, like N3384.
NGC 7457 is another example of a {\it pseudobulge} 
(see \S 5.3) in our sample, based on photometry and kinematics.
Tomita \etal (2000) find a point-like, bluish nucleus in
the WFPC2 ($V,I$) data, and Peletier \etal (1999) find a young stellar
population in the bulge.
Michard \& Marchal (1994) detect a small bar.

Lauer \etal (1991) report on the WFPC surface brightness
profile and find it unresolved and steep.
The improved {\it HST} WFPC2 imaging reveals a surface
brightness profile with a very shallow outer power
law; its $\beta = 1.05$ (Lauer 2002, private communication) is the
smallest in a sample of 29 early types.
However, the surface brightness profile in Andredakis  \etal (1995)
has a Sersic parameter $n \approx$ 6.2 which is higher than the values
$n \simlt $2.0 more typical of pseudobulges.
Most impressive is the strong surface brightness peak in
deconvolved WFPC2 F555W data (Lauer 2002, private communication) which jumps
4 magnitudes in the inner 0\farcs1.  Tomita \etal (2000) do not
interpret this as an AGN because of the lack of H$\alpha$ and
H$\beta$ emission reported by Ho \etal (1997).  However, Gebhardt
\etal (2003) find a central drop in calcium triplet equivalent widths 
and interpret the peak as AGN light.

\begin{figure*}[t]
\centerline{\psfig{file=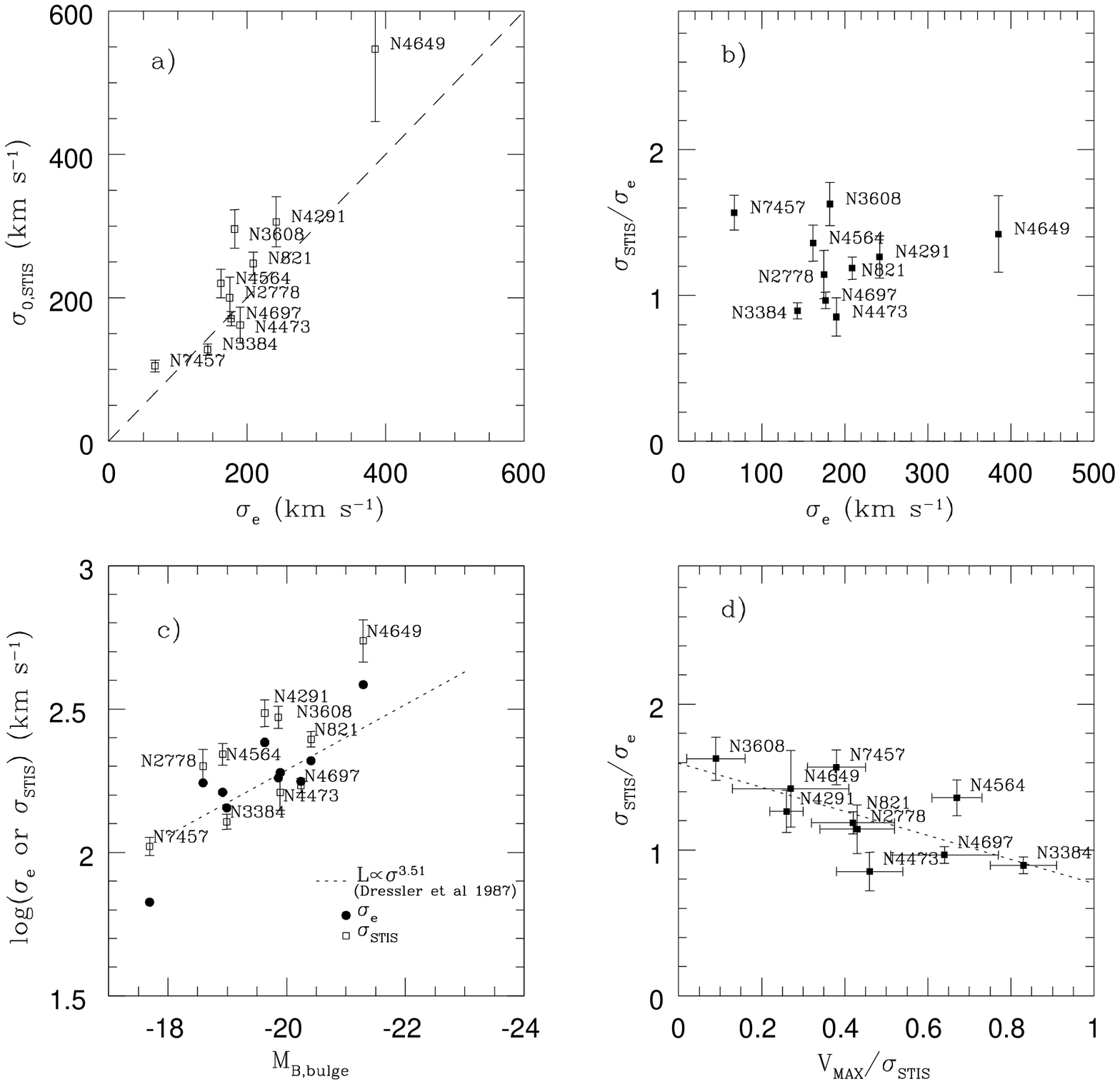,width=16cm,angle=0}}
\figcaption{Comparisons of high- and low-resolution velocity
dispersion measurements for our 10 target galaxies.  (a) Velocity
dispersion measured in the central STIS bin vs.  the effective
velocity dispersion (a luminosity-weighted dispersion from Modspec
data).  (b) Ratio of central STIS dispersion to the effective
dispersion vs.~effective velocity dispersion.  (c) The Faber-Jackson
relation for our ground-based dispersion, $\sigma_{e}$ (filled
circles), and our central STIS dispersion, $\sigma_{STIS}$ (open
circles).  The best-fit relation from Dressler \etal (1987) is shown
as a dashed line.  (d) Ratio of the centrally measured STIS velocity
dispersion to the effective dispersion, $\sigma_e$, versus the maximum
ground-based rotation divided by $\sigma_{STIS}$.
\label{sigplot}}
\end{figure*}

We find evidence for pseudobulge-like kinematics for NGC 7457.
Our Figure \ref{sigplot} demonstrates that the velocity dispersion falls 
low on the Faber-Jackson relation.  This result is robust for different global
values of $\sigma$ from Table \ref{kinresGND}.  
Our $v_{max}$/$\sigma$ values (Table 8)
are not especially large, however, given the ellipticity.

%%%%%%%%%%%%%%%%%%%%%%%%%%%%%%%%%%%%%%%%%%%%%%%%%%%%%%%%%%%%%%%%%%%%%%%%%%
% Fig 23
%%%%%%%%%%%%%%%%%%%%%%%%%%%%%%%%%%%%%%%%%%%%%%%%%%%%%%%%%%%%%%%%%%%%%%%%%%

%%%%%%%%%%%%%%%%%%%%%%%%%%%%%%%%%%%%%%%%%%%%%%%%%%%%%%%%%%%%%%%%%%%%%%%%%%
% Fig 24
\begin{figure*}[t]
\centerline{\psfig{file=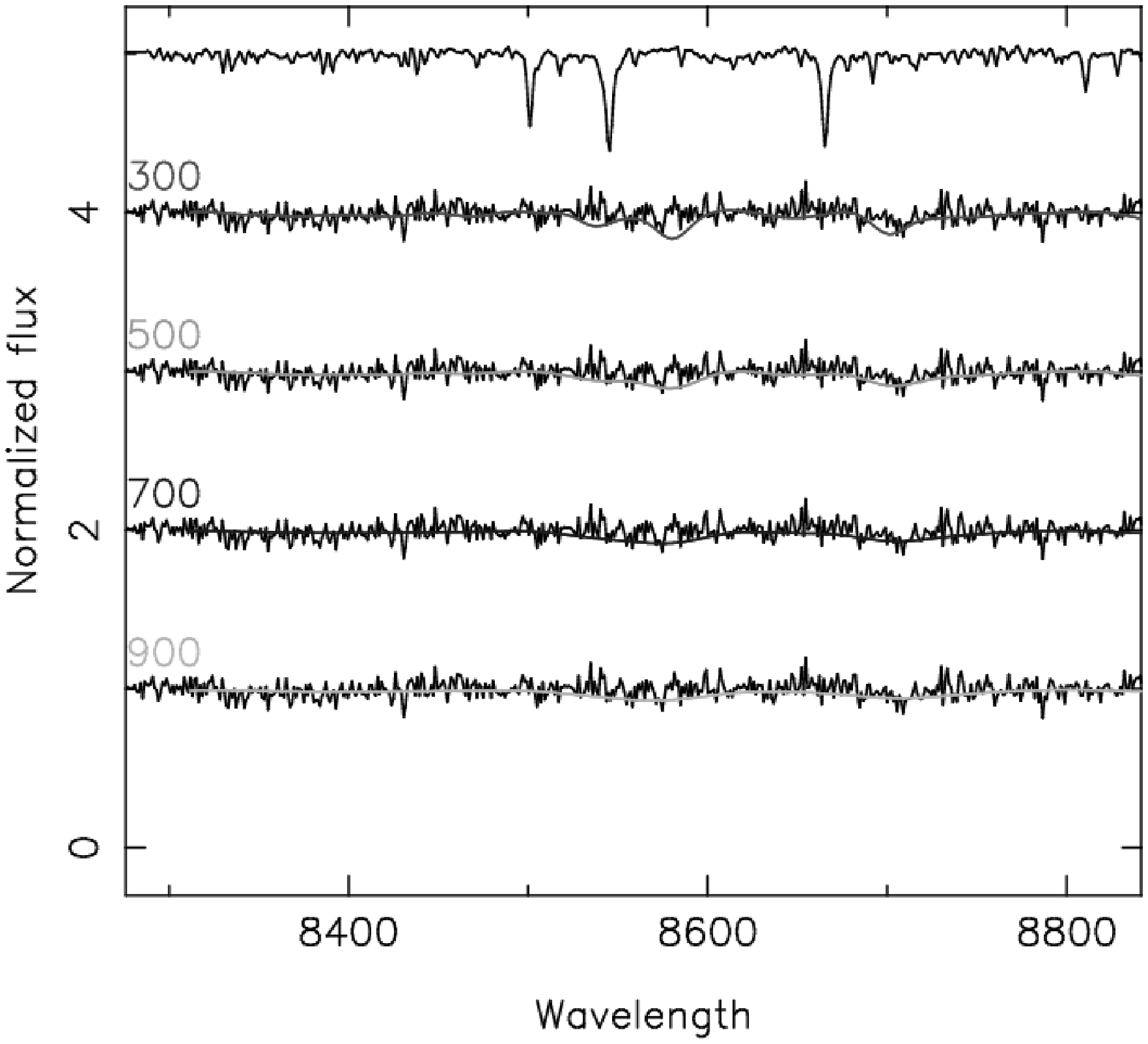,width=16cm,angle=0}}
\figcaption{Broadened template spectra overlaid upon the central, STIS
spectrum for NGC 4649.  The unbroadened template (HR6770) is shown on
top.  It is convolved with progressively broader functions with
$\sigma$=300, 500, 700, and 900\kmsc and then redshifted to 1110\kmsp
\label{N4649}}
\end{figure*}
%%%%%%%%%%%%%%%%%%%%%%%%%%%%%%%%%%%%%%%%%%%%%%%%%%%%%%%%%%%%%%%%%%%%%%%%%%

%%%%%%%%%%%%%%%%%%%%%%%%%%%%%%%%%%%%%%%%%%%%%%%%%%%%%%%%%%%%%%%%%%%%%%%%%%
% Fig 25
%%%%%%%%%%%%%%%%%%%%%%%%%%%%%%%%%%%%%%%%%%%%%%%%%%%%%%%%%%%%%%%%%%%%%%%%%%

\section{Discussion}

\subsection{Central Velocity Dispersions}

To what extent does the STIS central velocity dispersion
drive the detection of the nuclear black holes?
Modeling is required to fully answer this question; however,
a strong peak in dispersion in the central 0\farcs5 ($\sim$ 45 pc at $d=17$
Mpc)
is difficult to explain without a central dark mass (van der Marel 1994b).
In Figure \ref{sigplot}, we compare the central STIS dispersion to our
ground-based, ``effective" dispersion, $\sigma_e$ (i.e., the rms
velocity relative to the systemic velocity of the galaxy, averaged
over a slit width of 1\arcsec\ extending to $R_e$; G00).
In 7 out of 10 cases, the STIS central value is higher.
The most extreme differences are NGC 4649 (162\kmsc 42\%),
NGC 3608 (114\kmsc 55\%),
NGC 4564 (58\kmsc 37\%), NGC 4291 (64\kmsc 17\%), and NGC 7457 (38\kmsc
60 \%).
Three of these have core profiles, and
it appears that the core galaxies show the largest differences.
NGC 7457 has a 60\% increase from $\sigma_e$ to $\sigma_{STIS}$,
the greatest fractional rise in the entire sample.
Figure \ref{sigplot}b demonstrates
that the fractional increase does not correlate with increasing
dispersion.  Likewise, there is no correlation with $M_{B,bulge}$ (not
shown).
NGC 4473,
unlike the other core galaxies, has a $\sigma_{STIS}$ lower than
$\sigma_{e}$.
This adds to the growing list of differences between NGC 4473
and the cores (\S 5.6).
The other galaxies with lower $\sigma_{STIS}$ are
NGC 3384 and 4697, which are also disky.
However, NGC 4697 has a higher $\sigma_{STIS}$ than our
alternative ground-based dispersions $\sigma_{0,GB}$ and
$\ovl{\sigma}$ in Table \ref{kinresGND}.
Eight of our galaxies, including NGC 3384, show a local maximum at
$r=$0\farcs0 in
their STIS dispersion profile.
NGC 4473 shows only a slight central, local maximum, while NGC 4649
and NGC 4697 show local minima in their symmetrized $\sigma$
profiles.
Thus, the majority of galaxies (70 -- 80\%) show
a central peak at STIS resolution, and about 70\% have
a larger $\sigma_{0,STIS}$ than $\sigma_{e}$,
Only one galaxy, NGC 4697, does not obey either criterion.

In Figure \ref{sigplot}c, we show the Faber-Jackson (1976) relation
for our 10 galaxies.  The $\sigma_e$ is an adequate surrogate for
the traditional central dispersions measured through 
$\sim 2\times2$\arcsec\ apertures, 
but use of $\sigma_{STIS}$ adds scatter to the relation
presumably because of the strong dispersion gradient induced by
the central black hole.
We also find a marginal correlation between
the ratio $ \sigma_{STIS} / \sigma_e $ and the ratio
$ v_{max} / \ovl{\sigma}$, where $v_{max}$ is the maximum
rotation in the ground-based data and $\ovl{\sigma}$
is the average dispersion in Table \ref{kinresGND}.
Apparently, galaxies with more rotational support have
less of a dispersion peak at STIS resolution.
The probability that no correlation exists is $<$ 3\%
(using Cox regression and Kendall's Tau,
Isobe \etal 1986)
This correlation is present but not as strong if $ v_{max} / \ovl{\sigma}$
is replaced by ($v_{max} / \ovl{\sigma}$)* .

\subsection{Low Surface Brightness Giant: NGC 4649}

The galaxies in the BH literature, including our own sample, show a
preference
for high surface brightness in core galaxies since these are easier to
observe.
In fact, the measurement of dispersion profiles in core galaxies is so
difficult that most of the BH detections in these galaxies rely on
emission lines from rotating gas disks
[e.g., M87 (Harms 1994), NGC 4261 (Ferrarese \etal 1996),
M84 (Bower \etal 1997)].
In the case of the giant E3, IC1459, Cappellari \etal (2002) have
measured the BH mass using stellar kinematics as well as gas kinematics.
However, this galaxy does not have an especially faint surface
brightness, $\mu_{V,0.1''}\simeq 15.3$ (Carollo \etal 1997).
Ground-based stellar kinematics have provided a BH mass estimate for
M87 (Dressler \& Richstone 1990; van der Marel 1994a) but the
BH was not required to fit these data.
The preference for high surface brightness cores could conceivably
introduce a bias in plots of \mbhsig\ or $\mbh - L_{B}$.
Moreover, the measurement of BH mass using gas kinematics
gives a different answer than stellar kinematics in the
only giant where they have been compared (IC1459, Cappellari \etal 2002).
Thus, NGC 4649 is important as the only BH measurement
from stellar kinematics in a galaxy which is representative of
low-surface brightness cores.

Not surprisingly, we find that the LOSVD is particularly difficult to
measure
in NGC 4649.  We requested sufficient exposure
time to measure Ca lines in a galaxy with $\mu_{V,0\farcs1}$=15.9 mag arcsec$^{-2}$
and $\sigma \approx 350$\kmsp  However, this galaxy appears to
have a {\em very} large central velocity dispersion at STIS resolution.
This
broadens the lines to such an extent that their detection
is difficult even at this S/N (Figure \ref{N4649}).
Our MPL technique was able to measure an LOSVD
with $ \sigma > $500\kmsp  Naturally,
the upper limits are more weakly constrained than the lower limits.  The FCQ
technique gave more erratic results (not shown in Fig. \ref{kp4649}) for
these data.

\subsection{Gauss-Hermite Parameters}

Our parameterized stellar LOSVDs suggest properties of a galaxy's
orbital structure that are elucidated by the 
3-integral modelling.  This merits a comparison between the two.
% Our parameterization of the stellar LOSVDs allows a comparison
% to some results from the 3-integral modeling.
Gebhardt \etal (2003) find the tangential component
of the velocity dispersion tensor to be stronger in the central
bins of the galaxy than in the bins near $R_{e}/4$ (see Figure
\ref{h4plot}b).
The $h4$ parameter also indicates a tangential bias when
it takes a negative value (i.e., the LOSVD is boxy).
Indeed, we find that our ground-based $h4$ profiles usually
have a central dip (NGC 821 is a good example). Moreover,
the central, STIS$-$measured $h4$ parameters, although noisier, are also
predominantly negative.
To measure a more reliable central STIS $h4$, we averaged the
$h4$ values from the innermost 3 bins.
Figure \ref{h4plot} demonstrates that these values of $h4$
are negative in 9 out of 10 cases.
Individual cases are not significantly negative, but altogether
they suggest a real trend.  NGC 4473 and NGC 3608 are the most
significantly negative.  There is one outlier, NGC 4697, which
has a positive $h4$ and this pulls the weighted mean to
$-0.027\pm 0.01$. When this point is excluded, the weighted
mean is $-0.05\pm 0.01$.
We looked for correlations between the central $h4$ value and
other photometric and kinematic parameters.  The large uncertainties
in the $h4$ values and the small sample do not permit
any significant results.
A positive correlation is hinted at by Figures \ref{h4plot}a, and
\ref{h4plot}c.

It is puzzling that the disky elliptical, NGC 4697, shows positive central
$h4$
values while the other disky galaxies (NGC 821, 3384, 4564, and 4473) have
negative central $h4$ values.   One might search for other ways
in which NGC 4697 differs from the other four.  The most obvious difference
is its prominant dust disk at $r<3.5$\arcsec .  This disk is inclined
at 77\arcdeg\ and has a sharp outer cut-off and a less clear-cut inner
radius (Pinkney \etal 2003).   One would expect the same LOSVD
regardless of the presence of the dust disk, if the dust disk
is an infinitely thin screen centered within an axisymmetric galaxy.
However, one could invoke a disk with finite thickness to preferentially
obscure,
say, an equatorial cold disk component and thereby influence the LOSVDs.
A second possible scapegoat is the stellar disk itself, which
may be more (or less) prominent in the central STIS bins
of NGC 4697 than the other four disky galaxies.
One expects a superposition of two LOSVDs with the same
centroid but different widths to produce an LOSVD with
positive $h4$.  Thus, we may find positive values of $h4$ in
galaxies where a bulge and disk population overlap.
However, there is little evidence in ground-based kinematics
that NGC 4697 has a significantly different disk contribution
than the other galaxies.  In fact, NGC 4473 seems to have the
strongest contribution of light from a stellar disk
at $r<5$\arcsec\ : it was the
only galaxy that warranted the inclusion of a disk
component in the modeling (Gebhardt \etal 2003).
Finally, we see in Figure \ref{h4plot} that
NGC 4697 has the lowest mean $h3$ at $r>2$\arcsec\
of the entire sample.
This is another artifact of diskiness, or more correctly,
of strong rotation.  Unfortunately, the kinematic properties
measured at $r\simgt 1\farcs0$ do not necessarily
predict the properties at $r\simlt 0\farcs1$ where the
$h4$ values in question are measured.

The bottom three panels of Figure \ref{h4plot} complement the
investigation of Gauss-Hermite parameters.  First, Figure \ref{h4plot}d shows
our 10 galaxies plotted on a correlation found by BSG.  Our
galaxies do not follow the trend well, but this is mostly
because of NGC 4564.  The $v_{max}$/$\sigma$ in this figure
uses $\ovl{\sigma}$,
but
if we used $\sigma_{0,GB}$ or $\sigma_{e}$  (Table \ref{kinresGND}),
$v_{max}$/$\sigma$ would be 0.9 for NGC 4564 instead of 1.3.
NGC 4564 happens to have the most extreme variation between
$\overline{\sigma}$ and our other two ground-based $\sigma$ estimates
for reasons that are clear in Figure \ref{kp4564}.
Second, Figure \ref{h4plot}e demonstrates
that galaxies with strong rotational support tend to be disky.
This has been established with larger samples by BSG.
Third, Figure \ref{h4plot}f shows a
correlation between the asymptotic inner slope $\gamma$
of the best-fit {\it Nuker}-law surface brightness profile, and the
deviations of isophotes from ellipses.  This follows from
the results of, e.g., Faber \etal (1997) wherein the disky
galaxies (100\afour $>0$) tend to have power-law profiles and
the boxy-isophote galaxies (100\afour $<0$) tend to have core
profiles.  This figure is intended to underscore the peculiarity of
NGC 4473 as a ``disky core."

\subsection{Demographic Results}

Our sample of early-type galaxies includes a wide range of
velocity dispersions (70 to 385\kms ) and BH masses
($10^{6.5}$ to $10^{9.3}$ \Msun ), allowing us to address questions
of BH demographics (\S 1).  A word should be said about
selection effects near the high-mass end of the
\mbhsig\ relationship.
The measurement of LOSVDs becomes
increasingly more time consuming with decreasing surface brightness.
Therefore, a selection bias is introduced whereby,
given two core galaxies of the same luminosity, or the same dispersion,
one is compelled to select the one with the higher surface brightness.
We selected one challenging galaxy located in this region of
parameter space, NGC 4649, with $\mu_{0,V} = 15.9$ mag arcsec$^{-2}$.
Nevertheless, none of our cores, including NGC 4649, represent the lowest
surface brightnesses known in their luminosity range.
It is not certain whether the lowest surface brightness galaxies
obey the same demographic trends.

With the discovery of the \mbhsig\ relation, it is natural
to look for exceptions to the rule.   It appears
that the relation holds for Sa and Sb galaxies with
classical bulges, and even later types (Sc -- Sd) with pseudobulges
(Kormendy \etal 2002).
The Sc galaxy M33 appears exceptional in that it has a
BH mass upper limit significantly lower
than that predicted using the dispersion of its
nuclear star cluster (Gebhardt \etal 2001).
This galaxy, however, lacks a hot spheroidal component
and so it is debatable whether the \mbhsig\ relation
is applicable.

In Figure \ref{demplot2}, our ten BH masses are plotted against
velocity dispersion.  For just these galaxies,
we fit an \mbhsig\ relation of the form
\beq
\log_{10}(\mbh/\Msun )=\alpha + \beta \log_{10}(\sigma_{e}/200 \mbox{ km
s$^{-1}$}).
\eeq
The fitting procedure is identical to the one advocated by Tremaine \etal
(2002),
where we have assumed 5\% error in $\sigma_e$, and the uncertainty in
\mbh\ is the measured one combined with an extra 0.34 dex of intrinsic
dispersion, which gives a $\chi^2$ per degree of freedom of unity.
We find $\alpha = 8.06\pm0.13$ and $\beta = 3.67\pm0.70$.
Note that the BH masses for the galaxies used in this fit have been refined
since G00.  All galaxy distances are now taken from
Tonry \etal (2001),
and more dynamical models have been run to better
probe chi-square space (Gebhardt \etal 2003).
Nevertheless, the best fit is nearly indistinguishable from the G00 fit.
All of these same refinements to our sample are incorporated into the
preferred fit by Tremaine \etal (2002), which includes an additional
21 galaxies from the literature.
The most deviant point is for NGC 2778.
This faint galaxy has a lower S/N than most.
Also, its sphere of influence is unresolved.  Consequently, NGC 2778
has the least confident BH detection when modeled (Gebhardt \etal 2003).

%%%%%%%%%%%%%%%%%%%%%%%%%%%%%%%%%%%%%%%%%%%%%%%%%%%%%%%%%%%%%%%%%%%%%%%%%%
% Fig 26
\hskip -30pt{\psfig{file=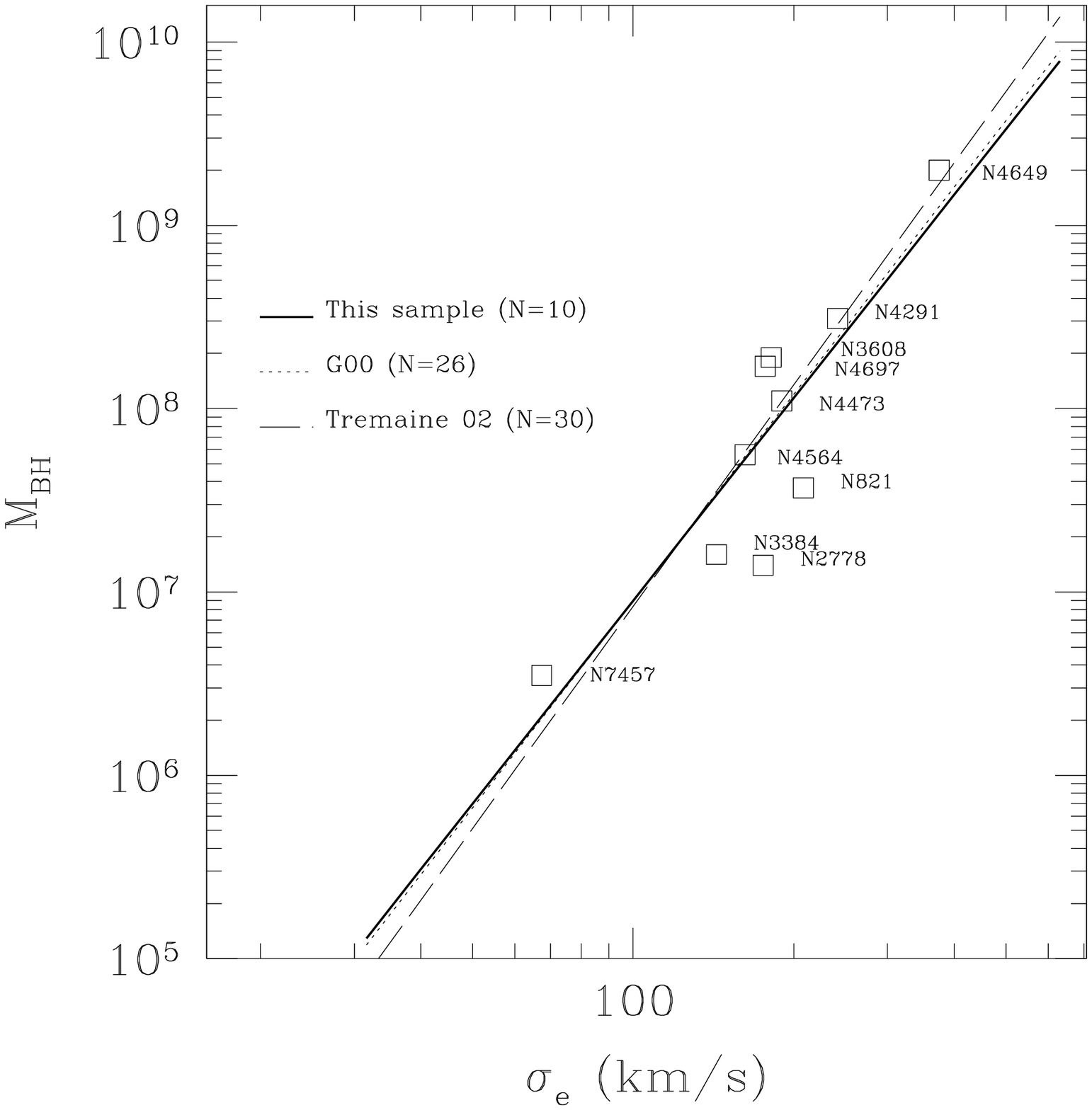,width=10cm,angle=0}}
\figcaption{Updated plot of \mbhsig\ for only the galaxies in this
paper (labelled squares).  The best-fit line to these 10 galaxies is
shown as a solid line.  We overlay the original fit by Gebhardt \etal
2000 (dotted line), and the preferred fit by Tremaine \etal 2002
(dashed line).
\label{demplot2}}
%%%%%%%%%%%%%%%%%%%%%%%%%%%%%%%%%%%%%%%%%%%%%%%%%%%%%%%%%%%%%%%%%%%%%%%%%%
\vskip 10pt

In hierarchical scenarios, the formation of bulges involves mergers
which are believed to have occurred during the
first few Gyr of the universe.  But what is the effect of a
merger occurring $\simlt$ 1 Gyr before the present?
Will the likely increase in bulge dispersion be accompanied
by the appropriate increase in BH mass?
We have inspected our galaxies for merger candidates
to see if they are systematically displaced on the \mbhsig\
relation.
Three of our galaxies show marginal evidence for recent
mergers: NGC 3608, 4473, and 4697.
NGC 3608 has a counter-rotating core and patchy dust.
NGC 4473 has unusual photometric properties that
may be related to a merging component (\S 5.6).
NGC 4697 has a dust ring which may have been deposited by a merger.
As can be seen in Figure \ref{demplot2}, all three of these fall reasonably
close to the \mbhsig\ relation.
Two of these, NGC 4697 and 4473, are among our most secure
BH detections.

\section{Conclusions}

We have presented long-slit spectroscopy from $STIS$ aboard $HST$ and
the MDM 2.4-m telescope that is used to derive black-hole
masses for 10 galaxies by Gebhardt \etal (2000).
Ninety percent of the galaxies have a centrally peaked velocity
dispersion and/or a higher dispersion from STIS than the ground-based
Modspec.
The galaxies with the strongest rotational support, as
quantified by $v_{MAX}$/$\sigma_{STIS}$, have the smallest dispersion
excess at STIS resolution.
When only our ten galaxies are used for a fit to \mbhsig , the slope is
3.68, which is similar to the value of 3.75 found by G00 for 26 galaxies.
We review individual galaxies and identify candidates
for recent merger: NGC 4697, 3608, and 4473.
These galaxies are {\em not} outliers in the \mbhsig\ relation.
We also identify pseudobulges in NGC 3384 and 7457
and find that they also obey the \mbhsig\ relation.
Finally, we find a trend toward flat-topped line-of-sight velocity
distributions
(i.e., negative values of $h4$) at the center of the galaxies,
implying a tangential bias in the stellar velocity dispersion.

\acknowledgements

We thank the TACs at {\it HST} and MDM for telescope time, and Bob Barr
for observing assistance.  This work was
supported by {\it HST} grants to the {\it Nuker} team, GO-6587 and
7388, and by NASA grant NAG 5-8238 to D.O.R..
Support to A.V.F. was provided by NASA grant NAG 5-3556 and
the Guggenheim Foundation.
Support for the {\it HST} proposal \# 7388 was provided by NASA through a grant
from the Space Telescope Science Institute, which is operated by
the Association of Universities for Research in Astronomy, Inc.,
under NASA contract NAS 5-26555.
This research used the NASA/IPAC Extragalactic Database
(NED) which is operated by the Jet Propulsion Laboratory,
Caltech, under contract with NASA.
This work also used NASA's
Astrophysical Data System and the Lyon-Meudon Extragalactic
Database Hypercat.

\clearpage
\begin{deluxetable}{lllllccrc}
\tablecaption{\sc Target Galaxies}
\tablewidth{0pt}
\tablehead{
\colhead{Name}   &
\colhead{Type\tablenotemark{a}} &
\colhead{$M_B$ \tablenotemark{b}} &
\colhead{Profile\tablenotemark{c}} &
\colhead{Dist.\tablenotemark{d}} &
\colhead{Scale\tablenotemark{e}} &
\colhead{$\mu_V$ \tablenotemark{f} }  &
\colhead{PA \tablenotemark{g} }  &
\colhead{$r_e$ \tablenotemark{h} }  \\
\colhead{NGC}    &
\colhead{    }  &
\colhead{mag }  &
\colhead{Type}    &
\colhead{Mpc }    &
\colhead{pc/$\arcsec$} &
\colhead{mag/$\sq\arcsec$ }  &
\colhead{deg }   &
\colhead{$''$ }
}
\startdata
821     & E6?   & -20.41 & $\backslash$             & 24.1 & 117 & 14.3 & 25
&  50  \\  % Dold=23
2778    & E     & -18.59 & $\backslash$             & 22.9 & 111 & 14.9 & 40
&  16  \\  % Dold=(31)
3384\tablenotemark{i} & SBs0- & -18.99 & $\backslash$  & 11.6 &  56 & 13.7 &
53 & 25   \\  % Dold=10.9
3608    & E2    & -19.86 & $\cap$                   & 23.0 & 111 & 14.7 & 75
&  34  \\  % Dold=20.8
4291    & E     & -19.63 & $\cap$                   & 26.2 & 127 & 14.9 &
110 & 17   \\  % Dold=22.0
4473    & E5   & -19.89 & $\backslash \cap$         & 15.7 &  76 & 14.7 &
100 & 26   \\  % Dold=14.7
4564    & E     & -18.92 & $\backslash$             & 15.0 &  73 & 14.1 & 47
& 20   \\  % Dold=16.8
4649    & E2    & -21.29 & $\cap$                   & 15.9 &  88 & 15.9 &
105 & 69   \\
\enddata
\tablenotetext{a}{Galaxy type from NED.}
\tablenotetext{b}{Blue absolute magnitude, $M_{B}$, calculated using surface
brightness fluctuation (SBF) distances.}
\tablenotetext{c}{Surface brightness profile type characterized as either
$\backslash$ (power-law) or $\cap$ (core).  Sources are Faber \etal (1997),
Byun \etal (1996, for NGC 4473), and Lauer (2002, private communication, for NGC
821, 2778, 4291). }
\tablenotetext{d}{Distances are from SBF measurements in Tonry \etal
(2000). }
\tablenotetext{e}{Spatial scale derived from col. 5.}
\tablenotetext{f}{Central ``$\mu_{V}$" is the $V-$band surface brightness
measured at $r=$0\farcs1 .}
\tablenotetext{g}{Position angle from the RC3.}
\tablenotetext{h}{Effective radius derived from the RC3, in arcseconds.}
\tablenotetext{i}{NGC 3384 also identified as NGC 3371.}
\label{c7targets}
\end{deluxetable}

\begin{deluxetable}{llllccll}
\footnotesize
\tablecaption{\sc Long-Slit Spectrograph Configurations}
\tablewidth{0pt}
\tablehead{
\colhead{Code}  &
\colhead{Instrument}   &
\colhead{Slit size}    &
\colhead{$\lambda_{cen}$}\tablenotemark{a} &
\colhead{$\lambda$-range \tablenotemark{a} } &
\colhead{Dispersion$^{-1}$ \tablenotemark{b}} &
\colhead{Comp. line $\sigma$ \tablenotemark{c}} &
\colhead{Spatial scale} \tablenotemark{d}  \\
\colhead{ }          &
\colhead{+ grating}    &
\colhead{$\arcsec \times \arcsec$ }    &
\colhead{ \AA }        &
\colhead{ \AA }    &
\colhead{\AA\ pix$^{-1}$}  &
\colhead{\AA\ (\kms) } &
\colhead{\arcsec ~pix$^{-1}$ }
}
\startdata
1 & STIS + G750M & 52$\times$0.1 & 8561 & 8275-8847 & 1.1089 & 0.45 (17.5) &
0.05071 \\
2 & STIS + G750M &52$\times$0.2 &  8561 & 8275-8847 & 1.1089 & 0.68 (23.8) &
0.05071 \\
3 & STIS + G750M &52$\times$0.1 &  6581 & 6295-6867 & 1.1130 & 0.37 (16.9) &
0.05071 \\
4 & Charlotte+831g/mm & 760$\times$0.8 & 8500 & 7801-9199 & 1.44 & 0.98
(35.) & 0.593 \\
5 & Wilbur+831g/mm &760$\times$0.9 &  8500 & 7580-9420 & 0.909 & 1.10 (39.)
& 0.371  \\
\enddata
\tablenotetext{a}{Central wavelength and wavelength range. STIS values taken
from Leitherer \etal (2001, pp. 231, 234). }
\tablenotetext{b}{Reciprocal dispersion was measured using our own
wavelength solutions.
The distribution of dispersion solutions for a given dataset had a
$\sigma \approx 0.00015$ \AA pix$^{-1}$.
The average dispersion given in the Handbook for G750M is 1.12 \AA
pix$^{-1}$. }
\tablenotetext{c}{Instrumental line widths measured by fitting
Gaussians to emission lines on comparison lamp exposures.
This gives an estimate of the instrumental
line width for {\em extended} sources.  We use $\sim$5 lines per
exposure,
and at least 5 measurements per line.
In the cases of the G750M setups, the comparison lamp exposures were
unbinned, while the galaxy spectra were binned by 2X along the dispersion
axis. The 2X binning increases the
measured widths by, e.g., $\sim$25\% for the 0\farcs1 slit,
and by $\sim$ 3\% for the 0\farcs2 slit at 8561 \AA .
Leitherer \etal (2001, p. 300) give these instrumental line widths for
point sources: $\sigma$=13.3, 15.0, 16.7\kms\ for the first three
G750M setups in this table. }
\tablenotetext{d}{The spatial scale along the slit is constant, but it
varies across the slit from one grating to the next.  It is 0.05597 \arcsec\ 
pix$^{-1}$ for G750M at 8561 \AA\, and 0.05465 \arcsec pix$^{-1}$ for
G750M at 6581 \AA\ (STIS ISR 98-23). }
\tablecomments{ Most of the STIS setups used a CCD readout with 2X
overbinning
along the dispersion, and a readnoise of $\sim$ 1 $e^{-}$pix$^{-1}$.
The Modspec setup values are given by the program ``modset" by J.
Thorstensen.
The slit widths, spectral and spatial resolutions varied for the Modspec
observations; characteristic values are shown here .
}

\label{specs}
\end{deluxetable}
\begin{deluxetable}{lllccc}
\tablecaption{\sc STIS Observations}
\tablewidth{0pt}
\tablehead{
\colhead{Name}   &
\colhead{Exposure}  &
\colhead{Obs. Date \tablenotemark{a} } &
\colhead{PA \tablenotemark{b} } &
\colhead{$\Delta$PA \tablenotemark{c} } &
\colhead{Setup \tablenotemark{d} }   \\
\colhead{NGC}     &
\colhead{seconds  } &
\colhead{ DDMMYY  } &
\colhead{ degrees } &
\colhead{ degrees } &
\colhead{Code }
}
\startdata
821  & 27478   & 131098,080999 & $-150$ &   5 & 1 \\
2778 & 15026     & 030699         &   50 &  10 & 1 \\
3384 & 27744   & 100598,160499 &   58 &   5 & 1 \\
3608 & 12950    & 140598         &   94 &  19 & 2 \\
4291 & 16013    & 210498         &  110 &   0 & 2 \\
4473 & 12840    & 041297         & $-100$ & $-20$ & 2 \\
4564 & 15060     & 080398         & $-143$ & $-10$ & 1 \\
4649 & 46467    & 090699         &   95 & $-10$ & 2 \\
4697 & 24820  & 270799,200899 &   68 &  $-2$ & 1,3 \\
7457 & 15282  & 240999         &  130 &   0 & 1 \\
\tableline
HR7576\tablenotemark{e} & 5x3.2s,5x3.2s & 140599  & $-173$ & \nodata & 1,2 \\
HR6770                  & 5x2.2s,5x3.2s & 060999  &  54 & \nodata & 1,2
\enddata
\tablenotetext{a}{The date at the end of the visit. }
\tablenotetext{b}{Position angle of the STIS slit.  From the FITS headers. }
\tablenotetext{c}{Difference between the STIS position angle and the RC3 PA. }
\tablenotetext{d}{Code for the spectrograph setup used (see Table \ref{specs}).}
\tablenotetext{e}{Template stars were HR7576 ($V$=5.03 mag, K3 III) and
HR6770 ($V$=4.64 mag, G8 III). }
\label{stisobs}
\end{deluxetable}

\begin{deluxetable}{llrrc}
\tablecaption{\sc Ground-based (M-D-M 2.4-m + Modspec) Observations}
\tablewidth{0pt}
\tablehead{
\colhead{Name}   &
\colhead{Obs. Date \tablenotemark{a} } &
\colhead{PA \tablenotemark{b} } &
\colhead{Exposure \tablenotemark{c}}  &
\colhead{Setup \tablenotemark{d} }   \\
\colhead{NGC}     &
\colhead{ MM/YY } &
\colhead{ \arcdeg } &
\colhead{(sec) } &
\colhead{Code }
}
\startdata
821   & 11/95 &  0 &  6$\times$1600  & 6 \\ % Charlotte at Mg. I see 11/95 Spectra of Mg at 0 and 90
821   & 11/95 & 85 &  5$\times$1600  & 6 \\ \hline
2778  & 11/97 &  0 & 23$\times$1200 & 5 \\       % Wilbur at CaT
2778  & 11/97 & 70 & 20$\times$1200 & 5 \\ \hline
3384  & 03/97 &  0 & 8$\times$1200  & 4 \\       % Charlotte at CaT.
3384  & 01/97 & 10 & 6$\times$1200  & 6 \\   % 1/97 Mg spectra, 6x1200
3384  & 03/97 & 70 & 7$\times$1200  & 4 \\
3384  & 01/97 & 88 & 6$\times$1200  & 6 \\
\enddata
\tablenotetext{a}{The month and year of observation. }
\tablenotetext{b}{The position angle of the slit relative to
the major axis of the galaxy. }
\tablenotetext{c}{Number of exposures $\times$ exposure length.}
\tablenotetext{d}{Code from Table \ref{specs} for the spectrograph setup
used.}
\label{gndobs}
\end{deluxetable}

\begin{deluxetable}{lcrrrrrr}
\tablecaption{\sc Nuker-law Profile fits}
\tablewidth{0pt}
\tablehead{
\colhead{NGC No. }   &
\colhead{Filter \tablenotemark{a} }  &
\colhead{I$_{b}$  \tablenotemark{b} } &
\colhead{r$_{b}$ \tablenotemark{c} } &
\colhead{$\alpha$  }  &
\colhead{$\beta$  }  &
\colhead{$\gamma$ \tablenotemark{d} }  &
\colhead{Source \tablenotemark{e} }
}
\startdata
 821 & H & 13.2   & 1.1   & 1.0   &  1.6   & 0.64 & Rav   \\ \hline
 % 821 & I & 14.6   & 0.7   & 0.4   &  1.8   & 0.16  & Nuk   \\
 % 821 & V & 16.1   & 0.7   & 0.4   &  1.9   & 0.01  & Nuk   \\ \hline
2778 & I & 14.5   & 0.3   & 0.3   &  1.8  & 0.30  & Nuk    \\
2778 & V & 15.1   & 0.1   & 0.4   &  1.6   & 0.16  & Nuk    \\ \hline
3384 & I & 15.1   & 2.9   & 11.2   & 1.8   & 0.65  & Nuk    \\
3384 & V & 16.3   & 2.9   & 12.0   & 1.8   & 0.64  & Nuk    \\ \hline
3608 & V & 15.5   & 0.3   & 1.1   &  1.3   & 0.00  & Fab    \\ \hline
4291 & H & 12.3   & 0.5   & 2.1   &  1.5   & 0.02  & Rav    \\
4291 & R & 15.1   & 0.6   & 1.4   &  1.6   & 0.00  & Rest   \\
\enddata
\tablenotetext{a}{Filter used: H=F160W(nic), I=F814W, R=F702W, V=F555W.}
\tablenotetext{b}{Surface brightness at the break radius. }
\tablenotetext{c}{Break radius: the radius of the transition between inner
and outer power-laws.}
\tablenotetext{d}{Inner power-law slope. }
\tablenotetext{e}{Source for fit parameters. Rav=Ravindranath \etal 2001;
Fab=Faber \etal 1997; Rest=Rest \etal 2001; Nuk=Lauer 2002, private
communication. }
\tablecomments{ These are fits to the surface brightness
profile along the major axis. }
\label{profiles}
\end{deluxetable}

\begin{deluxetable}{lrcr}
\tablecaption{\sc Bins for spectral extraction }
\tablewidth{0pt}
\tablehead{
\colhead{Name}   &
\colhead{r$_{cen}$($''$)}   &
\colhead{Range(pix)}   &
\colhead{Width(pix)}
}
\startdata
STIS CCD &  &    &   \\ \hline
cen & 0.000 & 511 - 511 & 1 \\
r1 & 0.051 & 512 - 512 & 1 \\
r2 & 0.101 & 513 - 513 & 1 \\
r3 & 0.177 & 514 - 515 & 2 \\
r4 & 0.304 & 516 - 518 & 3 \\
r5 & 0.507 & 519 - 523 & 5 \\
r6 & 0.811 & 524 - 530 & 7 \\
l1 & -0.051 & 510 - 510 & 1 \\
l2 & -0.101 & 509 - 509 & 1 \\
l3 & -0.177 & 507 - 508 & 2 \\
l4 & -0.304 & 504 - 506 & 3 \\
l5 & -0.507 & 499 - 503 & 5 \\
l6 & -0.811 & 492 - 498 & 7 \\ \hline
Wilbur  & &         &   \\ \hline  % Use 0.365"/pix
cen & 0.0 & 300 - 300 & 1 \\
r1 & 0.37 & 301 - 301 & 1 \\
r2 & 0.73 & 302 - 302 & 1 \\
r3 & 1.28 & 303 - 304 & 2 \\
r4 & 2.01 & 305 - 306 & 2 \\
r5 & 3.10 & 307 - 310 & 4 \\
r6 & 4.93 & 311 - 316 & 6 \\
r7 & 7.85 & 317 - 325 & 9 \\
r8 & 12.23 & 326 - 339 & 14 \\ \hline
Charlotte  & &         &   \\ \hline  % Use 0.59"/pix
cen & 0.0 & 300 - 300 & 1 \\
r1 & 0.59 & 301 - 301 & 1 \\
r2 & 1.48 & 302 - 303 & 2 \\
r3 & 2.95 & 304 - 306 & 3 \\
r4 & 4.72 & 307 - 309 & 3 \\
r5 & 7.38 & 310 - 315 & 6 \\
r6 & 11.80 & 316 - 324 & 9 \\
r7 & 18.29 & 326 - 336 & 11 \\
r8 & 28.32 & 337 - 359 & 23 \\
r9 & 44.25 & 360 - 390 & 31 \\
r10 & 69.03 & 391 - 443 & 52
\enddata
\label{bins}
\end{deluxetable}

\begin{deluxetable}{rlrrcrr}
\tablecaption{\sc Summary of STIS Kinematics}
\tablewidth{0pt}
\tablehead{
\colhead{Name}   &
\colhead{$\sigma_{0,STIS}$ \tablenotemark{a} }  &
\colhead{$v_{max} $ \tablenotemark{b} } &
\colhead{$v_{max}/\sigma_{0,STIS}$ \tablenotemark{c} } &
\colhead{$v_{grad}$ \tablenotemark{d} }  &
\colhead{$h4_{0}$ \tablenotemark{e} }  &
\colhead{$h4_{3}$ \tablenotemark{f} }
  \\
\colhead{NGC}  &
\colhead{\kms }  &
\colhead{\kms }   &
\colhead{  }   &
\colhead{\kms arcsec$^{-1}$ }    &
\colhead{  }   &
\colhead{  }
}
\startdata
821  & 248 (16) & 103 (23) & 0.42 (0.10) & 510 & $-$0.04  (0.05) & $-$0.05
(0.03) \\
2778 & 200 (29) & 61 (18) & 0.43 (0.09) & 129 & $-$0.15  (0.17) & $-$0.08
 (0.08) \\
3384 & 128 ( 8) & 85 (10) & 0.83 (0.08) & 414 & $-$0.04  (0.07) & $-$0.06
 (0.03) \\
3608 & 296 (27) & 28 (20) & 0.09 (0.07) & 113 & $-$0.13  (0.08) & $-$0.08
 (0.04) \\
4291 & 306 (35) & 74 (18) & 0.26 (0.04) & 410 & $-$0.00  (0.07) & $-$0.04
 (0.05) \\
4473 & 162 (25) & 67 (13) & 0.46 (0.08) & 278 & $-$0.06  (0.04) & $-$0.06
 (0.03) \\
4564 & 220 (20) & 90 (34) & 0.67 (0.06) & 218 & 0.00  (0.03) & $-$0.01
 (0.03) \\
4649 & 547 (101) & 145 (70) & 0.27 (0.14) & 403 & $-$0.07  (0.08) & $-$0.07
(0.04) \\
\enddata
\tablenotetext{a}{Dispersion measured in the central STIS bin with error
from
LOSVD fitting.}
\tablenotetext{b}{Maximum rotational velocity from symmetrized STIS data.
Points with large scatter were ignored in the case of NGC 821 and NGC
2778. }
\tablenotetext{c}{Ratio of maximum rotation (over all data) and central
STIS dispersion.}
\tablenotetext{d}{Velocity ``gradient" is measured as the maximum of
v17/0.17
or v30/0.3, where v17 is the symmetrized rotation velocity at 0.17\arcsec , and
v30 is the rotation at 0.30\arcsec . The entire rotation curve is used to define
the galaxy's systemic velocity, not the central bin.  }
\tablenotetext{e}{The value of h4 in the central STIS bin ($r$=0\farcs0).}
\tablenotetext{f}{The weighted mean of the 3 central h4 values ($r$=0\farcs0,
$r$=$\pm$0\farcs05, not symmetrized). }
\label{kinresSTIS}
\end{deluxetable}
\begin{deluxetable}{rrccrcccrr}
\scriptsize
\tablecaption{\sc Summary of Modspec Kinematics}
%\rotate
\tablewidth{0pt}
\tablehead{
\colhead{Name}   &
\colhead{ $\sigma_{0,GB}$ \tablenotemark{a} }  &
\colhead{ $\overline{\sigma}$ \tablenotemark{b} }  &
\colhead{ $\sigma_{e}$ \tablenotemark{c} }  &
\colhead{ $v_{max,GB} $ \tablenotemark{d} } &
\colhead{ ${v_{max}}/{\overline{\sigma}}$ \tablenotemark{e} } &
\colhead{(${v_{max}}/{\overline{\sigma}}$ )*  \tablenotemark{f} } &
\colhead{$\epsilon$  \tablenotemark{g} } &
\colhead{$h4_{0,GB}$ \tablenotemark{h} }  &
\colhead{$<h3>$ \tablenotemark{i} }
  \\
\colhead{NGC}  &
\colhead{\kms }  &
\colhead{\kms }  &
\colhead{\kms }  &
\colhead{\kms }   &
\colhead{  }   &
\colhead{ }   &
\colhead{ }   &
\colhead{ }   &
\colhead{  }
}
\startdata
 821 & 235 (~3) & 192 (21) & 209 &  74 ( 6) & 0.39 (0.05) & 0.54 & 0.34
& -0.07  (0.02) & -0.03 (0.04) \\
2778 & 202 (10) & 179 (52) & 175 &  85 (14) & 0.47 (0.16) & 0.85 & 0.24
& -0.06  (0.04) & -0.04 (0.07) \\
3384 & 134 (~2) & 119 (7) & 143 & 107 ( 8) & 0.90 (0.09) & 0.99 & 0.45
& -0.05  (0.01) & -0.04 (0.10) \\
3608 & 181 (~5) & 150 (12) & 182 &  24 ( 5) & 0.16 (0.04) & 0.31 & 0.21
& -0.08  (0.02) & -0.01 (0.03) \\
4291 & 262 (12) & 183 (18) & 242 &  79 (10) & 0.43 (0.07) & 0.71 & 0.27
& -0.04  (0.03) & -0.05 (0.02) \\
4473 & 183 (~7) & 197 (10) & 190 &  75 ( 4) & 0.38 (0.03) & 0.47 & 0.40
& -0.00  (0.02) & -0.11 (0.02) \\
4564 & 171 (~6) & 112 (17) & 162 & 147 ( 4) & 1.31 (0.20) & 1.18 & 0.55
& -0.01  (0.02) & -0.02 (0.04)  \\
4649 & 373 (~9) & 318 (37) & 375 &  93 (27) & 0.29 (0.09) & 0.65 & 0.16 &
0.00  (0.02) & -0.04 (0.02)  \\
\enddata
\tablenotetext{a}{Dispersion measured in central Modspec bin with error from
LOSVD fitting. }
\tablenotetext{b}{Dispersion averaged along the Modspec slit, major axis,
for $r >$ 2\farcs0, using symmetrized LOSVDs and an $\approx$ 1\farcs0 slit
width.
The error is the standard deviation of
averaged values.  Only three numbers are averaged for NGC 2778 (the 6
measurements
outside of $r=$2\farcs0 are symmetrized).
The other galaxies include 4$-$8 numbers in the average.  }
\tablenotetext{c}{ Effective dispersion: the rms velocity relative to the
systemic
velocity of the galaxy,
averaged over a slit width of $\sim$ 1\arcsec\ extending to $R_e$. }
\tablenotetext{d}{Maximum rotational velocity from symmetrized Modspec data.
Implausible maxima were discarded in the case of NGC 821, 4649, 4697 . }
\tablenotetext{e}{Ratio of maximum rotation to $\overline{\sigma}$.  The
$v_{max}$ comes from the ground-based data. }
\tablenotetext{f}{The ratio of $v_{max}$/$\overline{\sigma}$ to the
$v/\overline{\sigma}$ expected
for an oblate, isotropic rotator calculated as $\sqrt{\epsilon
/(1-\epsilon}$),
where $\epsilon$ is the isophote ellipticity. }
\tablenotetext{g}{Ellipticity $(1-b/a)$ from Bender (1988), Faber \etal
(1997)
(NGC 7457), Lauer \etal (1995) (NGC 3608), and Peletier \etal (1990) (NGC
2778).
We measured $\epsilon$ for NGC 4291 and 3384 using WFPC2 F555W images. }
\tablenotetext{h}{Fourth Gauss-Hermite coefficient from the $r$=0\farcs0 bin
of the major axis spectrum.  }
\tablenotetext{i}{Third Gauss-Hermite coefficient averaged over $r>2\farcs0
$
using the symmetrized, major axis LOSVD. The error is the standard deviation
of the mean. }
\label{kinresGND}
\end{deluxetable}

\begin{deluxetable}{lcrrrrrrrrrr}
\tabletypesize{\footnotesize}
\tablecaption{\sc Ground-based (M-D-M 2.4-m + Modspec) Kinematics}
\tablewidth{0pt}
\tableheadfrac{0.05}
\tablehead{
\colhead{Name}   &
\colhead{PA$_{rel}$\tablenotemark{a} }   &
\colhead{Sym?\tablenotemark{b} }   &
\colhead{r($\arcsec$)\tablenotemark{c} }   &
\colhead{$v$(\kms )\tablenotemark{d} }   &
\colhead{$\epsilon_v$  }   &
\colhead{$\sigma$ (\kms )\tablenotemark{e} }   &
\colhead{$\epsilon_{\sigma}$ }   &
\colhead{h3 \tablenotemark{f} }   &
\colhead{$\epsilon_{h3}$ }   &
\colhead{h4 \tablenotemark{g} }   &
\colhead{$\epsilon_{h4}$ }
}
\startdata
N821(Mg) & 0 & Yes &      0.0 &      $-$7. &    4. &  235. &    3. &   0.02 &
0.02 &  $-$0.07 &   0.02    \\
N821(Mg) & 0 & Yes &      0.6 &      25. &    7. &  231. &    6. &  $-$0.02 &
0.03 &  $-$0.01 &   0.03    \\
N821(Mg) & 0 & Yes &      1.2 &      51. &    6. &  215. &    6. &  $-$0.07 &
0.02 &   0.02 &   0.03    \\
N821(Mg) & 0 & Yes &      1.8 &      58. &    5. &  203. &    7. &  $-$0.07 &
0.03 &   0.05 &   0.03    \\
N821(Mg) & 0 & Yes &      2.7 &      59. &    5. &  204. &    5. &  $-$0.04 &
0.03 &   0.04 &   0.03    \\
N821(Mg) & 0 & Yes &      4.4 &      62. &    5. &  198. &    6. &  $-$0.04 &
0.03 &   0.04 &   0.03    \\
N821(Mg) & 0 & Yes &      7.1 &      71. &    5. &  187. &    7. &  $-$0.04 &
0.03 &   0.05 &   0.03    \\
N821(Mg) & 0 & Yes &     11.2 &      64. &    6. &  195. &    7. &   0.01 &
0.03 &   0.03 &   0.03    \\
\enddata
\tablenotetext{a}{Position angle of slit relative to the major axis.}
\tablenotetext{b}{If ``Yes", the spectra from the two sides of the galaxy
center have been symmetrized.}
\tablenotetext{c}{Radius, in arcseconds.}
\tablenotetext{d}{Line-of-sight velocity, in\kms , from fit to Equation 1
(see text).}
\tablenotetext{e}{Velocity dispersion, \kms, from fit to Equation 1 (see
text).}
\tablenotetext{f}{Third Gauss-Hermite coefficient.}
\tablenotetext{g}{Fourth Gauss-Hermite coefficient.}
\label{kintabGND}

\end{deluxetable}
%\tabletypesize{\tiny}
\tabletypesize{\footnotesize}
\begin{deluxetable}{llrrrrrrrrrr}
\tablecaption{\sc STIS-measured Kinematics from Ca II}
\tablewidth{0pt}
\tableheadfrac{0.01}
\tablehead{
\colhead{Name}   &
\colhead{PA\tablenotemark{a} }   &
\colhead{Sym?\tablenotemark{b} }   &
\colhead{r($\arcsec$)\tablenotemark{c} }   &
\colhead{$v$(\kms )\tablenotemark{d} }   &
\colhead{$\epsilon_v$  }   &
\colhead{$\sigma$ (\kms )\tablenotemark{e} }   &
\colhead{$\epsilon_{\sigma}$ }   &
\colhead{h3 \tablenotemark{f} }   &
\colhead{$\epsilon_{h3}$ }   &
\colhead{h4 \tablenotemark{g} }   &
\colhead{$\epsilon_{h4}$ }
}
\startdata
N821 & $-$150. & N &  0.00 &      $-$1. &   19. &  248. &   16. &   0.02 & 0.05 
&  $-$0.04 &   0.05 \\
N821 & $-$150. & N &  0.05 &    $-$65. &   19. &  273. &   21. &   0.06 &   0.05
&  $-$0.07 &   0.05  \\
N821 & $-$150. & N &  0.10 &   $-$101. &   20. &  244. &   20. &   0.05 &   0.05
&  $-$0.02 &   0.07  \\
N821 & $-$150. & N &  0.17 &    $-$81. &   15. &  186. &   12. &   0.04 &   0.03
&   0.00 &   0.05  \\
N821 & $-$150. & N &  0.30 &    $-$47. &   24. &  177. &   21. &   0.10 &   0.07
&   0.02 &   0.08  \\
N821 & $-$150. & N &  0.50 &    $-$47. &   32. &  206. &   19. &   0.05 &   0.07
&  $-$0.08 &   0.08  \\
N821 & $-$150. & N &  0.82 &   $-$143. &   51. &  233. &   47. &   0.09 &   0.15
&   0.04 &   0.16  \\
N821 & $-$150. & N & $-$0.05 &       5. &   20. &  229. &   16. &  $-$0.08 & 0.05 
&  $-$0.05 &   0.05 \\
\enddata
\tablenotetext{a}{Position angle of STIS slit in degrees.}
\tablenotetext{b}{If ``Y", the spectra on the two sides of the galaxy center
have been symmetrized.}
\tablenotetext{c}{Radius, in arcseconds.}
\tablenotetext{d}{Line-of-sight velocity, in\kms , from fit to Equation 1.}
\tablenotetext{e}{Velocity dispersion, \kms, from fit to Equation 1.}
\tablenotetext{f}{Third Gauss-Hermite coefficient.}
\tablenotetext{g}{Fourth Gauss-Hermite coefficient.}
\label{kintabSTIS}

\end{deluxetable}

\end{document}